\documentclass[trackchanges]{aastex701}
\usepackage{float}
\usepackage{amsmath} 
\usepackage{amssymb}
\usepackage{bm}
\usepackage{subcaption}
\usepackage{url}
\usepackage{natbib}
\usepackage{hyperref}
\usepackage{hypernat}

\newcommand{\vect}[1]{\bm{#1}}
\newcommand{\mat}[1]{\mathbf{#1}}

\begin{document}

\title{A Core-Collapse Supernova Neutrino Parameterization with Enhanced Physical Interpretability}

\correspondingauthor{Guoliang~L\"u and Xuefei~Chen}
\email{guolianglv@xao.ac.cn; cxf@ynao.ac.cn}

\author[0009-0007-9418-2632]{Haihao Shi}
\affiliation{
Xinjiang Astronomical Observatory, Chinese Academy of Sciences, Urumqi 830011, China
}
\affiliation{
School of Astronomy and Space Science, University of Chinese Academy of Sciences, Beijing 101408, China}
\email{shihaihao@xao.ac.cn}

\author[0009-0009-8188-5632]{Zhenyang Huang}
\affiliation{
Xinjiang Astronomical Observatory, Chinese Academy of Sciences, Urumqi 830011, China
}
\affiliation{
School of Astronomy and Space Science, University of Chinese Academy of Sciences, Beijing 101408, China}
\email{huangzhenyang@xao.ac.cn}

\author[0009-0004-7473-1727]{Junda Zhou}
\affiliation{Yunnan Observatories, Chinese Academy of Sciences,
             Kunming, Yunnan 650216, China}
\affiliation{
School of Astronomy and Space Science, University of Chinese Academy of Sciences, Beijing 101408, China}
\email{zhoujunda@ynao.ac.cn}

\author[0000-0002-3839-4864]{Guoliang~L\"u}
\email{guolianglv@xao.ac.cn}
\affiliation{School of Physical Science and Technology, Xinjiang University, Urumqi 830046, China}
\affiliation{
Xinjiang Astronomical Observatory, Chinese Academy of Sciences, Urumqi 830011, China
}

\author[0000-0001-5284-8001]{Xuefei~Chen}
\email{cxf@ynao.ac.cn}
\affiliation{International Centre of Supernovae, Yunnan Key Laboratory of Supernova Research, Yunnan Observatories, Chinese Academy of Sciences, Kunming 650216, China}
\affiliation{Key Laboratory for Structure and Evolution of Celestial Objects, Chinese Academy of Sciences, Kunming 650216, China}
\affiliation{
School of Astronomy and Space Science, University of Chinese Academy of Sciences, Beijing 101408, China}








\begin{abstract}
We introduce a novel parameterization of supernova neutrino energy spectra with a clear physical motivation. Its central parameter, $\tau(t)$, quantifies the characteristic thermal‐diffusion area during the explosion. When applied to the historic SN1987A data, this parameterization yields statistically significant fits and provides robust constraints on the unobserved low-energy portion of the spectrum. Beyond this specific application, we demonstrate the model's power on a suite of 3D core-collapse supernova simulations, finding that the temporal evolution of $\tau(t)$ distinctly separates successful from failed explosions. Furthermore, we constrain the progenitor mass of SN 1987A to approximately 19 solar masses by applying Smoothed Isotonic Regression, while noting the sensitivity of this estimate to observational uncertainties. Moreover, in these simulations, $\tau(t)$ and the gravitational-wave strain amplitude display a strong, synergistic co-evolution, directly linking the engine’s energetic evolution to its geometric asymmetry. This implies that the thermodynamic state of the explosion is imprinted not only on the escaping neutrino flux, but also recorded in the shape of the energy spectrum. Our framework therefore offers a valuable tool for decoding the detailed core dynamics and multi-messenger processes of future galactic supernovae.

\end{abstract}



\section{introduction}

The predicted rate of core-collapse supernovae (CCSN) within our Galaxy is estimated to be a few per century \citep{Cappellaro1993,1994ApJS...92..487T,Rozwadowska2020OnTR}. Despite this, the latest decade-long search by the IceCube Neutrino Observatory reported no such events \citep{Abbasi_2024}, a null result that underscores the importance of continuous readiness for the next galactic supernova. When a CCSN occurs, approximately 99\% of its gravitational binding energy is released in the form of neutrinos \citep{Huang_2021}. \textcolor{black}{Detecting these neutrino emissions provides key insights into the dynamics of supernova explosions \citep{burrows2021core,Guo:2023sbt,Obergaulinger:2021omt,Nagakura:2020qhb}, the neutrino mass hierarchy \citep{Huang:2023aob,esteban2025nufit}, limits on neutrino masses \citep{PhysRevLett.58.1906,Lu:2014zma,katrin2025direct}, and neutrino oscillations \citep{maki1962remarks,pagliaroli2010neutrino}.} The neutrino energy spectrum, in particular, serves as a powerful probe of the internal dynamics of the collapsing star, as its shape is forged by the varying depths and thermal conditions where different neutrino flavors decouple. Therefore, developing robust models to interpret these spectra is of paramount importance for maximizing the scientific return from a future detection.

In contemporary supernova neutrino studies, it is widely accepted that the spectra can be well approximated by the Keil--Raffelt--Janka (KRJ) parameterization, also referred to as the ``quasi-thermal distribution'' \citep{Keil_2003, Vitagliano:2019yzm, mirizzi2016supernova}:

\begin{equation}
\varphi(E)\propto A_{\text{KRJ}} \left(\frac{E}{\langle E \rangle}\right)^\alpha e^{-(\alpha+1) \frac{E}{\langle E \rangle}},
\end{equation}
where \( A_{\text{KRJ}} \) is the normalization factor, \( \langle E \rangle \) is the average energy, and \( \alpha \) is a phenomenological parameter. This parameterization is valuable, especially as $\alpha$ quantifies the degree to which the spectrum deviates from a thermal distribution. However, it faces significant limitations. Primarily, for the purpose of extracting the real-time explosion mechanism, the parameter $\alpha$ is purely phenomenological and lacks a direct quantitative mapping to an underlying physical model.

\textcolor{black}{In this article, we address these limitations by proposing a new analytical framework for the CCSN neutrino spectrum motivated from the first-principles physics of energy diffusion.} We model the complex energy transport within the post-shock region as a diffusion problem and, from its solution, derive an analytical form for the resulting energy spectrum. The key advantage of our framework is that it imbues the spectral parameters with clear physical meaning. Specifically, the spectral shape is primarily governed by a single parameter, $\tau$, which represents the integrated thermal diffusion area. This approach provides a direct physical basis for the phenomenological "pinching" parameter $\alpha$ of previous empirical models, thus creating a clear pathway to map an observed spectrum back to the thermodynamic conditions of the supernova engine. \textcolor{black}{While crucial neutrino oscillation effects (e.g., Mikheyev-Smirnov-Wolfenstein effect \citep{smirnov2003msw,PhysRevD.17.2369,Lai:2020sdv}}, collective oscillations \citep{Duan:2010bg,Mirizzi:2015eza}) are not explicitly included, our model is intentionally designed to characterize the final, observed spectrum rather than to simulate neutrino transport. Its purpose is to extract the source's underlying physical evolution from this post-oscillation signal. A deeper theoretical study connecting this extracted evolution back to the specifics of oscillation physics remains a compelling topic for future work.

In \autoref{Theoretical Framework}, we present the theoretical framework for deriving our spectral form. Subsequently, in \autoref{Applications and Results}, we demonstrate the applications of this framework, including a fit to the SN 1987A event and an analysis of modern 3D CCSN simulations. We discuss the strengths and limitations of our model, along with its future outlook, in \autoref{Discussion}, and provide a summary in \autoref{Conclusion}.

\section{Theoretical Framework}\label{Theoretical Framework}

Phenomenological models, such as the widely used KRJ parameterization \citep{Keil_2003}, are standard tools for analyzing neutrino energy spectra from core-collapse supernovae (CCSNe). While the KRJ model is effective and provides a meaningful physical insight, extracting deeper information requires a clearer understanding of the parameters of the neutrino energy spectrum. 

Here, we introduce a new framework that derives the spectral form from first principles by projecting the energy transport in the supernova environment onto a thermal diffusion process. This approach is physically grounded in the central role of neutrinos. As the most weakly interacting particles, neutrinos have the longest mean free path and therefore play an important role in transport. Their interactions with the medium strongly depend on the local thermodynamic conditions. In the extreme densities of the supernova interior, their mean free path can become much smaller than the core region, in which case they behave diffusively (“trapped”, contributing to thermal conductivity, shear viscosity, etc.). In the outer, less dense regions, they transition to behaving radiatively (“free streaming”, carrying off energy and lepton number) \citep{Alford:2025jtm}.

Based on this dominant diffusive behavior within the core, and motivated by the physical essence of a CCSN—an instantaneous release of enormous energy from a compact region \citep{Kneller:2014oea,2025MNRAS.541..116I,1946JApMM..10..241S}—we adopt an idealized representation, modeling the event as a point-source explosion embedded in a uniform medium, in the spirit of the piston \citep{Young:2006kq} or thermal-bomb approaches \citep{10.1093/mnras/stac3239}. Although this uniform-medium assumption is a simplification of the complex stellar structure, our aim is not to construct a detailed numerical simulation but to capture the dominant thermodynamic behavior and derive an analytic expression for the resulting energy spectrum. Crucially, when this expression is applied to describe observational or simulated data, the parameter for the medium's thermal conductivity is promoted from a constant to a dynamic, data-driven quantity. In this formulation, the neutrino spectral shape naturally emerges from the global energy distribution of the explosion, linking spectral parameters directly to physical quantities of the supernova interior. This approach therefore provides a more physically grounded alternative to previous phenomenological models and establishes a framework for analyzing observed neutrino spectra to infer the underlying physical conditions of core-collapse supernovae.

The energy distribution resulting from an instantaneous point-source injection is governed by the diffusion equation \citep{Macdonald1977}:

\begin{equation}
\frac{\partial E(x, t)}{\partial t} = \kappa' \frac{\partial^2 E(x, t)}{\partial x^2}, \quad x \in \mathbb{R}, \; t > 0
\end{equation}

Here, $\kappa'$ is the thermal diffusivity, defined as $\kappa' = \frac{\kappa}{\rho c_p}$, where $\kappa = \frac{-j}{\nabla T}$, with $j$ denoting the heat flux and $\nabla T$ the temperature gradient. \textcolor{black}{For clarity, although $\kappa'$ has a well-defined physical meaning, our fitting procedure is data-driven. In this setting, the contributions of the physical quantities entering $\kappa'$ are mutually degenerate in their observational impact, so the data alone do not allow us to disentangle the individual parameters or to perform a direct sensitivity analysis. Achieving this will require forward simulations with richer physical detail and greater fidelity.}

The initial condition describes an instantaneous release of total energy $Q$ at $x = 0$ \citep{HARRIS2014195} :

\begin{equation}
E(x, 0) = Q \, \delta(x),
\end{equation}

where $\delta(x)$ is the Dirac delta function, representing a localized energy injection at $t = 0$.
The goal of the subsequent work is to determine the spatiotemporal energy distribution as the system evolves.

After undergoing some involved calculations, we can obtain the spectral form for each time slice as well as the form of the integrated spectrum:

\begin{equation}\label{main}
\varphi(E,t) = Q \times \underbrace{A \left( \frac{E}{\langle E \rangle} \right)^{\frac{\tau(t) m  \langle E \rangle}{E^*}}}_{\text{state function}} \underbrace{\exp\left[ -\tau(t) \left( \frac{E}{\langle E \rangle^*} \right)^\beta \times \begin{cases} 
1 & \text{time-sliced spectrum} \, (\beta=2), \\ 
m & \text{time-integrated spectrum} \, (\beta=1). 
\end{cases} \right]}_{\text{distribution function}}
\end{equation}

Here, $Q$ denotes the explosion energy of the scenario under consideration, $A$ is the normalization constant of the energy spectrum, $E$ is the neutrino energy, and $\langle E\rangle$ is its mean value. The quantities $E^{*}$ and ${\langle E\rangle}^{*}$ represent the corresponding dimensionless forms. We introduce the parameter $\tau(t)=\int_{0}^{t}\kappa'(s)\,ds$, which physically corresponds to the characteristic thermal‑diffusion area, i.e., the square of the diffusion length $\ell_{\rm diff}^{2}=\kappa' t$; a larger $\tau$ indicates more efficient heat transport \citep{ahtt6e}. This parameter replaces the $\alpha$ parameter previously adopted in the Monte Carlo–based KRJ parameterization. Notably, the spectrum exhibits the common physical structure of a state function multiplied by an energy distribution function.

The distinction between the time-integrated and time-sliced spectra arises from the asymptotic reduction of the Gaussian-mixture exponential tail to a single exponential decay \citep{Grushka1972}. In our model, $m$ denotes the mass of particles that conduct energy within the medium. Although $m$ is fixed in the idealized model, its value is otherwise arbitrary; for computational convenience, we set $m=1$ while retaining dimensional units to ensure dimensional consistency of the equations. \textcolor{black}{The detailed computational procedures are provided in the \autoref{PA}}. 

For the KRJ parameterization, when the shape parameter is approximately 1.3, the spectrum reduces to a Fermi–Dirac distribution \citep{Keil_2003}. In contrast, our model does not exhibit this behavior. This difference arises from its physical motivation: our framework is designed to describe global thermodynamic variations driven by energy transport in CCSNe, whereas a Fermi–Dirac distribution implies a fully thermalized state. Weak interaction timescales in the core are much shorter than the dynamical timescale. This allows neutrinos to reach an thermalized state in regions where they are trapped by the dense medium, such as inside the neutrinosphere, but this condition does not hold across the entire supernova environment \citep{kotake2006explosion,mezzacappa2020physical}. This limitation explains why \autoref{main} cannot reduce to a Fermi–Dirac distribution.

\section{Applications and Results}\label{Applications and Results}

In this section, we apply the theoretical framework developed above to both observational data and state-of-the-art numerical simulations. We first validate our model by fitting the neutrino spectrum from the landmark event SN1987A. Subsequently, we leverage our model as a diagnostic tool on a large suite of 3D core-collapse supernova simulations to uncover novel physical correlations and explore its potential for multi-messenger astronomy.


\subsection{Fitting the SN1987A Spectrum}\label{Fitting the SN1987A Spectrum}

\textcolor{black}{\autoref{SN1987A_event} in \autoref{PB} shows the time–energy event data recorded during the SN 1987A burst \citep{dos2022understanding}. We integrate this dataset over time, explicitly accounting for the energy-dependent uncertainty of each detector. This uncertainty is shown in \autoref{PB} \autoref{Detector_Resolution_with_energy} and is reflected in the error bars of \autoref{spectrum_fit_with_errorbars}}

For spectral comparison, we adopt the non-parametric reconstruction of the SN 1987A neutrino spectrum from \citet{Yuksel:2007mn}, which is derived solely from detector response parameters and is therefore free of model-dependent assumptions. Because this reference spectrum is provided as a smooth curve without statistical error bars, we combine it with the energy uncertainties independently derived from the raw event data to construct a hybrid dataset suitable for quantitative analysis. In this part of the discussion, we use the same number spectrum as \citet{Yuksel:2007mn}; consequently, the normalization parameter $Q$ in this section directly corresponds to the total neutrino number.

The best-fit spectrum obtained was compared with the reference SN1987A neutrino spectrum:

\begin{figure}[H]
\centering
\includegraphics[width=1\linewidth]{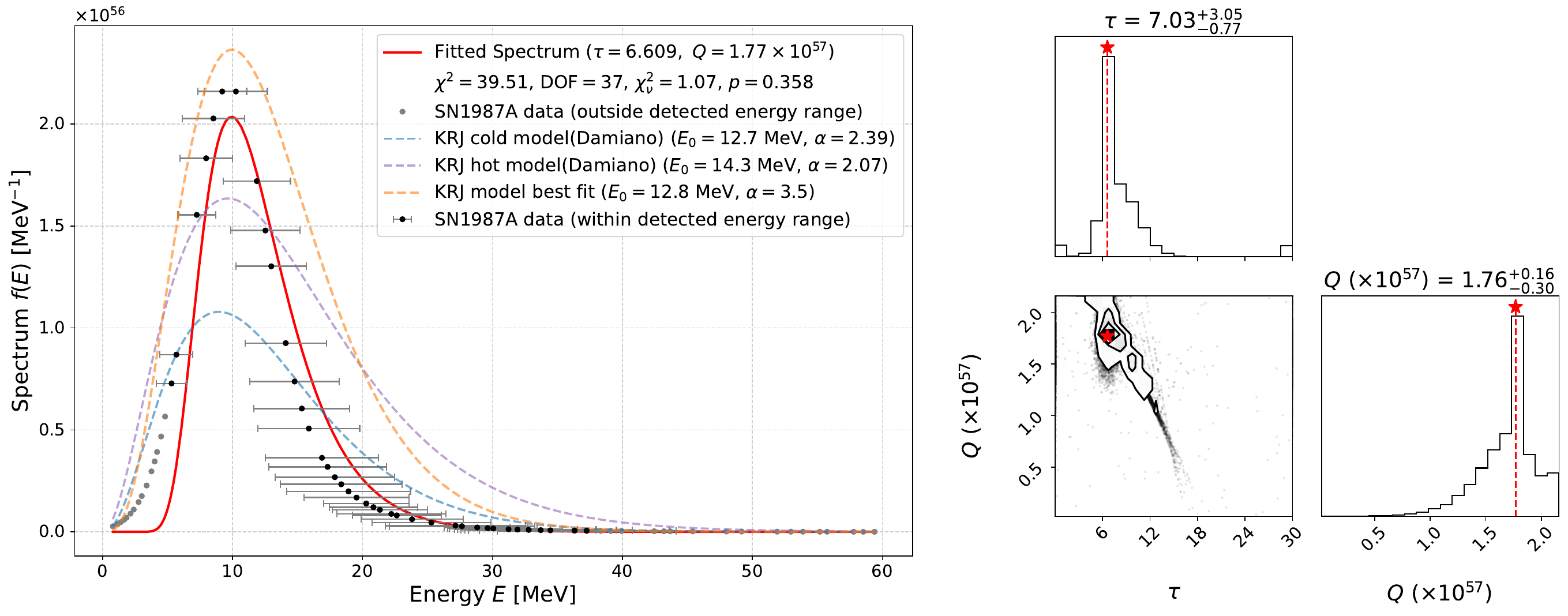}
\caption{
\textcolor{black}{Left: SN1987A detected spectrum (black) with our best-fit model (red solid line), obtained using a Genetic Algorithm with Empirical Bayesian Guided Evolution to enable rapid convergence within a vast and largely unexplored parameter space (computational details are provided in the \autoref{PB})}. The fit shows good agreement with the data and is statistically significant. For comparison, we also plot the cold (green dashed) and hot (purple dashed) KRJ parameterized models from \citet{Fiorillo:2022cdq}, as well as the KRJ model best fit curve obtained using the KRJ form and the same optimization method as the red line.
Right: Corner plot of the posterior distributions of the parameters $\tau$ and $Q$, derived from the same optimization process. The red star marks the best-fit values, and the distributions indicate that $\tau$ is well constrained, while $Q$ has a relatively broader uncertainty range with a negative correlation between them.
}

\label{spectrum_fit_with_errorbars}
\end{figure}



The fitted spectrum is statistically consistent with the measurements. Our best-fit model yields a chi-squared value of $\chi^2 = 39.51$ for 37 degrees of freedom, giving a reduced chi-squared of $\chi^2_\nu = 1.07$ and a corresponding $p$-value of 0.358. For a significance level of $\alpha = 0.01$, the critical rejection threshold is $\chi^2 > 59.89$ \citep{NIST_eHandbook}; since our computed value lies well below this threshold, we cannot reject the null hypothesis, indicating that the model provides a statistically acceptable fit. Within the observational energy range, the model shows good agreement with the data, particularly in the mid‑ to high‑energy region, where it captures deviations from a thermal equilibrium spectrum especially well. However, it deviates at lower energies ($\lesssim 5~\mathrm{MeV}$), where no observational data are available and, consequently, no uncertainties can be assigned. This deviation could stem from detector threshold effects or, more intriguingly, from an intrinsic deficit of low‑energy neutrinos—a testable prediction of our model that, thanks to advances in coherent elastic antineutrino–nucleus scattering detection \citep{Ackermann:2025obx}, may soon be within reach of future sensitive observations. For comparison, we include several benchmark spectra in \autoref{spectrum_fit_with_errorbars}. First, we plot the reference models from \citet{Fiorillo:2022cdq}: a "cold" model (blue dashed line) with total energy $E_{\mathrm{tot_{\text{cold}}}} = 1.98\times 10^{53}~\mathrm{erg}$, mean energy $\langle E_{\text{cold}}\rangle = 12.7~\mathrm{MeV}$, and $\alpha_{\text{cold}} = 2.39$; and a "hot" model (purple dashed line) with $E_{\mathrm{tot_{\text{hot}}}} = 3.93\times 10^{53}~\mathrm{erg}$, $\langle E_{\text{hot}}\rangle = 14.3~\mathrm{MeV}$, and $\alpha_{\text{hot}} = 2.07$. The performance of the cold/hot model in characterizing data points as the standard spectrum is not as good compared to the red solid line (our work). When calculating, note that $E_{\mathrm{tot}}$ is the sum of all neutrino flavors, and averaging should be performed before calculating the energy spectrum of a specific neutrino.

In addition, we perform our own best-fit of the KRJ model to the data, using the same optimization method as for our main result (red solid line). This yields a best-fit curve (orange dashed line) with parameters $E_{\mathrm{tot}} = 3.97\times 10^{53}~\mathrm{erg}$, $E_{0} = 12.8~\mathrm{MeV}$ and $\alpha = 3.5$. We find that while the KRJ fit is reasonable near the mean energy, it deviates significantly at both the low- and high-energy tails of the spectrum. It is important to note that the judgment of the fitting effect depends on the way the energy spectrum is reconstructed. Here, choosing a non-parametric reconstruction based on detector performance as the benchmark is a robust option, but this does not mean it is absolutely correct. This choice is made in situations with insufficient data.

\subsection{Statistics Based on 3D Core-Collapse Supernova Simulation Data}

Having demonstrated our model's validity against the archetypal SN1987A event, we now turn to a comprehensive set of 3D simulations to systematically explore the physical information encoded in its parameters. To ensure parameter reliability, we apply the fitting procedure described in \autoref{main} to each spectrum at every time step, yielding cosine similarities greater than 0.99 and median relative errors of approximately -5\%, thereby providing confidence in the robustness of these spectral parameters. For SN1987A, where observational uncertainties are well defined, we adopted a standard $\chi^2$ test to quantify the statistical consistency of the fit. In contrast, for the 3D simulations—where data are noiseless, error bars are absent, and the sample size is enormous—we instead employed cosine similarity and relative error as computationally efficient proxies to ensure robustness of the spectral parameterization. This error is far smaller than that of the observational fit in \autoref{spectrum_fit_with_errorbars}, where instrumental uncertainties, noise, and unknown physics explain the larger SN1987A fitting error without affecting the conclusions of \autoref{Fitting the SN1987A Spectrum}. 

In this work, we separately analyze the spectral parameters of $\bar{\nu}_e$ and $\nu_e$ from the dataset due to their distinct observational and physical relevance. In \autoref{linear relationship} and \autoref{Constraining the Progenitor Mass of SN1987A}, we focus on $\bar{\nu}_e$ because their energy spectra can be directly compared with the rare observational data from SN 1987A, enabling a joint analysis that constrains model parameters.In contrast, \autoref{Physical Insights from Parameter Evolution} and \autoref{Co-evolution} focus on $\nu_e$ spectral parameters, which are more tightly coupled in time to the internal dynamics of core-collapse supernovae—as evidenced by the evolution of angle-averaged $\nu_e$ luminosities in \citet{Choi:2025igp}—thus serving as effective probes of the explosion's hydrodynamical evolution.

\subsubsection{Approximate linear relationship between $\tau(t)$ and Q}\label{linear relationship}

\begin{figure}[H]
    \centering
    \begin{subfigure}[b]{0.45\textwidth}
        \centering
        \includegraphics[width=\linewidth]{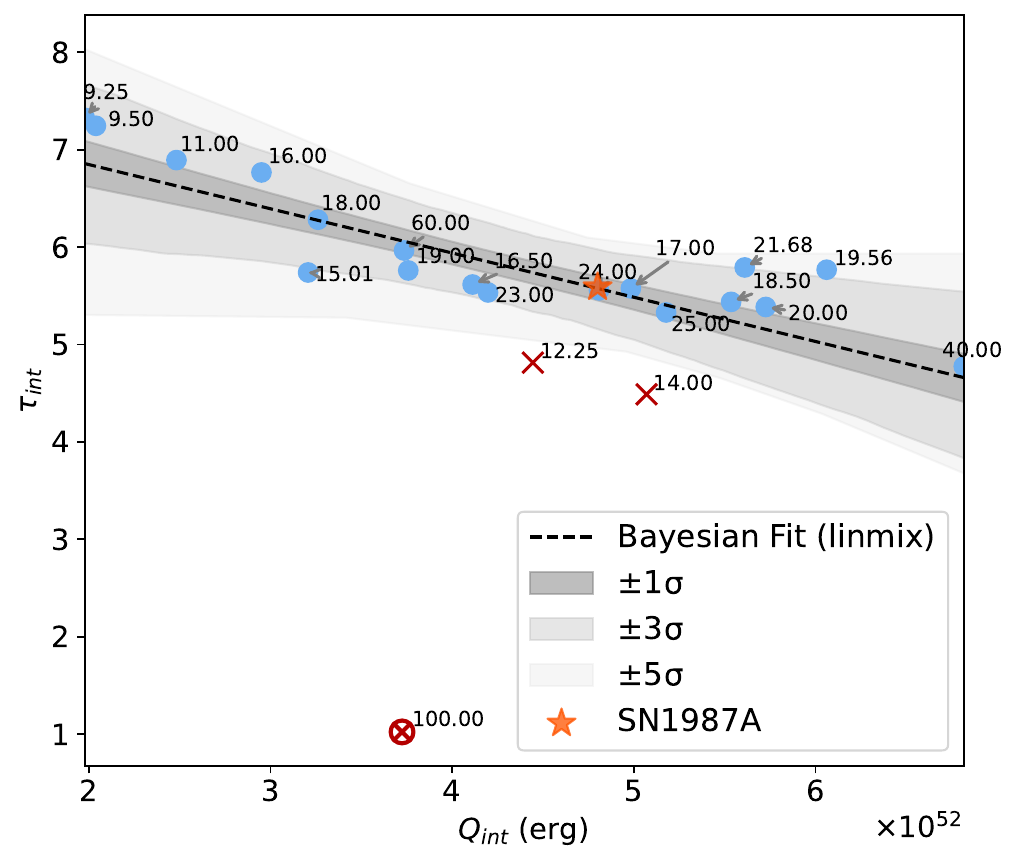}
        \caption{}
        \label{fig:3a}
    \end{subfigure}
    \begin{subfigure}[b]{0.45\textwidth}
        \centering
        \includegraphics[width=\linewidth]{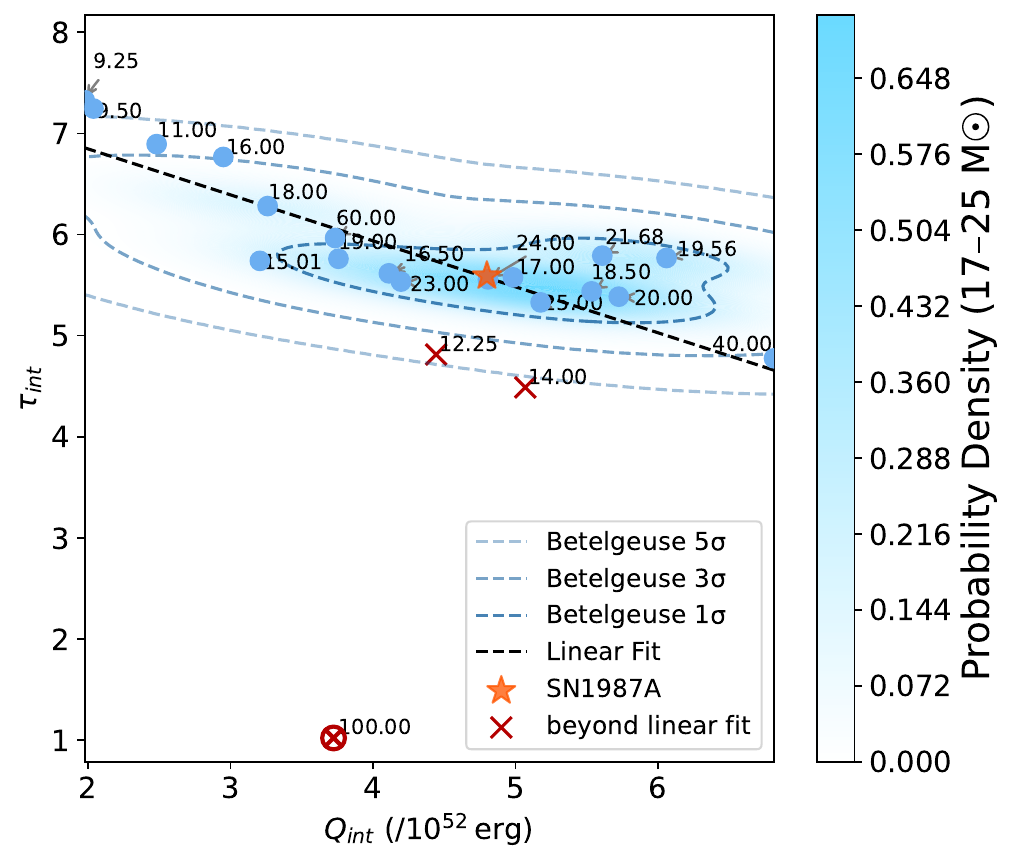}
        \caption{}
        \label{fig:3b}
    \end{subfigure}
\caption{
Horizontal axis: total neutrino energy released within 2\,s of explosion, $Q_{\mathrm{int}} = \int_0^{2\,\mathrm{s}} Q(t)\,\mathrm{d}t$; 
vertical axis: time-integrated spectral parameter, $\tau_{\mathrm{int}} = \int_0^{2\,\mathrm{s}} \tau(t)\,\mathrm{d}t$. 
Numerals denote progenitor masses in $M_\odot$. 
(a) Shaded bands indicate the statistical uncertainty of the linear regression fit. 
(b) Dashed contours correspond to the 1$\sigma$, 3$\sigma$, and 5$\sigma$ credible regions in probability density, delineating the parameter space associated with Betelgeuse’s plausible progenitor mass range. 
Each contour represents the projected confidence boundary in its neutrino spectral parameters for a future explosion.
}

    \label{fig:trends}
\end{figure}

Using the 3D core-collapse supernova simulation data from \citet{Choi:2025igp}, we fit the resulting spectra with our model to extract the key parameters $Q$ and $\tau$ for each progenitor model, aiming to assess whether these physically motivated parameters can distinguish exploding from non-exploding progenitors. The simulated spectra are integrated over the first two seconds (with 0.01 s resolution) to obtain the distribution shown in \autoref{fig:3a}. This two-second integration window was chosen specifically to capture the most dynamic phase of the spectral evolution; including the subsequent, more quiescent phase was found to dilute the strength of the underlying linear correlation.

Blue points denote models well described by the linear fit, crosses mark non‑exploding progenitors, and circled crosses indicate shock revival followed by failure. The linear relation, obtained through a Bayesian \texttt{linmix} regression \citep{2007ApJ...665.1489K,LinMix_python}, captures the exploding models within a $5\sigma$ credible interval and thus delineates a parameter regime predictive of explosion outcomes (\textcolor{black}{see \autoref{fig:corner} in \autoref{PC} for detailed fit parameters}). The Bayesian linear regression converges robustly and yields a slope posterior of $\beta = -3.69^{+0.77}_{-0.79}$,
which is approximately $5.4\sigma$ excluded from zero, confirming a highly significant negative linear correlation. The fitted relation is:

\begin{equation}\label{line}
\tau_{int} = \left(7.76_{-0.39}^{+0.38}\right) - \left(4.56_{-0.83}^{+0.86}\right)\times 10^{-53}\,Q_{int}.
\end{equation}

Because the two parameters differ greatly in magnitude, $Q$ is divided by a characteristic scale in the fitting to balance their orders of magnitude. Although only the parameters of the $\bar{\nu}_e$ energy spectrum are discussed here, we have also performed the same calculations for the $\nu_e$ energy spectrum, which similarly follows a similar linear relationship; this will not be elaborated on here. 

\autoref{line} presents the empirical relation between the energy deposition within the first two seconds of collapse, $Q_{\rm int}$, and the time-integrated thermal diffusion scale, $\tau_{\rm int}$, for successfully exploding models. The intercept mainly acts as a dimensional baseline and has limited physical interpretability: it corresponds to the diffusion efficiency required when $Q_{\rm int}\!\to\!0$, which is clearly an unrealistic limit.
The negative slope encodes the trade-off between energy deposition and transport: within the fitted interval, larger energy release allows explosion at systematically lower $\tau_{\rm int}$, the numerical value of the slope characterizes the intensity of this trade-off, and the specific physical meaning requires further discussion in conjunction with microscopic transport and precursor structure. This linear relation delineates the “explodability region”: models with high energy but insufficient transport (too small $\tau_{\rm int}$) fail to explode, whereas those with modest energy but highly efficient transport (larger $\tau_{\rm int}$) can still succeed.
Importantly, explosions lying significantly above the regression band are not expected: along the evolutionary trajectory in parameter space, any attempt to overshoot this boundary would necessarily pass through the allowed region and trigger explosion earlier. Hence, our conclusion is confined to the parameter range sampled by current models, without extrapolation beyond it.

For SN 1987A, the limited temporal resolution of the observed neutrino events prevents a reliable reconstruction of the time-dependent spectral parameter $\tau(t)$, and thus makes it difficult to directly estimate the integrated quantity $\Sigma\tau$ required by the trend shown in \autoref{fig:trends}. Instead, we evaluate the linear relation at the observed total energy $Q = 4.8 \times 10^{52}$ erg for SN 1987A (This value is obtained from integrating the baseline data shown in \autoref{spectrum_fit_with_errorbars}), yielding a corresponding vertical-axis value of approximately 5.6. 

Considering the possible progenitor mass range of Betelgeuse ($17$–$25~M_\odot$) \citep{dolan2016evolutionary}, we filter the data and perform a kernel density estimation (KDE) analysis (\autoref{fig:3b}). This analysis yields only weak constraints on Betelgeuse’s spectral parameter space, reflecting the limited sampling of progenitor masses in the current dataset and underscoring the need for additional data to strengthen the statistical interpretation.

\subsubsection{Constraining the Progenitor Mass of SN1987A}\label{Constraining the Progenitor Mass of SN1987A}

We constrain the progenitor mass of SN 1987A by comparing its effective spectral parameter with results from 3D CCSN simulations. It is worth noting that the SN 1987A data point lies remarkably close to the simulated model with a 24 $M_\odot$ progenitor in the parameter space shown in \autoref{fig:trends}. While this proximity is suggestive, it does not warrant a direct inference of equal progenitor masses, given the observational uncertainties, intrinsic scatter in the simulations, and potential degeneracies in the parameter space. To obtain a more robust quantitative constraint, we compute for each model the product of two spectral parameters, which represents the effective area over which the thermal flux, carrying explosion energy $Q$, acts within time $t$ on the diffusion scale $\ell_{\rm diff}$; larger values indicate more efficient heat transport. We observe a marked difference in the parameter distributions for progenitors with masses above and below $40\,M_\odot$, suggesting a potential threshold where the details of the explosion mechanism may change. Given that spectroscopic analyses place SN 1987A in the lower-mass regime, we therefore focus our analysis on this subset.

\begin{figure}[H]
  \centering
  \includegraphics[width=0.8\linewidth]{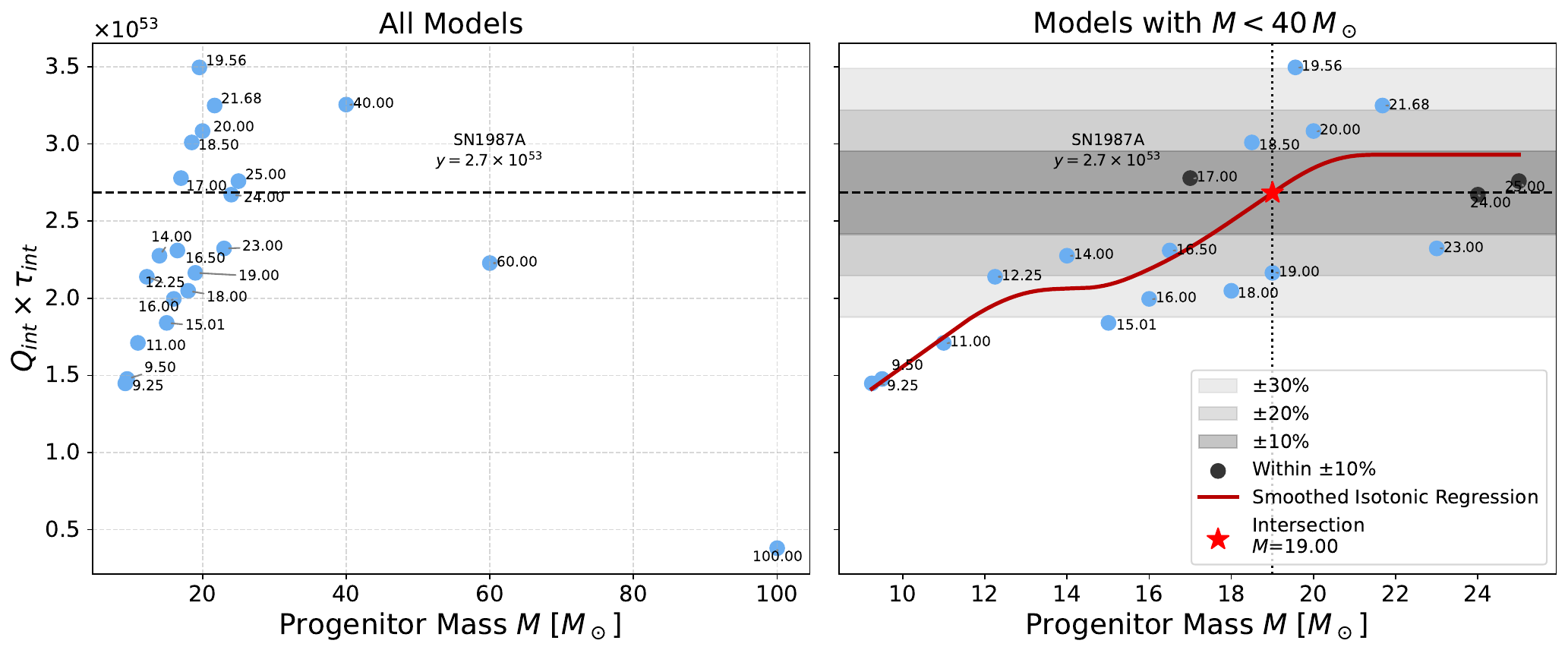}
\caption{
Left: Spectral parameter versus progenitor mass, with the dashed line marking SN~1987A’s reference value. Models below and above $40\,M_\odot$ exhibit distinct distributions. 
Right: Smoothed Isotonic Regression applied to the low-mass models (red curve), with its intersection with SN~1987A’s value (red star) yielding an estimated progenitor mass of $\sim17\,M_\odot$. Shaded regions indicate 10\%, 20\%, and 30\% parameter uncertainties.
}

  \label{fig:Estimated_Mass_SN1987A}
\end{figure}

To estimate the progenitor mass of SN 1987A within our framework, we first convert the parameter $Q$ used in \autoref{linear relationship} from a neutrino-number-based quantity to one consistent with the 3D simulation data, namely the neutrino luminosity in erg units, yielding a value of approximately $5 \times 10^{52} \mathrm{erg}$. Applying Smoothed Isotonic Regression \citep{Isotonic} to the low-mass subset then gives a best-estimate progenitor mass of $19M_\odot$. With a 10\% parameter uncertainty, the inferred mass range is consistent with stellar-evolution estimates \citep{1989ARA&A..27..629A,2016ApJ...821...38S}, while a 30\% uncertainty broadens the allowed range to nearly encompass all available models, highlighting the critical role of measurement precision in constraining progenitor properties. \textcolor{black}{It should be emphasized that this result is only applicable to the distribution obtained under the current simulated data scenario, as different research groups can produce different results for the same parent star in their simulations.}

We employ Isotonic Regression to model the relationship between the derived physical parameter and progenitor mass because the simulation data exhibit a distinct non-linear trend, with a noticeable inflection around $40\,M_\odot$. While piecewise linear regression could capture this behavior, it requires specifying a break point and imposes rigid assumptions on the functional form. In contrast, Isotonic Regression is a non-parametric approach that adaptively follows the monotonic trend in the data without presuming a specific model, thereby preserving the intrinsic structure of the underlying physical process. For the numerical implementation, we made use of the \texttt{IsotonicRegression} module from the Python package \texttt{scikit-learn} \citep{scikit-learn} together with the \texttt{Statsmodels} package \citep{seabold2010statsmodels}.

\subsubsection{Physical Insights from Parameter Evolution}\label{Physical Insights from Parameter Evolution}

\begin{figure}[H]
    \centering
    \begin{subfigure}[b]{0.45\textwidth}
        \centering
        \includegraphics[width=0.85\linewidth]{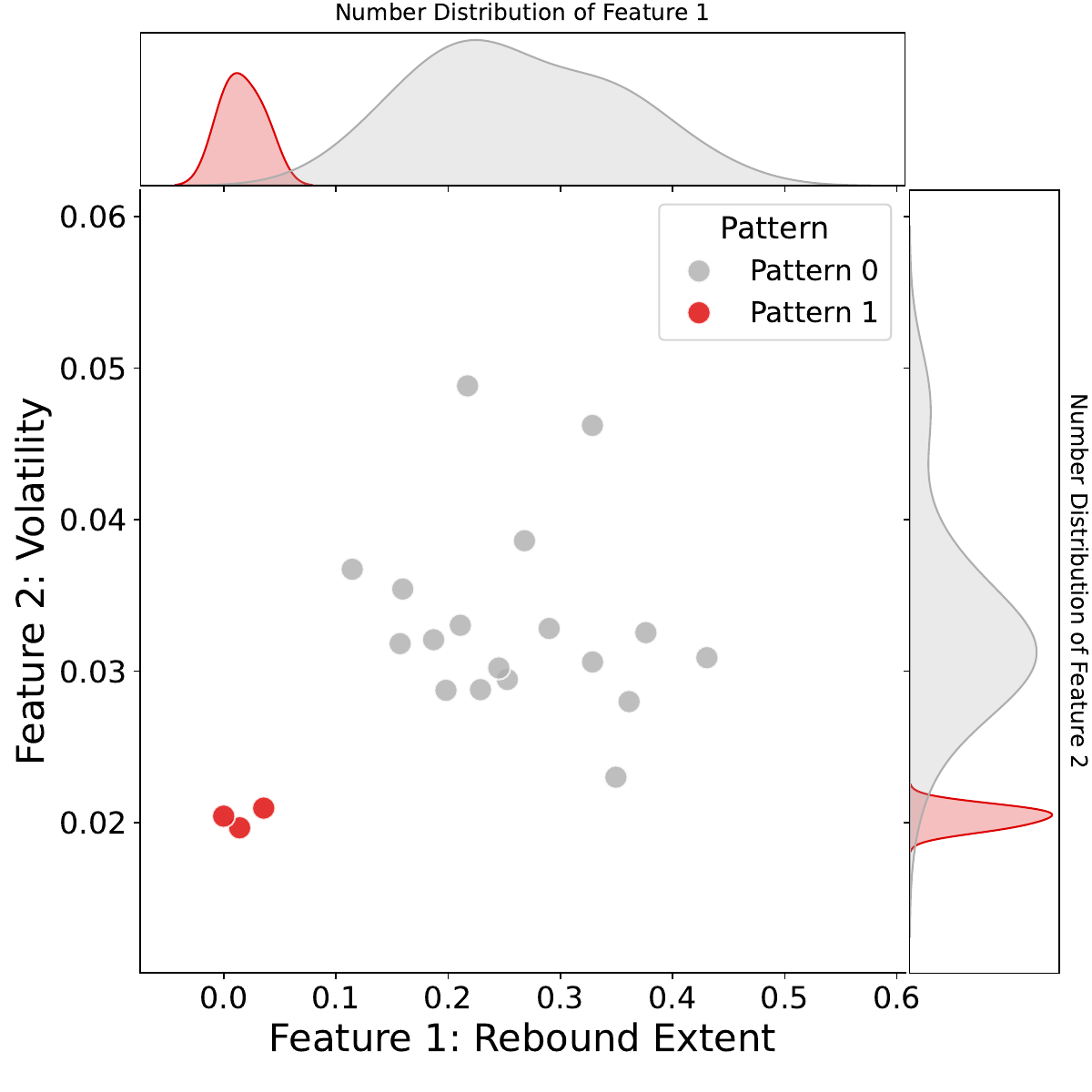}
        
        \caption{}
        \label{cluster:a}
    \end{subfigure}
    \begin{subfigure}[b]{0.45\textwidth}
        \centering
        \includegraphics[width=0.85\linewidth]{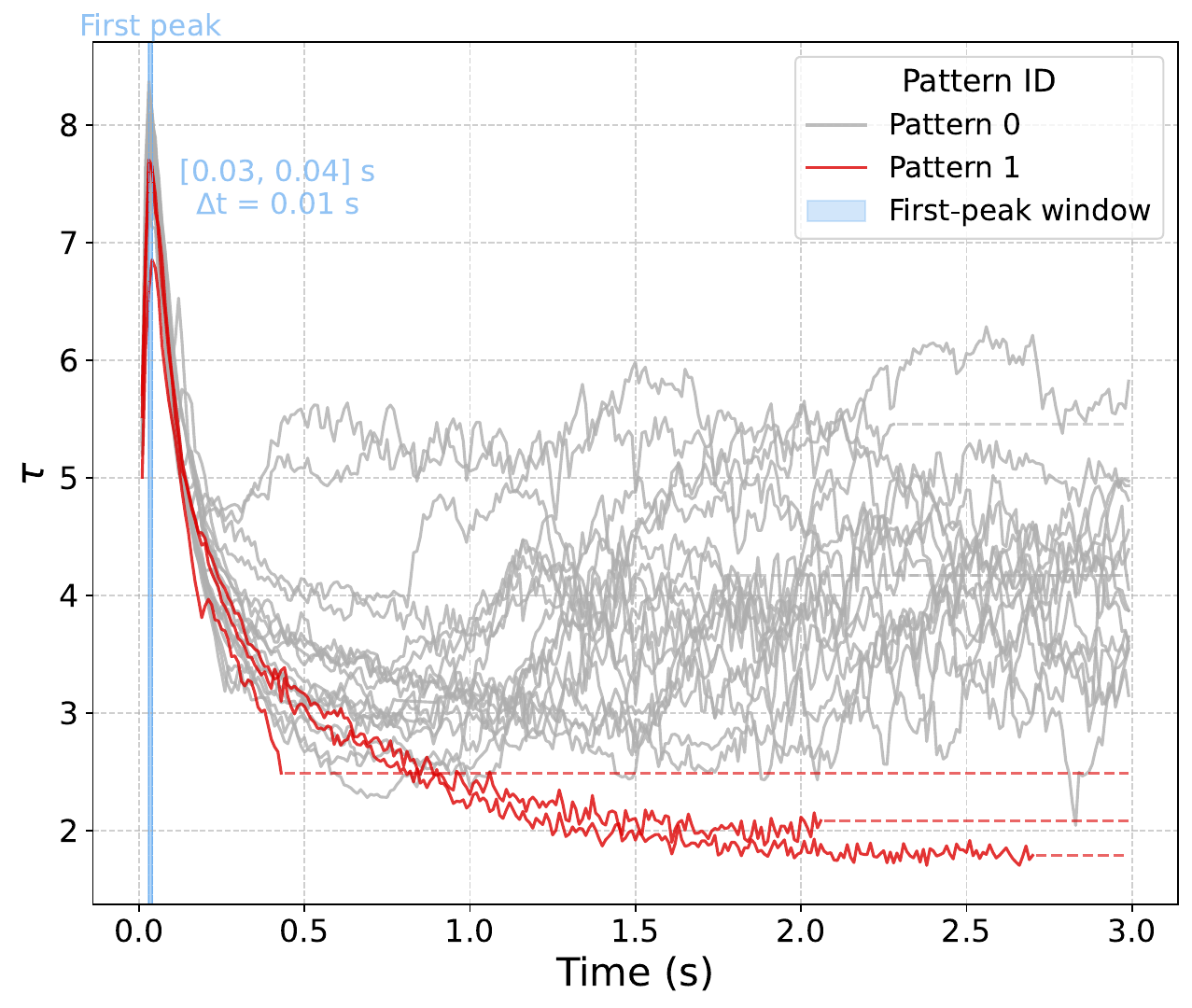}
        \caption{}
        \label{cluster:b}
    \end{subfigure}
\caption{The temporal evolution of $\tau$ serves as a robust diagnostic of explosion outcome. (a) Unsupervised clustering based on engineered features of the $\tau(t)$ curves—specifically their volatility and rebound extent—separates successful (Pattern 0) and failed (Pattern 1) explosions into distinct clusters. The two KDE plots show the number distributions of Feature 1 and Feature 2 for different clustering patterns.
 (b) Representative $\tau(t)$ curves for each cluster, with blue shading indicating the time window of the first pronounced peak. Dashed curve segments are not physically meaningful, representing constant extrapolation beyond the simulation endpoint for visual clarity.}
    \label{fig:cluster}
\end{figure}

While the classification capability demonstrated above validates our model, its greater potential lies in using neutrinos as unique messengers to probe real-time core physics through parameter evolution. Analyzing the time series of $\tau$, $Q$, gravitational-wave amplitude, and mean energy reveals distinct behaviors: in successful supernovae, $\tau$ rapidly recovers to a high, fluctuating level after the initial outburst, whereas in failed cases it declines monotonically. Considering the significant advantages of unsupervised learning in astronomical data pattern recognition \citep{FOTOPOULOU2024100851}, we employ its clustering algorithms to analyze these temporal features. As shown in \autoref{fig:cluster}, this method can clearly distinguish between successful and failed supernova explosions. In contrast to the analyses presented in \autoref{linear relationship} and \autoref{Constraining the Progenitor Mass of SN1987A}, our focus now shifts to the time-dependent evolution of $\nu_e$ spectral parameters during the first three seconds post-bounce. This temporal window is particularly significant, as the $\nu_e$ signal is most tightly coupled to the core's internal dynamics, and it isolates the crucial pre-cooling phase of the proto-neutron star, where the explosion mechanism is most active \citep{Janka2017}.

We apply an unsupervised clustering algorithm to the post-explosion evolution of the energy transport timescale, $\tau(t)$, to classify our simulated models. The classification is driven by two derived temporal features: a late-time rebound, defined as the normalized signal difference between the final and minimum states, and its volatility, the standard deviation of the signal's first-order difference. This method robustly separates the models into successful (Pattern 0) and failed (Pattern 1) explosions. In \autoref{cluster:a}, we present a JointGrid visualization that simultaneously shows the feature space and the marginal distributions of the clustered data. The central scatter plot depicts the joint distribution of Feature 1 (rebound extent) and Feature 2 (volatility), with points colored by clustering pattern. The KDE plots above and to the right illustrate the distribution profiles of Feature 1 and Feature 2, respectively, revealing how each feature is distributed across different patterns and enabling direct comparison of their statistical characteristics. Failed explosions are quantitatively identified by near-zero rebound and low volatility, consistent with continuous, monotonic accretion. In contrast, successful explosions occupy a distinct region of the feature space, characterized by high rebound and volatility.

\textcolor{black}{\autoref{fig:Heartbeat} in \autoref{P:Parameter Evolution} shows the frequency-domain characteristics of the evolving neutrino spectral shape parameter, $\tau$. Panel (a) displays the power spectral density (PSD) for each progenitor mass, stacked along the y-axis.} A striking dichotomy is immediately apparent: models corresponding to successful explosions exhibit rich spectral structures with significant power in the wide-frequency domain (approximately 5-45 Hz). In stark contrast, models that fail to explode appear as quiescent, dark bands with a notable deficit of power, as highlighted by the red boxes. To quantify this visual difference, the lower panels show the mean spectra for these two distinct patterns. Pattern 0, representing the average of the successful explosions, is characterized by significant, broadly distributed power across the entire frequency range shown. Conversely, Pattern 1, the average of the failed events, exhibits a near-zero power floor. This stark visual and quantitative separation reveals a strong correlation between the spectral properties of the $\tau$ parameter's evolution and the final outcome (success or failure) of the core-collapse simulation.


The pronounced dichotomy in these quantitative features suggests that the post‑explosion evolution of $\tau(t)$ serves as a “fossil record” of the preceding shock‑revival dynamics. We hypothesize that the strong rebound and sustained fluctuations observed in successful explosions reflect hydrodynamic conditions seeded by large‑scale turbulence during shock revival, whereas the monotonic decay and weak variability in failed cases indicate the absence of such a violent transition. Moreover, the interval from the first peak to the rebound marks the critical window transitioning from the stalled accretion shock phase to renewed shock heating. Notably, the first pronounced peak in all $\tau(t)$ curves emerges within a narrow time window (0.03–0.04 s), indicating that the initial neutrino outburst follows a characteristic timescale largely independent of progenitor mass or structure. This feature likely corresponds to the neutronization burst of $\nu_e$, which lasts for $\mathcal{O}(10)$ ms—the timescale over which the shock wave crosses the region around the neutrinosphere shortly after core bounce.

These contrasting patterns motivate two speculative but physically motivated interpretations: (1) a transition in the dominant energy transport regime—from a prompt, transient burst to a turbulence-supported dynamic equilibrium; and (2) the relaxation of a self-organized critical system returning to marginal stability following a large-scale release event. While these interpretations remain conjectural, they suggest that post‑explosion temporal diagnostics could provide indirect indications of the otherwise difficult to observe mechanisms governing shock revival in CCSNe.



\subsubsection{Co-evolution of $\tau(t)$ and gravitational-wave strain amplitude}\label{Co-evolution}

In CCSNe, we identify a clear co-evolutionary relationship between the diagnostic parameter $\tau$ and the total gravitational-wave (GW) amplitude $\mathrm{amp} = \sqrt{h_{+}^{2} + h_{\times}^{2}}$. To characterize this dynamic relationship, we adopt a state-space approach based on the Unscented Kalman Filter (UKF) \citep{882463}. Given the strongly nonlinear nature of the physical processes in supernova explosions (e.g., fluid dynamics), the UKF excels at capturing the underlying nonlinear dynamical trends of fluctuating time series, enabling accurate state estimation and making it particularly well-suited for such problems \citep{882463}. The key principle is to bypass direct approximation of the nonlinear functions and instead employ a deterministic sampling strategy: a set of representative sample points, known as sigma points, is selected around the current state to fully capture the characteristics of its probability distribution (e.g., mean and covariance) \citep{1271397}. These points are then directly propagated through the state-transition model, thereby yielding a more accurate estimation of the new state and its uncertainty. Unlike the various smoothing techniques commonly used in data processing, the UKF is more focused on capturing the dynamic evolutionary trend. As a result, the resulting trend line does not necessarily pass exactly through all data points. This is because it is not merely a “smoother” curve, but rather a combination of system dynamics–based empirical prediction and measurement statistics–based error correction, which iteratively yields the optimal state/parameter estimates. This explains why some of the processed trend lines do not perfectly intersect all the data points.  For the numerical implementation, we made use of the \texttt{UnscentedKalmanFilter} module from the Python package \texttt{filterpy} \citep{Labbe2015KalmanBayesian}.

We define a model-based UKF derivative correlation ($\rho'_{\mathrm{UKF}}$), which quantifies the correlation between the first time derivatives (i.e., evolution rates) of the two signals. This metric is extracted directly from the final state covariance matrix of the UKF. As established by optimal state estimation theory, the final covariance matrix $\mathbf{P}$ provides the estimates for the variances of the state variables on its diagonal (e.g., $P_{1,1}$) and the covariances between them on its off-diagonals (e.g., $P_{1,3}$) \citep{Simon2006Kalman}. Therefore, the correlation coefficient is computed from its standard statistical definition as:
\begin{equation}
\rho'_{\mathrm{UKF}} = \frac{\mathbf{P}_{1,3}}{\sqrt{\mathbf{P}_{1,1} \cdot \mathbf{P}_{3,3}}}, 
\end{equation}
where $\mathbf{P}$ denotes the final state covariance matrix, and indices $1$ and $3$ correspond to the velocity components of $\mathrm{amp}$ and $\tau$, respectively. As a robust metric, its value lies in the range $[-1, 1]$, with values closer to $1$ indicating stronger positive correlation and those closer to $-1$ indicating stronger negative correlation. \textcolor{black}{The computational details are provided in the \autoref{P:Parameter Evolution}.}

\begin{figure}[H]
    \centering
    \begin{subfigure}[b]{0.60\textwidth}
        \centering
        \includegraphics[width=\linewidth]{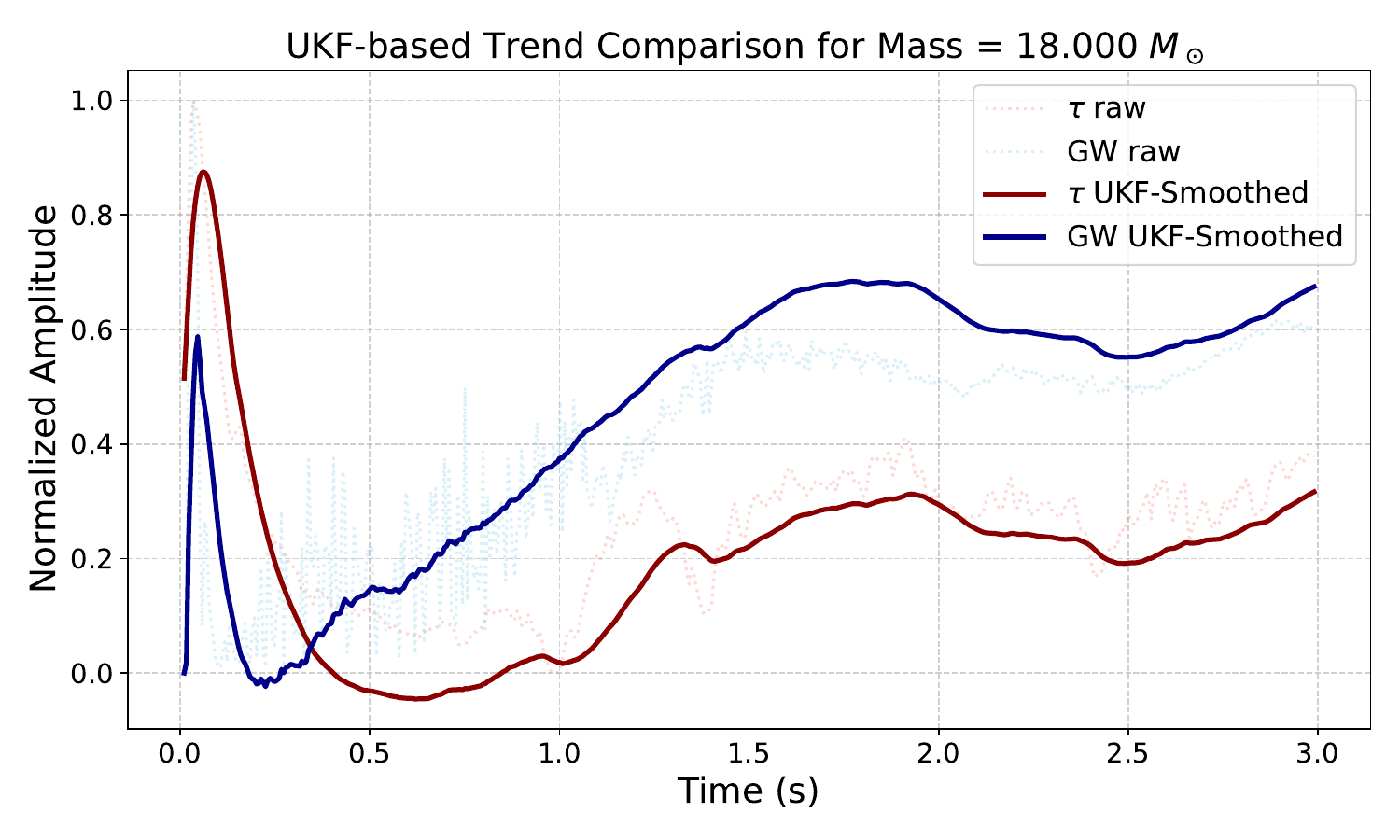}
        \caption{}
        \label{fig:a}
    \end{subfigure}
    \begin{subfigure}[b]{0.35\textwidth}
        \centering
        \includegraphics[width=\linewidth]{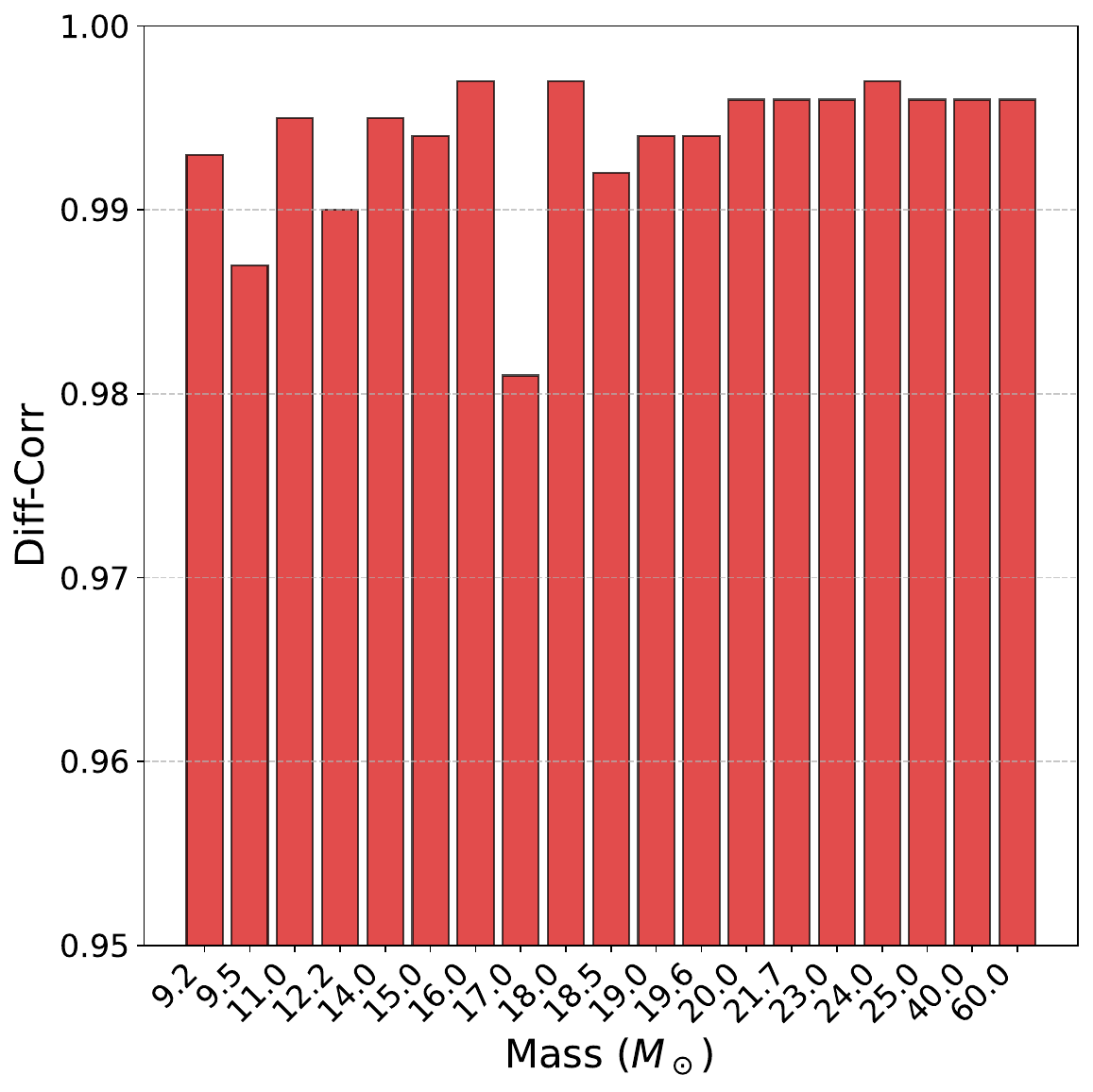}
        \caption{}
        \label{corrfig:b}
    \end{subfigure}
\caption{(a) Figure demonstrates an example of model parameter evolution curves for a progenitor star with a mass of 18 solar masses. \textcolor{black}{Additional models are presented in the \autoref{P:Parameter Evolution}.} (b) Figure illustrates the diff-corr parameters for $\tau$ and gravitational wave amplitude across various models. Notably, all of the models exhibit extremely high correlation.}
    \label{fig:corr}
\end{figure}

Our analysis reveals that, across all progenitor models studied, the evolution rates of these two quantities exhibit a significant and consistent strong positive correlation, with $\rho'_{\mathrm{UKF}}$ exceeding $0.98$ in all cases (see \autoref{corrfig:b}). This result indicates that the UKF’s dynamic modeling capability has uncovered an almost perfect underlying synchrony between the evolution rates of the two physical processes. Such near-perfect synchrony suggests that, regardless of whether the explosion ultimately succeeds or fails, the growth and decay rates of $\tau$ and $\mathrm{amp}$ are tightly coupled. This behavior is likely driven by large-scale, non-spherical hydrodynamic instabilities (e.g., convection, SASI) in the post-shock region, in which $\tau$ tracks the energy efficiency of the central engine while $\mathrm{amp}$ captures the asymmetry of its geometry. However, it should be noted that we find that the characteristic thermal diffusion region $\tau(t)$ exhibits a strong correlation with the gravitational-wave signal only in the low-frequency domain (as revealed after filtering). A detailed frequency-resolved analysis shows that above 100 Hz this correlation becomes insignificant. This result may indicate that gravitational waves at different frequencies originate with richer physical details.

This co-evolution suggests a potentially valuable multi-messenger diagnostic: real-time observation of supernova neutrino spectra, which closely reflect the system’s energy distribution, could provide indirect inference of the otherwise hidden energy transfer processes during collapse. The correlated evolution of neutrino emission and gravitational waves, also proposed in \citet{Choi:2025igp}, is corroborated here from a different perspective. The characteristic $\sim$ $10^{2}$Hz gravitational-wave signal from core-collapse supernovae falls squarely within the most sensitive band of the current LIGO-Virgo-KAGRA (LVK) network \citep{LVKObservationalScience2025} and will be a prime target for next-generation observatories such as the Einstein Telescope \citep{hild2011sensitivity} and Cosmic Explore \citep{evans2023cosmic}. Furthermore, detection and reconstruction algorithms for gravitational waves produced by CCSNe have also been widely proposed \citep{2020PhRvD.102d3022C,10.1093/mnras/stae604,2024arXiv240111635M}. This opens a direct observational window to verify the tight, physically-motivated correlation between neutrino emission, as traced by our parameter $\tau(t)$, and the GW amplitude. Due to the stochastic nature of gravitational waves produced by CCSNe \citep{2018ApJ...861...10M}, creating a comprehensive template bank is exceptionally challenging. Consequently, we cannot use the matched-filter method—a technique effectively employed for searching for signals from compact binary coalescences—to search for the GW signal from CCSNe \citep{1999PhRvD..60b2002O,McIver2015}. Recent years have seen the development of alternative methods for waveform search and reconstruction—such as principal component analysis \citep{2009CQGra..26j5005H,2009PhRvD..80j2004R}, dynamic time warping \citep{2019PhRvD..99l3012S}, and deep learning \citep{2020PhRvD.102d3022C} —but searches for CCSNe GW signals have so far yielded no significant detections \citep{LIGOScientific:2025abo}. However, our conclusions demonstrate the potential for future constraint searches of gravitational wave signals using neutrinos from CCSNe, as we've spotted what appears to be a secret handshake between them. The main challenge remains that both messengers are notoriously quiet.

We further identify a striking temporal coincidence between the onset of enhanced gravitational‑wave recovery and abrupt changes in the mean neutrino energy, characterized by a sudden drop followed by amplified fluctuations. This behavior is plausibly linked to the neutrino delayed‑heating mechanism, wherein charged‑current interactions deposit energy in the gain region and thereby revive the stalled shock. \textcolor{black}{A detailed discussion of this connection, while beyond the primary scope of this work, is provided in the \autoref{PF}.}

\section{Discussion}\label{Discussion}

We have introduced a new phenomenological parametrization of the neutrino energy spectrum and demonstrated its applicability to both SN 1987A observational data and state‑of‑the‑art 3D core‑collapse supernova simulations. A key finding is that the temporal evolution of the spectral parameter $\tau$ exhibits a clear correlation with explosion‑related multimessenger signals, such as gravitational waves, and serves as a diagnostic to distinguish between successful and failed supernova explosions. In addition, as discussed in \autoref{Constraining the Progenitor Mass of SN1987A}, applications such as progenitor mass inference based on this parametrization are inherently dependent on the precision and uncertainty of the available observations.

\textcolor{black}{Although our work primarily focuses on the neutrino energy spectra produced in core-collapse supernovae, it is also necessary to verify whether our formalism is applicable in broader contexts, such as those arising from Neutron-Star Common-Envelope Systems and those in systems that experience binary interactions before core collapse.}

\textcolor{black}{Here, we test our energy spectrum form against the Neutrino Signals from Neutron-Star Common-Envelope Systems provided in \cite{PhysRevLett.134.181003}, and use the KJR parametrization as a comparison.}

\begin{figure}[H]
\centering
\includegraphics[width=0.7\linewidth]{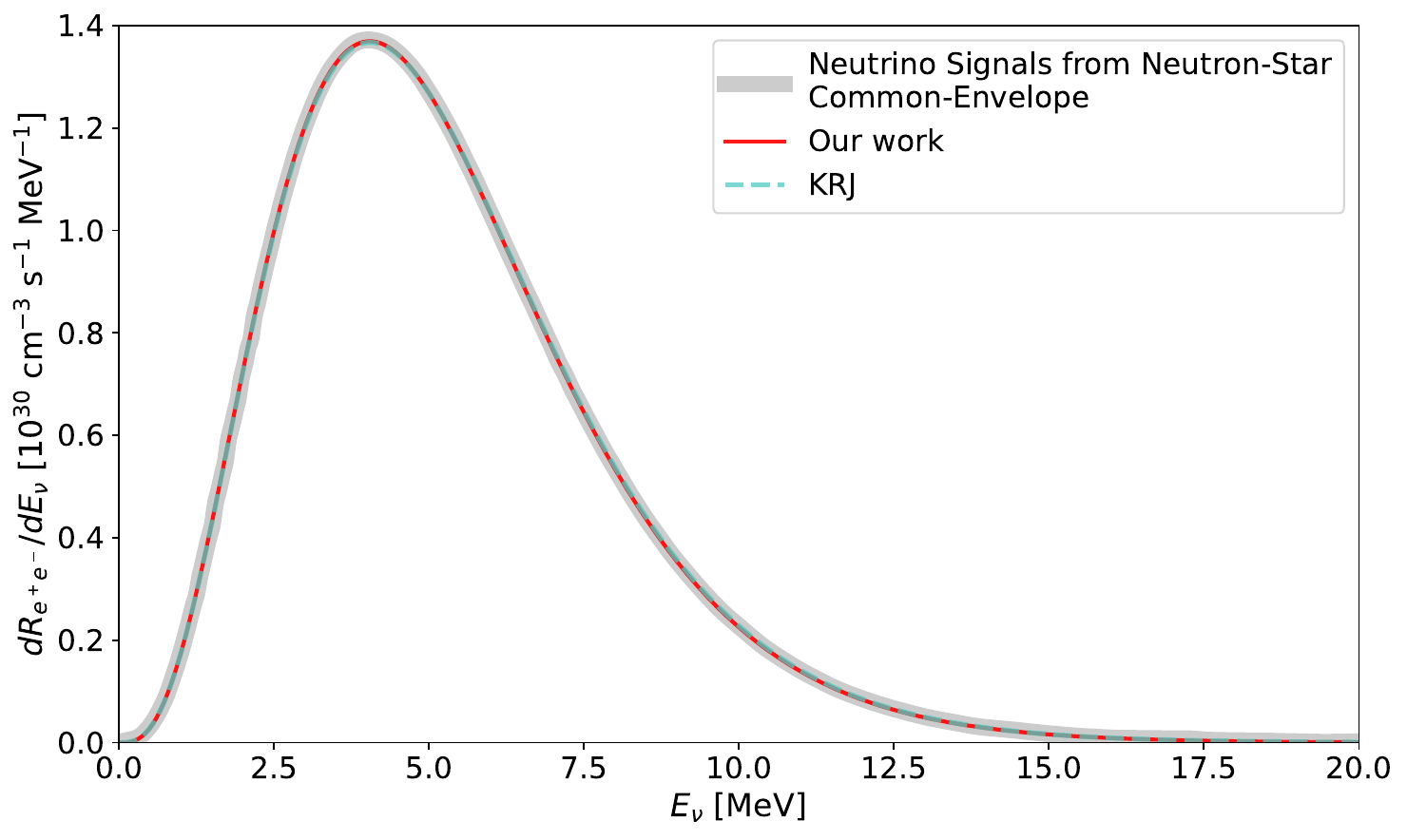}
\caption{\textcolor{black}{The grey thick line shows the neutrino spectrum due to thermal positron annihilation from \cite{PhysRevLett.134.181003}, at temperatures typical in a super-Eddington common-envelope event ($1.5 \times 10^{10}\text{K}$). The line is bolded to avoid obscuring the line style. The red solid line is the fit to our spectrum, and the blue-green dashed line is the fit using the KRJ parameterization. I used the Python package \texttt{emcee} for MCMC to search for parameters in both methods. See \autoref{corner_combined} for the CORNER plot.}}
\label{spectrum_comparison}
\end{figure}

\begin{figure}[H]
\centering
\includegraphics[width=1\linewidth]{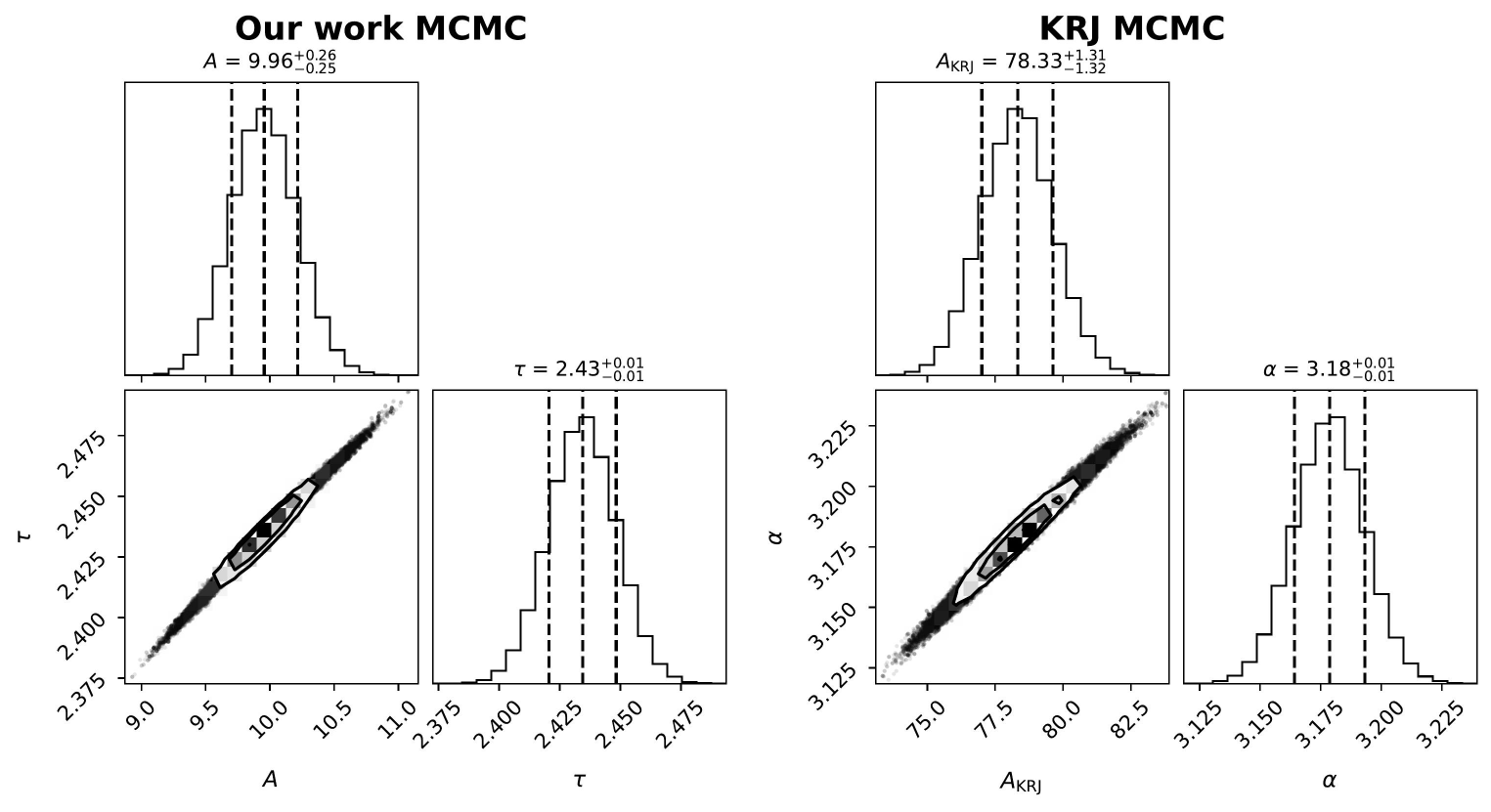}
\caption{\textcolor{black}{The left panel shows the joint posterior distribution of our spectral line parameters, and the right panel shows the joint posterior distribution of the KRJ parameterized fit.}}

\label{corner_combined}
\end{figure}

\textcolor{black}{Based on \autoref{spectrum_fit_with_errorbars}, \autoref{spectrum_comparison}, and \autoref{corner_combined}, it can be seen that our proposed spectral fitting method performs better when the average energy is high (e.g., CCSNe), but when the average energy is low (e.g., super-Eddington common-envelope event), the fitting results of the two methods are almost the same and both fit very effective ($\textbf{R}^{2} > 0.9999$). Numerically, our spectral shape can transition to KRJ parametrization with lower average energy. However, since \autoref{spectrum_comparison} and \autoref{corner_combined} are only case studies, we still cannot definitively say whether they can work robustly outside the CCSNe scenario, but there is no doubt that they have potential.} 

\textcolor{black}{For CCSNe that have undergone binary interactions and lost most of their outer layers, forming stripped-envelope progenitor stars, such progenitors are thought to systematically differ from \lq\lq classical\rq\rq single-star, red-supergiant-like progenitors in terms of their pre-supernova CO-core structure, density gradients, and compactness, and therefore in their ability to explode \citep{patton2020towards}. These structural differences should consequently impact the accretion history $\dot{M}(t)$, the neutrino luminosity $L_\nu(t)$, the evolution of the average neutrino energy and spectral pinching, and thus likely influence the time evolution of $\tau(t)$ and the location in the $\tau_{int}-Q_{int}$ plane. However, it's important to note that our spectral model doesn't directly simulate energy spectra from first principles. Therefore, it can only qualitatively estimate the scenario where binary interactions have occurred and most of the outer layers have been lost. As can be anticipated, the evolution of the parameter $\tau$ across the entire time series will still follow the classification in \autoref{cluster:b}: successful bursts will exhibit a rebound after falling from a peak, while failed bursts will show a monotonic overall trend. The only differences lie in the time of reaching the peak, the time of rebound, and the magnitude. As for the relationship between $\tau_{int}$ and $Q_{int}$, since we are considering the integral energy released during an outburst here, the evolution of the progenitor star will not significantly impact the energy released in terms of magnitude \citep{reed2020total}. Therefore, we expect this relationship to remain largely unaffected.}

\textcolor{black}{Finally, the relationship between the $\tau$ parameter, compactness $\xi_M$ \citep{O'Connor_2011}, and Ertl's two-parameter criterion ($M_{4}-\mu_{4}$) \citep{ertl2016two} is clearly noteworthy. These theoretical indicators, which describe the physical conditions of the stellar core before collapse, fundamentally determine the \lq\lq difficulty\rq\rq of the explosion, the dynamics after core bounce, the efficiency of neutrino heating, and the final explosion energy. The parameter $\tau$, on the other hand, is an observational parameter extracted from spectral data after the explosion, encoding the final outcomes of the explosion such as explosion energy, the mass of the ejecta, and velocity structure. Therefore, the potential connection between $\tau$ and these progenitor indicators is essentially a key link between \lq\lq theoretical causes\rq\rq (progenitor structure) and \lq\lq observational consequences\rq\rq (explosion spectrum). That said, our current dataset does not include objects with independently constrained $\xi_M$  or ($M_4, \mu_4$) parameters. A quantitative calibration of $\tau = f(\xi_M, M_4, \mu_4)$ would require a sample (or simulations) matching inferred progenitor core metrics with post-explosion observables. Therefore, we leave this direct empirical mapping for future work.}

This parametrization is, by construction, phenomenological. When additional microphysical effects (e.g., fast neutrino flavor conversions) also influence the spectrum, $\tau$ remains an efficient descriptor of the final spectral form; however, the physical interpretation of its temporal evolution becomes ambiguous. In particular, variations in $\tau$ cannot be uniquely attributed to changes in the high‑density core environment versus flavor‑oscillation dynamics. A key challenge for future work is therefore to disentangle these different contributions. Addressing this issue will require a robust multimessenger analysis framework, supported by next‑generation neutrino detectors with high energy resolution, large event statistics, and coordinated observations with gravitational‑wave and other channels. Owing to the inherent limitations of the phenomenological model, this parameterization may not accurately capture complex phenomena such as spectral splits induced by neutrino fast flavor conversions, which require more sophisticated model designs and dedicated simulations. Several related efforts based on theoretical analyses \citep{fiorillo2024theory,PhysRevD.109.063021}, numerical simulations \citep{PhysRevD.109.083019,PhysRevD.106.043011}, and data‑driven approaches \citep{shi2025application,abbar2024physics} have already been proposed. Building upon these developments, addressing such effects will constitute a central focus of our future work. In addition, whether this paradigm can be extended to other phenomena such as novae \citep{2016PASP..128e1001S,Luo_2025}, luminous red novae \citep{2019A&A...630A..75P,PhysRevLett.134.181003}, and fast blue optical transients \citep{2023ApJ...949..120H,guarini2022neutrino} is also a question worthy of investigation.



\section{Conclusion}\label{Conclusion}

In this work, we have developed and validated a novel, physically motivated parameterization for the neutrino energy spectrum from core-collapse supernovae. Our analysis demonstrates that the central parameter, $\tau(t)$, serves as a powerful and multifaceted diagnostic of the explosion dynamics. By applying our framework to state-of-the-art three-dimensional simulations, we have shown that the evolution of $\tau(t)$ not only provides a clear discriminant between successful and failed explosions but also exhibits a strong, synergistic co-evolution with the gravitational-wave signal, directly linking the engine's energetics to its asymmetry. Furthermore, the successful application of our model to the SN1987A neutrino dataset confirms its practical utility, yielding statistically significant fits and robust constraints on the spectrum’s low‑energy region. Although derived quantities—such as progenitor mass estimates—remain sensitive to observational precision and uncertainties, our framework still provides robust physical insight into the explosion dynamics. \textcolor{black}{It should be emphasized that in this work, $\tau$ is not intended to disentangle flavor-conversion microphysics; incorporating detailed oscillation dynamics is deferred to future studies.}

In summary, this parameterization provides a robust and versatile tool for interpreting the rich information encoded in supernova neutrino signals. As multi-messenger astronomy advances with next-generation facilities, our framework offers a direct pathway to decode the complex interplay of hydrodynamics, neutrino physics, and gravitational radiation from deep within the supernova core. While we have highlighted the future challenge of disentangling the contributions of various physical processes in complex scenarios, this defines a clear and compelling scientific objective for future synergistic analyses. 

\begin{acknowledgments}

Special thanks to Adam Burrows and David Vartanyan for their help and guidance in using the 3D simulation data. I would like to thank Shuai Zha, Qiyu Yan and  Xuwei Zhang for the useful discussions during the research process. This work received support from the National Natural Science Foundation of China under grants 12288102, 12373038, 12125303, 12090040/3, and U2031204; the Natural Science Foundation of Xinjiang No. 2022TSYCLJ0006; the science research grants from the China Manned Space Project No. CMS-CSST-2021-A10; the National Key R\&D Program of China No. 2021YFA1600401 and No. 2021YFA1600403; the Natural Science Foundation of Yunnan Province Nos. 202201BC070003 and 202001AW070007; the International Centre of Supernovae, Yunnan Key Laboratory No. 202302AN360001; and the Yunnan Revitalization Talent Support Program $-$ Science \& Technology Champion Project No. 202305AB350003.

\end{acknowledgments}

\newpage

\appendix



\section{Spectrum in the Instantaneous Point-Source Diffusion Approximation}\label{PA}

The core-collapse supernova (CCSN) fundamentally involves the rapid release of gravitational binding energy—on the order of $10^{53}\,\mathrm{erg}$—as the stellar iron core collapses on a timescale of seconds. Approximately 99\% of this energy is emitted as neutrinos within $\sim 10$s, while only $\sim 1\%$ ($\sim 10^{51}\,\mathrm{erg}$) is deposited to power the explosion. This energy release is highly localized, occurring within a region of tens of kilometers in diameter and nuclear-scale density, and is intrinsically transient. To capture the dominant thermodynamic behavior of this core process, we employ an idealized model: a thermodynamic system that is macroscopically infinitesimal—effectively treated as a point—yet microscopically contains a vast number of particles, ensuring well-defined thermodynamic variables such as pressure, temperature, and internal energy. 

\begin{figure}[H]
  \centering
  \includegraphics[width=0.7\linewidth]{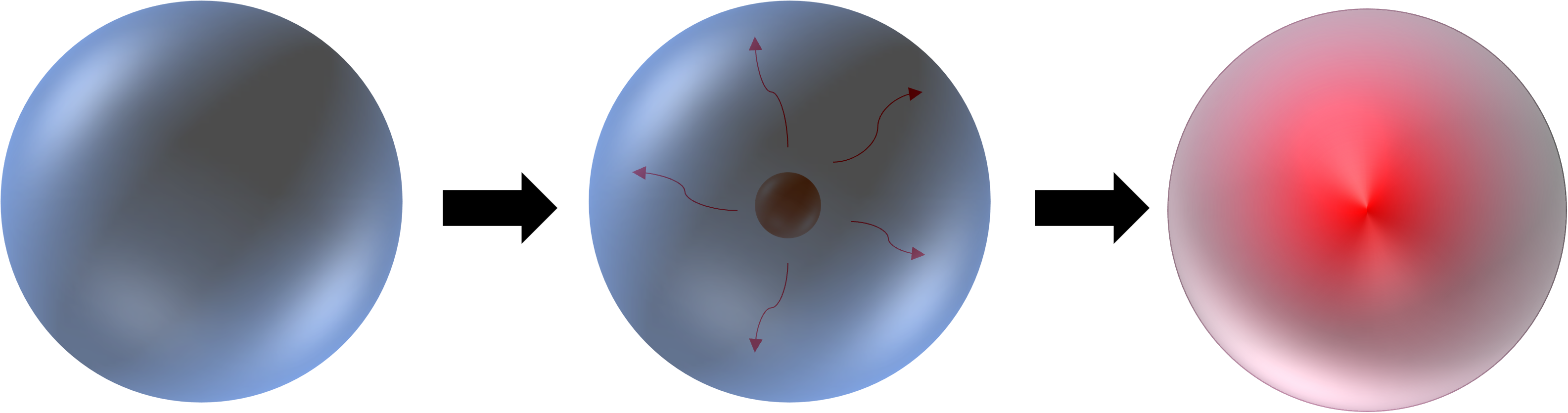}
  \caption{The three evolutionary stages of the toy model. Left: The system is initially in a uniform state of thermal equilibrium. Middle: At $t=0$, energy is instantaneously injected at the center as a point source and begins to diffuse outward. Right: After sufficient evolution, the energy has diffused throughout the system, which then reaches a new state of thermal equilibrium.}\label{toy_m}
\end{figure}

It should be emphasized that our model serves as an equivalent framework, primarily aimed at providing a physically intuitive and interpretable parameterization of the neutrino energy spectrum. It is not intended to reproduce the full complexity of the supernova explosion. Nevertheless, as demonstrated in the latter part of this work, its validity lies in directly linking observational data to the model’s key parameters, thereby enabling an effective characterization of the neutrino spectra throughout the various evolutionary stages of the supernova.

\subsection{Solution for an Instantaneous Point Source Diffusion}
In our model, the subsequent evolution of the energy density $E(x,t)$ in one-dimensional space is governed by the standard diffusion equation:
\begin{equation}
\frac{\partial E(x, t)}{\partial t}
= \kappa' \frac{\partial^2 E(x, t)}{\partial x^2},
\quad x \in \mathbb{R}, ; t>0 .
\end{equation}
Here, $\kappa' > 0$ is the thermal diffusivity, with dimension $[\kappa'] = \mathrm{m^2/s}$, determined by the medium’s thermal conductivity $\kappa$, density $\rho$, and specific heat capacity $c_p$ through $\kappa' = \kappa/(\rho c_p)$.
The thermal conductivity $\kappa$ itself is defined by Fourier’s law of heat conduction, $j = -\kappa \nabla T$, where $j$ is the heat flux $( [j] = \mathrm{W\,m^{-2}} )$ and $\nabla T$ is the temperature gradient $( [\nabla T] = \mathrm{K\,m^{-1}} )$, giving $[\kappa] = \mathrm{W\,m^{-1}\,K^{-1}}$.

This equation is subject to an initial condition describing the instantaneous injection of energy:
\begin{equation}
    E(x, 0)=Q \cdot \delta(x)
\end{equation}
This condition signifies that a total energy $Q$ is concentrated at the origin $x=0$ at the moment $t=0$.

By the linearity of the diffusion equation, the solution can be expressed as $E(x, t) = Q \cdot G(x, t)$. The problem is thus reduced to finding the Green's function, $G(x, t)$, for the equation. The Green's function represents the fundamental solution for a unit energy injection ($Q=1$) and must satisfy:
\begin{equation}
    \frac{\partial G(x, t)}{\partial t}=\kappa' \frac{\partial^2 G(x, t)}{\partial x^2}, \quad G(x, 0)=\delta(x)
\end{equation}

To solve for this Green's function, we employ the method of Fourier transforms. Applying a Fourier transform with respect to the spatial variable $x$, we define:
\begin{equation}
    \tilde{G}(k, t) = \int_{-\infty}^{\infty} G(x, t) e^{-i k x} d x
\end{equation}
Crucially, we consider that in our physical model, the thermal diffusivity may evolve with time, i.e., $\kappa'=\kappa'(t)$. The diffusion equation is thus converted into a first-order ordinary differential equation for the Fourier amplitude $\tilde{G}(k, t)$:
\begin{equation}
    \frac{d \tilde{G}(k,t)}{d t} = -\kappa'(t) k^2 \tilde{G}(k,t)
\end{equation}
The initial condition $G(x, 0)=\delta(x)$ becomes $\tilde{G}(k, 0)=1$ in the Fourier domain. This is a separable ODE, and its integration subject to the initial condition yields:
\begin{equation}
    \tilde{G}(k, t) = \exp\left(-k^2 \int_0^t \kappa'(s) ds\right)
\end{equation}
It is natural to define an "intrinsic diffusion time" or "effective time", $\tau(t)$, which represents the cumulative effect of diffusion up to time $t$:
\begin{equation}
    \tau(t) \equiv \int_0^t \kappa'(s) ds
\end{equation}
With this definition, the solution in Fourier space takes the elegant form:
\begin{equation}
    \tilde{G}(k, t) = e^{-k^2 \tau(t)}
\end{equation}
Finally, the solution in real space is recovered via the inverse Fourier transform:
\begin{equation}
    G(x, t) = \frac{1}{2 \pi} \int_{-\infty}^{\infty} e^{-k^2 \tau(t)} e^{i k x} d k
\end{equation}
This is a standard Gaussian integral, which evaluates to:
\begin{equation}
    G(x, t)=\frac{1}{\sqrt{4 \pi \tau(t)}} \exp \left(-\frac{x^2}{4 \tau(t)}\right)
\end{equation}
Multiplying this fundamental solution by the total energy $Q$ gives the final energy distribution we seek:
\begin{equation}\label{solution}
    E(x, t)=\frac{Q}{\sqrt{4 \pi \tau(t)}} \exp \left(-\frac{x^2}{4 \tau(t)}\right)
\end{equation}

To obtain the energy spectrum, we need to transform from real space to energy (momentum) space:

Let

\begin{equation}
\tilde{E}(k, t) = \int_{-\infty}^{\infty} E(x, t) \, e^{-i k x} \, \mathrm{d}x
\end{equation}

Given

\begin{equation}
E(x, t) = \frac{Q}{\sqrt{4 \pi \tau(t)}} \exp\left(-\frac{x^2}{4 \tau(t)}\right),
\end{equation}

substituting yields

\begin{equation}
\tilde{E}(k, t) = \int_{-\infty}^{\infty} \frac{Q}{\sqrt{4 \pi \tau(t)}} \exp\left(-\frac{x^2}{4 \tau(t)}\right) e^{-i k x} \, \mathrm{d}x.
\end{equation}

Since the Fourier transform of a Gaussian remains Gaussian:

\begin{equation}
\int_{-\infty}^{\infty} \exp(-a x^2) e^{-i k x} \, \mathrm{d}x = \sqrt{\frac{\pi}{a}} \exp\left(-\frac{k^2}{4a}\right), \quad \Re(a) > 0,
\end{equation}

with $a = \frac{1}{4\tau(t)}$, we have

\begin{equation}
\int_{-\infty}^{\infty} \exp\left(-\frac{x^2}{4\tau(t)}\right) e^{-i k x} \, \mathrm{d}x = \sqrt{4 \pi \tau(t)} \exp\left[-\tau(t) k^2\right].
\end{equation}

Multiplying by the prefactor $\frac{Q}{\sqrt{4 \pi \tau(t)}}$ gives

\begin{equation}
\tilde{E}(k, t) = Q \exp\left[-\tau(t) k^2\right].
\end{equation}

Here, $k$ represents the energy wavenumber. 

\textcolor{black}{To obtain the energy spectrum, we map \(k\) to \(E\) \emph{on shell} and include the Jacobian associated with the change of variables:
\begin{equation}
\varphi(E,t)\ \propto\ g(E)\,\tilde{E}\!\big(k(E),t\big).
\end{equation}
Here \(k(E)\) is determined by the dispersion relation \(\omega=\omega(k)\) for plane waves \(e^{i(kx-\omega t)}\) together with the quantum relations \(E=\hbar\omega\) and \(p=\hbar k\) in natural units we set $\hbar= c = 1$. For a free particle,
\begin{equation}
k(E)=\sqrt{E^2-m^2}\quad(\text{relativistic}),\qquad 
k(E)=\sqrt{2mE}\quad(\text{non-relativistic}).
\end{equation}
The function \(g(E)\) is the density-of-states (Jacobian) factor; it follows from conservation under the change of variables, and its explicit form and derivation are provided elsewhere in the manuscript and are not repeated here.}

\textcolor{black}{Substituting \(k(E)\) back into \(\tilde{E}(k,t)=Ae^{-\tau(t)k^2}\) gives
\begin{equation}
\varphi(E,t)\ \propto\ g(E)\,\exp\!\big[-\tau(t)\,k(E)^2\big].
\end{equation}
Hence,
\begin{equation}
\text{relativistic limit:}\quad k(E)\simeq E\ \Rightarrow\ \varphi(E,t)\propto g(E) \exp[-\tau(t)E^2],
\end{equation}
\begin{equation}
\text{non-relativistic limit:}\quad k(E)=\sqrt{2mE}\ \Rightarrow\ 
\varphi(E,t)\propto g(E)\,\exp\!\Big[-2m\,\tau(t)\,E\Big].
\end{equation}}

Consider the normalized form:
\textcolor{black}{
\begin{equation}
\varphi(E,t) \propto
\begin{cases}
A\times g(E) \exp\left[-\tau(t) (\frac{E}{{\langle E \rangle}^{*}})^{2}\right], & \text{relativistic case} \\[6pt]
A\times g(E) \exp\left[-\tau(t) m \frac{E}{{\langle E \rangle}^{*}}\right], & \text{non-relativistic case}
\end{cases}
\end{equation}
}
Here, $A$ is the normalization constant, and ${\langle E \rangle}^{*}$ is a scaled mean energy determined to ensure that $g_{(E)} \times P(E,t)$ is properly normalized. Numerical evaluation yields ${\langle E \rangle}^{*} \approx 0.6 \times {\langle E \rangle}$, where ${\langle E \rangle}^{*}$ is dimensionless. Note that the time dependence in $\varphi(E,t)$ arises solely from $\tau(t)$ and does not couple to other parts of the derivation. Note that the term 'relativistic' in this context refers to the rate of energy transfer. For all subsequent calculations, we adopt the relativistic form of $\varphi(E,t)$.

To achieve a form resembling the Keil-Raffelt-Janka (KRJ) parameterization \citep{Keil_2003},

\begin{equation}
\varphi(E) \propto A_{\mathrm{KRJ}}\left(\frac{E}{\langle E \rangle}\right)^\alpha \exp\left[-(\alpha+1) \frac{E}{\langle E \rangle}\right],
\end{equation}

\textcolor{black}{One must multiply $P(E,t)$ (the average occupation per quantum state) by the density of available states near energy $E$. This yields the number of available states × occupation probability = actual occupied states. $P(E,t)$ can be obtained from $\varphi(E,t)$.}

In the KRJ model, the state density is $g_{(E)} = \left(\frac{E}{\langle E \rangle}\right)^\alpha$. However, to ensure dimensionless exponents in our model, the form needs to be modified. Two options are considered: 1. $g_{(E)} = \left(\frac{E}{\langle E \rangle}\right)^{\frac{\alpha {\langle E \rangle}}{{ E }^{*}}}$ ; 2. $g_{(E)} = \left(\frac{E}{\langle E \rangle}\right)^{\frac{\alpha E}{{\langle E \rangle}^{*}}}$. Numerical and convergence analyses show that the second choice leads to super-exponential growth at high $E$, producing unphysical spectra. Therefore, the first form is adopted:

\begin{equation}
g_{(E)} = \left(\frac{E}{\langle E \rangle}\right)^{\frac{\alpha {\langle E \rangle}}{{E }^{*}}}.
\end{equation}

The final form is then:

\begin{equation}
\varphi(E, t) = A \times g_{(E)} \times P(E, t).
\end{equation}



Corresponding to the KRJ parameterization, we set $\alpha = \tau(t)m$. Here, $E^{*}$ is the dimensionless neutrino energy and $\langle E \rangle$ is the mean energy. For convenience, we hereafter set $m=1$ without explicitly writing it, noting that its dimensional role remains but different numerical values lead to the same outcome after spectrum normalization. The chosen functional form is motivated primarily by mathematical convenience rather than physical interpretation, a limitation likely reflecting our incomplete understanding of state functions in non-equilibrium conditions rather than any inherent physical inconsistency. Notably, for time-integrated spectra, the “exponential tail of Gaussian mixtures’’ asymptotically reduces to a single exponential form:

\begin{equation}
\varphi(E, t) = A \times E^{\frac{{\langle \alpha \rangle} {\langle E \rangle}}{E^{*}}} \times \exp\left(-\frac{\alpha E}{{\langle E \rangle}^{*}}\right).
\end{equation}

 A brief explanation \citep{Grushka1972} is as follows:

Consider writing the spectrum in the form:

\begin{equation}
P(E , \tau) = E^{\tau / E} \exp\left(-\tau \left(\frac{E}{\langle E \rangle^*}\right)^2\right),
\end{equation}

mixed over an exponential weighting distribution

\begin{equation}
w(\tau) = \lambda e^{-\lambda \tau}, \quad (\tau \geq 0),
\end{equation}

leading to the integral

\begin{equation}
I(E) = \int_0^{\infty} E^{\tau / E} \exp\left(-\tau a E^2\right) \lambda e^{-\lambda \tau} \, d\tau, \quad a = \frac{1}{\left(\langle E \rangle^*\right)^2}.
\end{equation}

This integral has the form of an Exponentially Modified Gaussian (EMG) distribution. Through appropriate transformations, it can be written as

\begin{equation}
I(E) = \frac{\lambda}{2 \sqrt{\pi a} E} \exp\left[\lambda\left(\mu + \frac{\lambda}{2a} - E\right)\right] \operatorname{erfc}\left(\frac{\mu + \lambda/a - E}{\sqrt{2/a} \, E}\right),
\end{equation}

where the complementary error function is defined as

\begin{equation}
\operatorname{erfc}(x) = 1 - \operatorname{erf}(x) = \frac{2}{\sqrt{\pi}} \int_x^{\infty} e^{-t^2} \, dt.
\end{equation}

Let

\begin{equation}
z = \frac{\mu + \lambda/a - E}{\sqrt{2/a} \, E} \equiv \frac{E - E_0}{\sigma_E E},
\end{equation}

where

\begin{equation}
E_0 = \mu + \frac{\lambda}{a}, \quad \sigma_E = \sqrt{\frac{2}{a}}.
\end{equation}

The integral then takes the form

\begin{equation}
I(E) = \underbrace{\frac{\lambda}{2 \sqrt{\pi a} E} \exp\left[\lambda\left(\mu + \frac{\lambda}{2a} - E\right)\right]}_{C E^{-1} e^{-\lambda E}} \operatorname{erfc}(z).
\end{equation}

For $z \rightarrow +\infty$, there is the standard asymptotic expansion:

\begin{equation}
\operatorname{erfc}(z) \sim \frac{e^{-z^2}}{z \sqrt{\pi}} \left(1 - \frac{1}{2 z^2} + \cdots \right).
\end{equation}

Substituting $z = \frac{E - E_0}{\sigma_E E}$, we have

\begin{equation}
z^2 = \frac{(E - E_0)^2}{\sigma_E^2 E^2} = \frac{E^2 - 2 E_0 E + E_0^2}{(2 / a) E^2} = \frac{a}{2}\left(1 - 2 \frac{E_0}{E} + \frac{E_0^2}{E^2}\right).
\end{equation}

Thus,

\begin{equation}
e^{-z^2} = \exp\left[-\frac{a}{2}\right] \exp\left[a \frac{E_0}{E}\right] \exp\left[-\frac{a}{2} \frac{E_0^2}{E^2}\right].
\end{equation}

Also,

\begin{equation}
\frac{1}{z} = \frac{\sigma_E E}{E - E_0} = \sqrt{\frac{2}{a}} \frac{E}{E - E_0} \approx \sqrt{\frac{2}{a}}\left(1 + \frac{E_0}{E} + \cdots\right).
\end{equation}

Combining these, we get

\begin{equation}
I(E) \sim \underbrace{C E^{-1}}_{\propto E^{-1}} \exp(-\lambda E) \times \frac{1}{\sqrt{\pi}} \frac{e^{-z^2}}{z} \propto E^{-1} \exp(-\lambda E - z^2) \times \frac{1}{z}.
\end{equation}

By substituting the asymptotic forms of $z^2$ and $1/z$, and neglecting $O(1/E^2)$ and higher-order corrections, we obtain

\begin{equation}
I(E) \propto E^{-1} \exp(-\lambda E) \exp\left[-\frac{a}{2} + a \frac{E_0}{E}\right] \times \sqrt{\frac{2}{a}}\left(1 + \frac{E_0}{E}\right).
\end{equation}

The constant factors such as $\exp(-a/2)$ and $\sqrt{2/a}$ can be absorbed into the normalization constant $A$.

For any positive $E$ and constant $C$, we can use the identity

\begin{equation}
\exp\left(\frac{C}{E}\right) = \exp\left(\frac{C}{E} \ln E\right) \times \exp\left(-\frac{C}{E} \ln E + \frac{C}{E}\right) \approx E^{\frac{C}{E}},
\end{equation}

where in the last step we neglect subleading corrections of order $O\left(\frac{\ln E}{E}\right)$.

In our expression, let

\begin{equation}
C = a E_0,
\end{equation}

then

\begin{equation}
\exp\left(a \frac{E_0}{E}\right) \approx E^{\frac{a E_0}{E}}.
\end{equation}

Moreover,

\begin{equation}
E^{-1}\left(1 + \frac{E_0}{E}\right) = \exp(-\ln E) \exp\left[\ln\left(1 + \frac{E_0}{E}\right)\right].
\end{equation}

For large $E$, we have the approximation $\ln\left(1 + E_0 / E\right) \approx E_0 / E$. Therefore,

\begin{equation}
E^{-1}\left(1 + \frac{E_0}{E}\right) \approx \exp\left(-\ln E + \frac{E_0}{E}\right) = E^{-1} \exp\left(\frac{E_0}{E}\right) \approx E^{-1 + \frac{E_0}{E}}.
\end{equation}

Combining terms yields

\begin{equation}
I(E) \propto E^{\frac{a E_0}{E}} \times E^{-1 + \frac{E_0}{E}} \times \exp(-\lambda E) = E^{\frac{a E_0 + E_0}{E} - 1} \exp(-\lambda E).
\end{equation}

We define

\begin{equation}
\langle\alpha\rangle \langle E\rangle = a E_0 + E_0, \quad \alpha / \langle E\rangle = \lambda,
\end{equation}

and absorb the extra $-1$ into the normalization constant $A$. This simplifies the expression to

\begin{equation}
I(E) \approx A \, E^{\frac{\langle\alpha\rangle \langle E\rangle}{E^*}} \exp\left(-\frac{\alpha E}{\langle E\rangle^*}\right).
\end{equation}

The introduction of m is to ensure that the exponent remains dimensionless, and the effect of the specific value of can be eliminated simply by adjusting the normalization coefficient. Therefore, this is an operation that is not rigorous but effective.

Thus, we arrive at the approximate form of the integrated spectrum used in our analysis:

\begin{equation}
\varphi(E, t) = A \times \left( \frac{E}{\langle E \rangle} \right)^{\frac{\langle\alpha\rangle \langle E\rangle}{E^*}} \times \exp\left(-\frac{\alpha E}{\langle E\rangle^*}\right).
\end{equation}

\section{Fitting the SN1987A Spectrum}\label{PB}

\begin{figure}[H]
\centering
\includegraphics[width=0.8\linewidth]{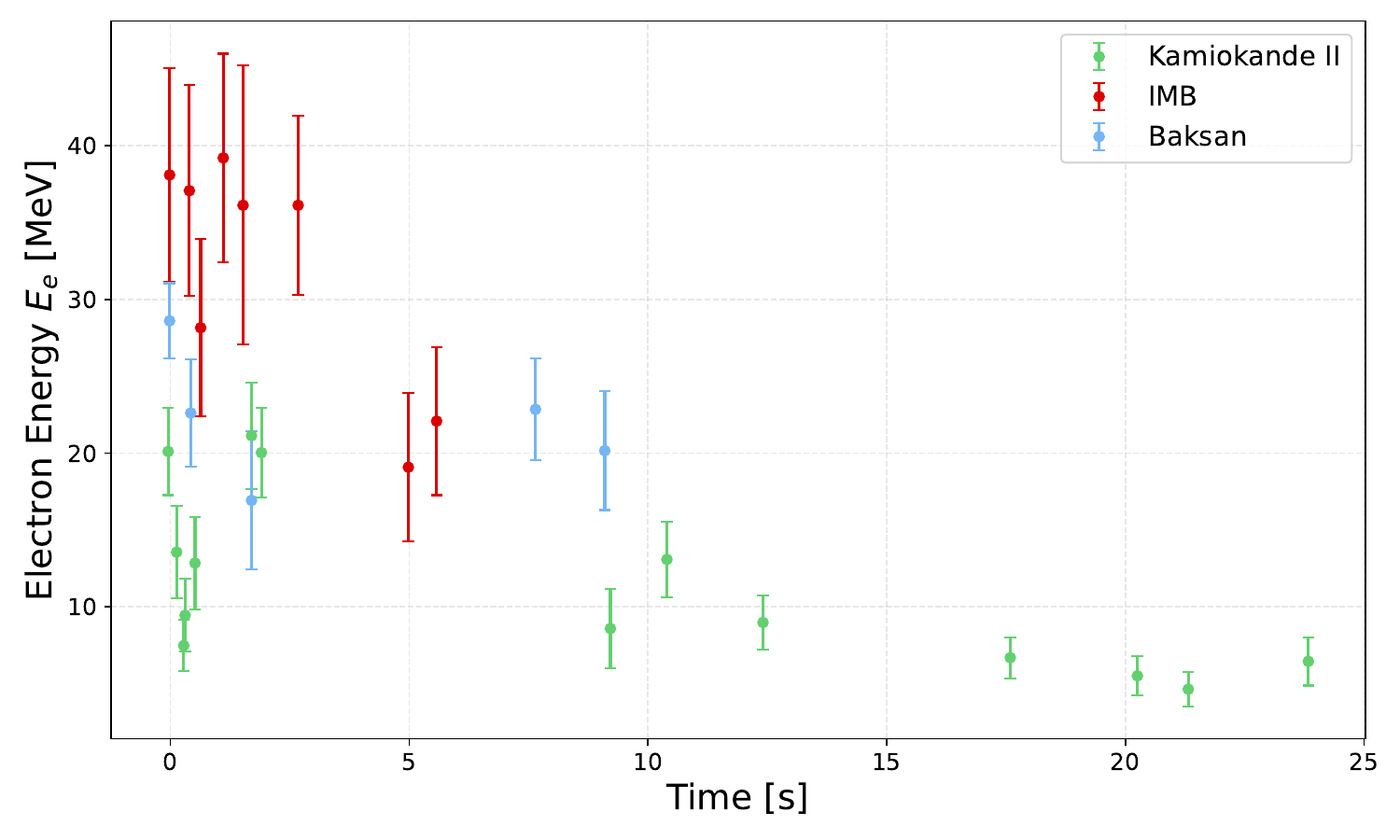}
\caption{\textcolor{black}{Neutrino events from SN1987A detected by the Kamiokande-II (green), IMB (red), and Baksan (blue) experiments. The plot displays the measured energy of each event versus its relative arrival time, with the first detected event defined as t=0. The vertical error bars represent the measurement uncertainty for the energy of each event.}}\label{SN1987A_event}
\end{figure}

\autoref{SN1987A_event} shows the time-energy data recorded in the SN1987A event \citep{dos2022understanding}.

We perform time integration while accounting for the energy-dependent resolution of each detector:

\begin{figure}[H]
\centering
\includegraphics[width=0.6\linewidth]{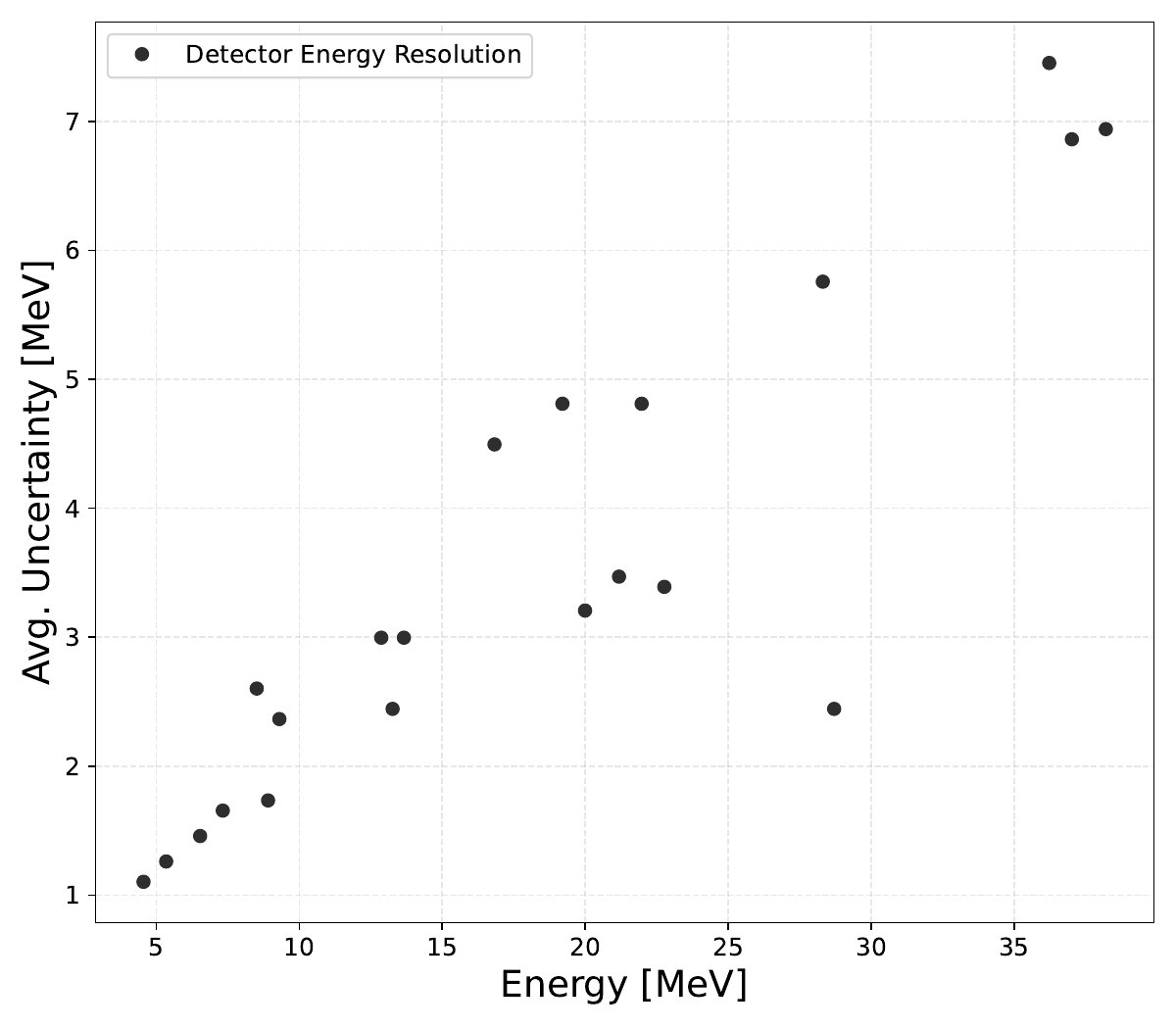}
\caption{\textcolor{black}{The combined energy resolution function of the three detectors. This data derived was generated by merging the total data from all three experiments, binning the data by energy, and then calculating the average measurement uncertainty within each energy bin.}}\label{Detector_Resolution_with_energy}
\end{figure}

\citet{Yuksel:2007mn} reconstructed the neutrino spectrum from SN 1987A using a non-parametric statistical approach that relies solely on detector parameters. We therefore adopt this spectrum as the reference for comparison with our model. It should be noted that we use the same number spectrum data as in \citep{Yuksel:2007mn}, and thus the resulting $Q$ here corresponds to the total neutrino number.


To accurately infer the key parameters of our theoretical model from observed spectral data $f_{\text{obs}}(E)$ and to overcome the tendency of traditional optimization methods to get trapped in local minima, we designed and implemented a hybrid optimization framework based on a Genetic Algorithm (GA), guided by Empirical Bayes principles.

First, we preprocess the observed spectral data by restricting the energy range to $E < 80$ to reduce high-energy noise, and compute the mean energy $E_{\text{mean}}$ for use in scaling the model.

Our theoretical spectral function model $P(E; \tau, Q)$ describes the energy distribution using two free parameters: an effective timescale parameter $\tau$ that shapes the spectrum and a normalization constant $Q$ representing the total energy or flux. The model is defined as:

\begin{equation}
P(E; \tau, Q) = Q \cdot \frac{f_{\text{shape}}(E; \tau)}{\int f_{\text{shape}}(E'; \tau) \, dE'}
\end{equation}

where the shape function $f_{\text{shape}}(E; \tau)$ includes a power-law factor $g(E)$ and an exponential cutoff:

\begin{equation}
f_{\text{shape}}(E; \tau) = g(E) \cdot \exp\left(-\tau \frac{E}{\langle E \rangle}\right).
\end{equation}

Here, the power-law term is designed as an energy-dependent function $g(E) = (E/E_{\text{scale}})^{\alpha(E)}$ with $\alpha(E) = (\tau - 1) E_{\text{scale}} / E$, providing flexibility to capture low-energy behavior. $E_{\text{scale}}$ is a characteristic energy scale calculated from the data, and $\langle E \rangle$ is the mean of $E$ over the effective energy range, used to nondimensionalize the exponential term.

The optimization framework is built on a genetic algorithm maintaining a population of parameter pairs $\theta_i = (\tau_i, Q_i)$. In each generation, individuals are evaluated using a loss function that quantifies the discrepancy between the predicted and observed spectra (fitness).

The core innovation of our algorithm lies in avoiding blind crossover and mutation. Instead, we construct a dynamic memory of parameter posterior distribution. Specifically, we collect all historical individuals $\{\theta_i\}$ and their corresponding loss values $\{L_i\}$. Using an exponential weighting scheme $w_i = \exp(-L_i/T)$ (with $T$ as a temperature parameter), we approximate an empirical Bayes posterior distribution $p(\theta | \text{Data})$ as a multivariate Gaussian $\mathcal{N}(\mu_{\text{post}}, \Sigma_{\text{post}})$. The posterior mean $\mu_{\text{post}}$ and covariance $\Sigma_{\text{post}}$ are computed as weighted averages over historical samples.

The next generation of the population is generated through a hybrid process:

\begin{enumerate}
\item \textbf{Elitism}: The best-performing individuals from the previous generation are retained directly in the new generation.
\item \textbf{Bayesian Sampling}: With high probability (e.g., 70\%), new individuals are sampled from the empirical posterior $p(\theta | \text{Data})$. This step focuses the search efficiently on known high-quality regions of parameter space.
\item \textbf{Crossover and Mutation}: The remaining individuals are generated via standard genetic crossover and mutation operations to maintain diversity and avoid premature convergence.
\end{enumerate}

In implementation, we use a population size of 100 and evolve over 2000 generations. The parameter search ranges for $\tau$ and $Q$ are constrained based on physical considerations and data scaling. Through this iterative "evaluation–learning–sampling" process, the algorithm converges to an optimal parameter set $(\tau^*, Q^*)$ that best describes the observed data.

The best-fit spectrum obtained was compared with the reference SN1987A neutrino spectrum:

\begin{figure}[H]
\centering
\includegraphics[width=0.7\linewidth]{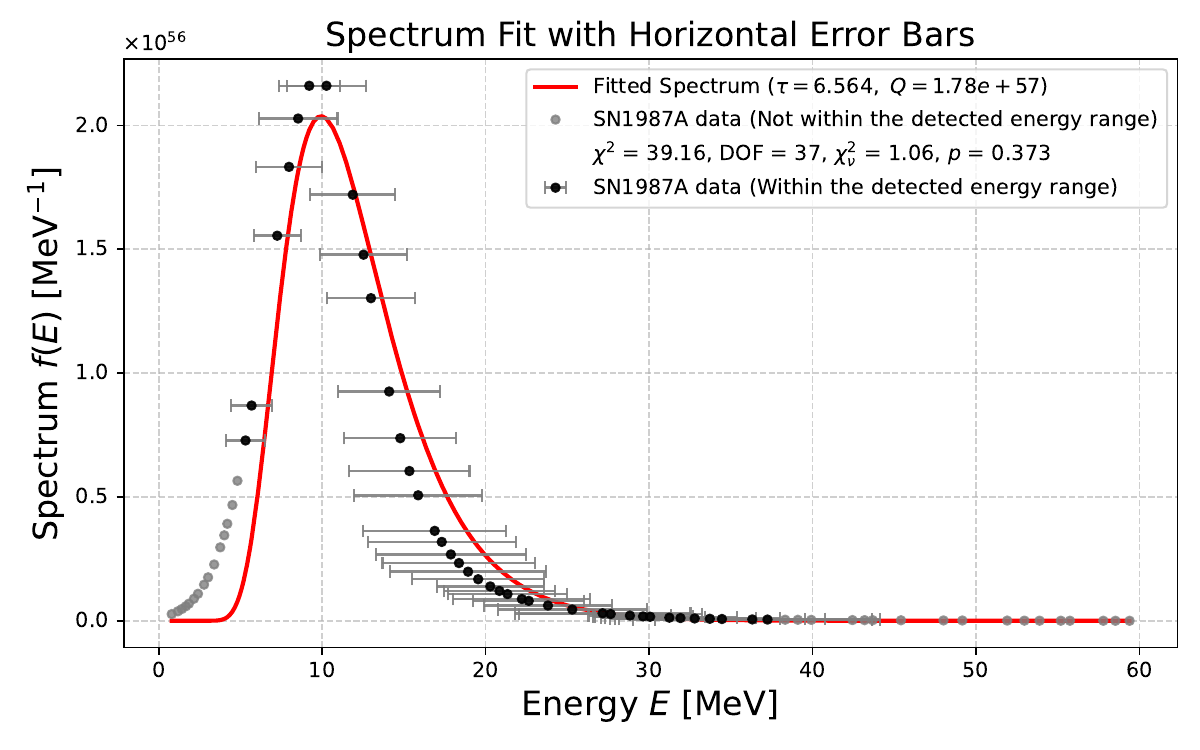}
\caption{Our model shows good agreement with the SN1987A detected spectrum, statistically significant. This may indicate that, in core-collapse supernovae, the spectral shape is governed primarily by the dynamics of the explosion rather than the detailed nuclear processes. To avoid accidental overfitting of a single random initialization, we did not fix the random number seed in multiple calculations. Therefore, there are slight differences between the displayed indicators and the statistical indicators in the main text; this difference is within the acceptable statistical fluctuation range, and the conclusions remain consistent.}
\label{supp:spectrum_fit_with_errorbars}
\end{figure}

\begin{figure}[H]
\centering
\includegraphics[width=0.6\linewidth]{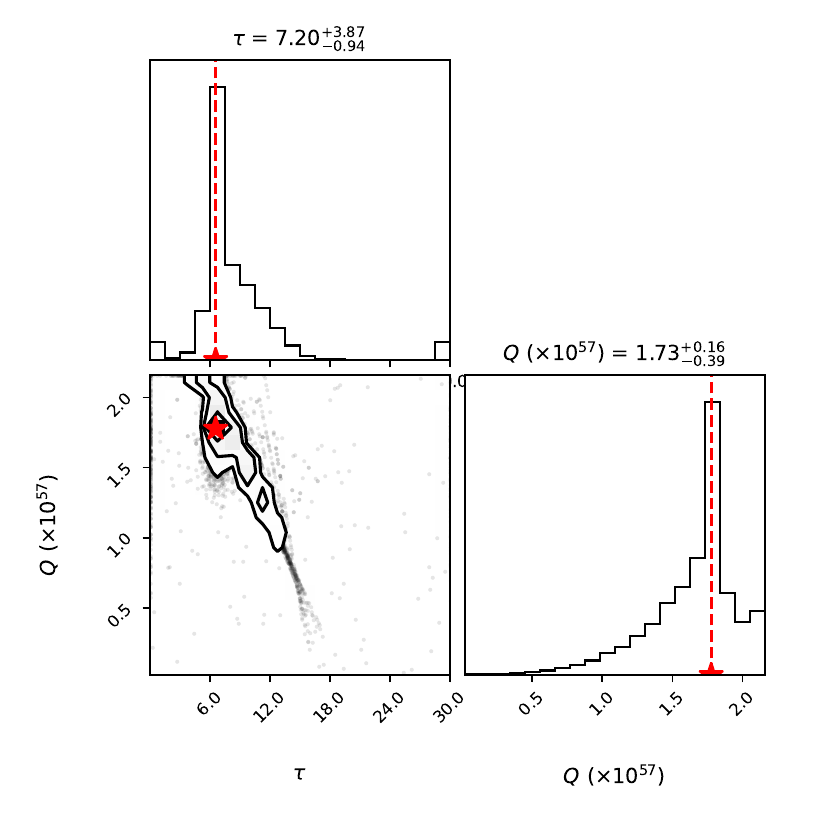}
\caption{
    \textbf{Corner plot} showing the posterior distributions of the parameters $\tau$ and $Q$ obtained using Genetic Algorithm with Empirical Bayesian Guided Evolution. The red star marks the best-fit values of $\tau = 6.5933$ and $Q = 1.7709 \times 10^{57}$. The plot shows that $\tau$ is well-constrained, with an uncertainty range of $[6.26, 7.68]$, while $Q$ has a larger uncertainty range of $[1.52 \times 10^{57}, 2.02 \times 10^{57}]$. The lower-left corner plot demonstrates the negative correlation between $\tau$ and $Q$, as seen from the shape of the joint distribution. This analysis indicates that the optimized parameters $\tau$ and $Q$ are well-constrained by the data, and the method has successfully converged to a global optimum.
}
\label{corner-fig}
\end{figure}

\begin{table}[H]
\centering
\begin{tabular}{lcc}
\hline
\textbf{Test Item} & \textbf{Value} & \textbf{Description} \\
\hline
Significance level $\alpha$ & 0.01 & Predefined threshold \\
Rejection region & $\chi^2 > 59.89$ & Critical value corresponding to $\alpha = 0.01$ \\
Computed $\chi^2$ & 39.16 & Goodness-of-fit statistic \\
$p$-value & 0.373 & Probability corresponding to $\chi^2$ \\
Conclusion & Null hypothesis not rejected, statistically significant & $\chi^2$ does not fall in the rejection region \\
\hline
\end{tabular}
\caption{Chi-squared goodness-of-fit test results ($\alpha = 0.01$)}
\label{tab:chi2_test}
\end{table}

The fitted spectrum is consistent with the measurements within their experimental uncertainties in the observationally constrained energy range. However, it deviates significantly at lower energies ( $\lesssim 5 \mathrm{MeV}$ ), a discrepancy that requires validation with future data.

%
%
%
%
%
%

\section{Statistics Based on 3D Core-Collapse Supernova Simulation Data}\label{PC}

Using the data provided in \citep{Choi:2025igp}, we fit the resulting spectra with our model to extract the parameters $Q$ and $\tau$ for each supernova model. Our first goal is to examine whether a parameter range can help distinguish between exploding and non-exploding progenitors. We integrate the data over the first two seconds (with 0.01-second resolution), obtaining:

\begin{figure}[H]
  \centering
  \includegraphics[width=0.55\linewidth]{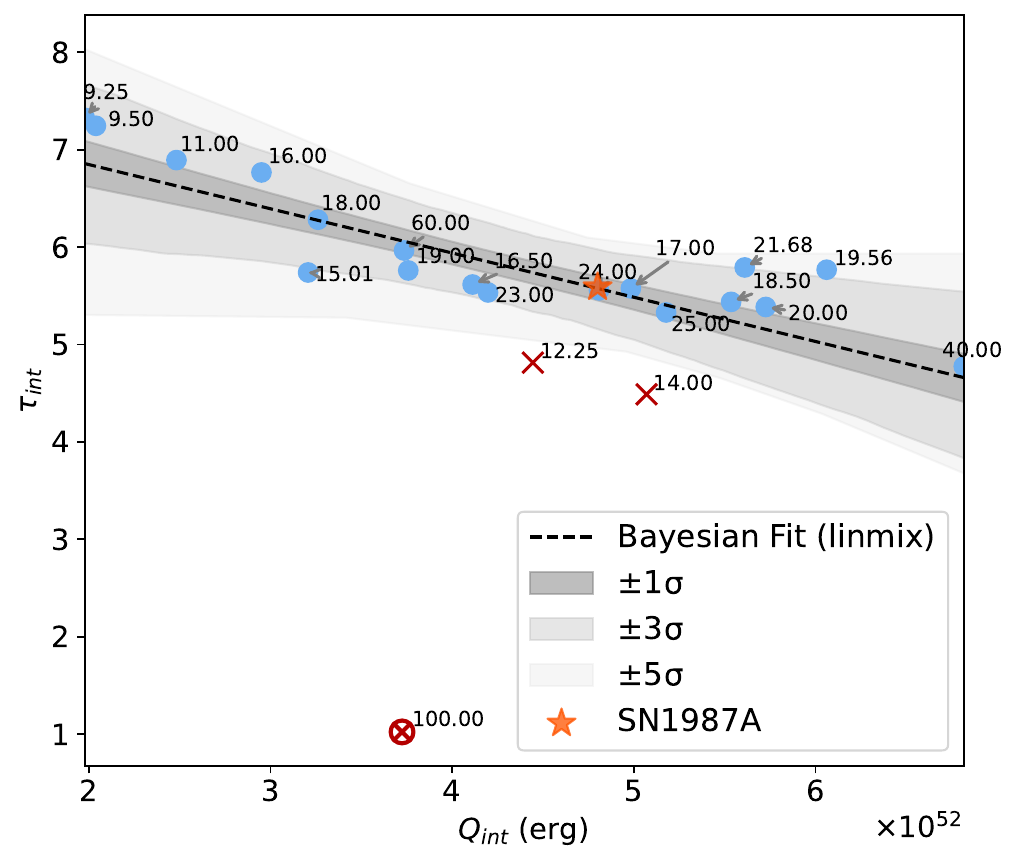}
  \caption{The horizontal axis represents the total energy released by each model within 2 seconds of explosion, while the vertical axis represents the sum of $\tau(t)$ over the same period. The numbers in the figure indicate the progenitor masses in units of solar mass.}\label{trend_envelope}
\end{figure}

\begin{figure}[H]
  \centering
  \includegraphics[width=0.55\linewidth]{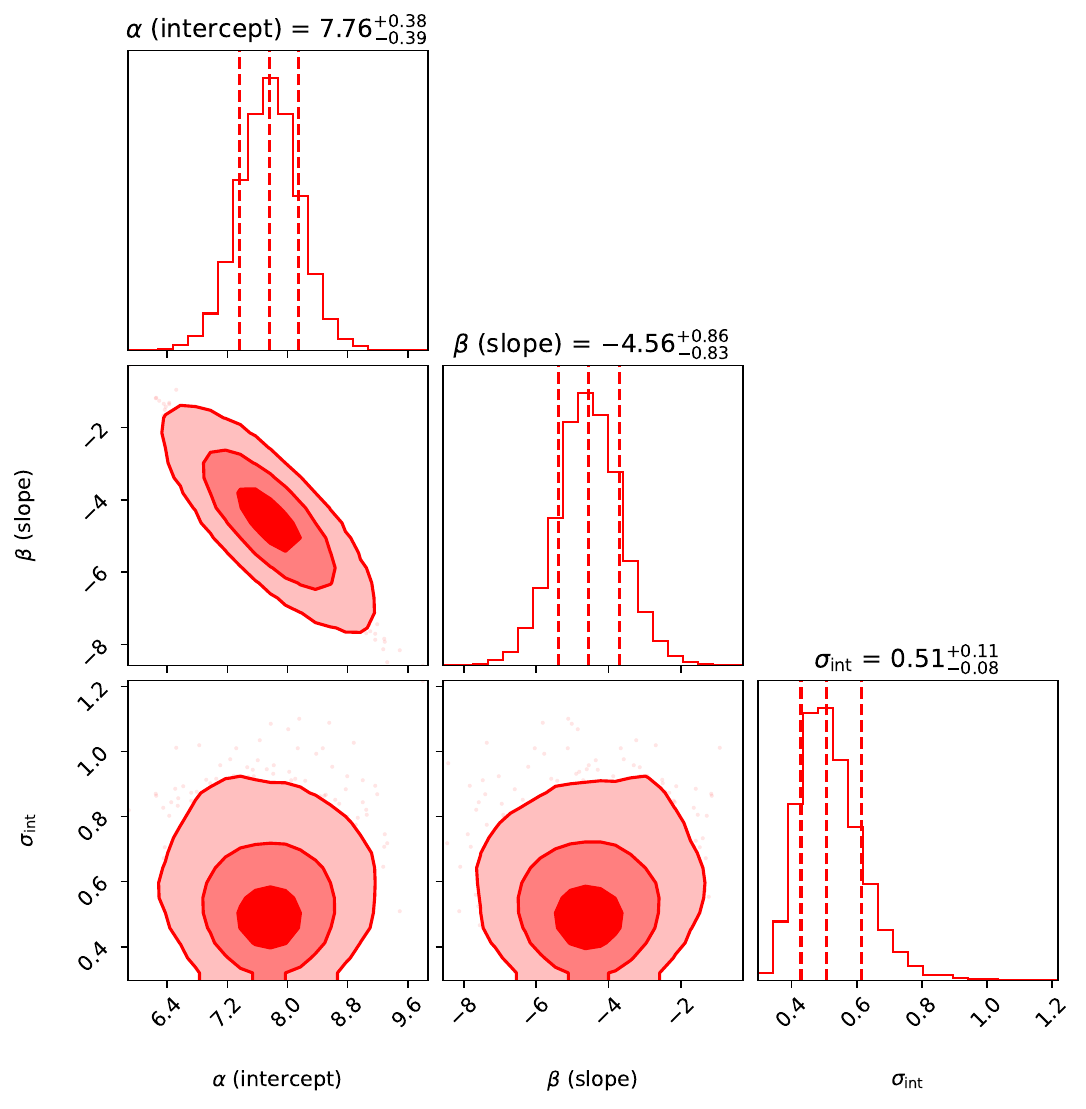}
\caption{
Corner plot of the posterior distributions for the intercept $\alpha$, slope $\beta$, 
and intrinsic scatter $\sigma_{\mathrm{int}}$ from the Bayesian \texttt{linmix} fit. 
Diagonal panels show marginalized one-dimensional histograms with median values and 
68\% credible intervals (dashed lines). Off-diagonal contours correspond to the 1$\sigma$, 3$\sigma$, and 5$\sigma$ credible regions. The fit converges well with a significantly nonzero slope, 
indicating a robust linear correlation: $\beta = -4.56^{+0.86}_{-0.83}$, $\alpha = 7.76^{+0.38}_{-0.39}$, and $\sigma_{\mathrm{int}} = 0.51^{+0.11}_{-0.08}$.}

  \label{fig:corner}
\end{figure}

The Bayesian linear regression with \texttt{linmix} converges well, yielding a significantly negative slope ($\beta = -4.56^{+0.86}_{-0.83}$), corresponding to a $\sim5.4\sigma$ detection away from zero. This indicates a robust negative linear correlation between the variables, with an intercept of $\alpha = 7.76^{+0.38}_{-0.39}$ and an intrinsic scatter of $\sigma_{\mathrm{int}} = 0.51^{+0.11}_{-0.08}$. The posterior distributions are unimodal and nearly Gaussian, further confirming the stability and reliability of the fit. The posterior of the intrinsic scatter is nearly flat across the allowed range, indicating weak constraints from the data; \autoref{fig:corner} therefore quote an upper limit on $\sigma_{int}$ rather than a point estimate. This suggests that the current dataset cannot tightly constrain any additional physical scatter, and that more data or smaller measurement uncertainties would be required to resolve its detailed distribution. The fitted relation is:
\begin{equation}
\tau_{int} = \left(7.76_{-0.39}^{+0.38}\right) - \left(4.56_{-0.83}^{+0.86}\right)\times 10^{-53}\,Q_{int}.
\end{equation}

The numbers next to each point indicate the progenitor mass (in solar masses). Blue points represent cases well described by a linear model, crosses indicate models that did not explode, and circled crosses represent failed explosions after initial shock revival. Our linear model captures the successfully exploding models within a 5-sigma interval, effectively defining a region predictive of explosion outcomes.

Note that the "sigma interval" here refers to the 'standard error of the mean prediction'. For a given $x = x_0$, the standard error of the predicted mean $\hat{y}(x_0)$ is:

\begin{equation}
SE[\hat{y}(x_0)] = s \cdot \sqrt{\frac{1}{n} + \frac{(x_0 - \bar{x})^2}{\sum_{i=1}^{n}(x_i - \bar{x})^2}},
\end{equation}

where the first term $\frac{1}{n}$ represents uncertainty in the sample mean, and the second term reflects greater prediction uncertainty for points far from the mean. The sigma interval thus indicates the plausible vertical shift of the regression trend line.

For SN 1987A, the limited temporal resolution of the observed neutrino events prevents a reliable reconstruction of the time-dependent spectral parameter $\tau(t)$, and thus makes it difficult to directly estimate the integrated quantity $\Sigma\tau$ required by the trend shown in \autoref{fig:trends}. Instead, we evaluate the linear relation at the observed total energy $Q = 4.8 \times 10^{52}$ erg for SN 1987A, yielding a corresponding vertical-axis value of approximately 5.6. 

Considering the mass range of Betelgeuse ($17 \sim 25$ solar masses) \citep{dolan2016evolutionary}, we filter the data and perform KDE analysis:

\begin{figure}[H]
  \centering
  \includegraphics[width=0.8\linewidth]{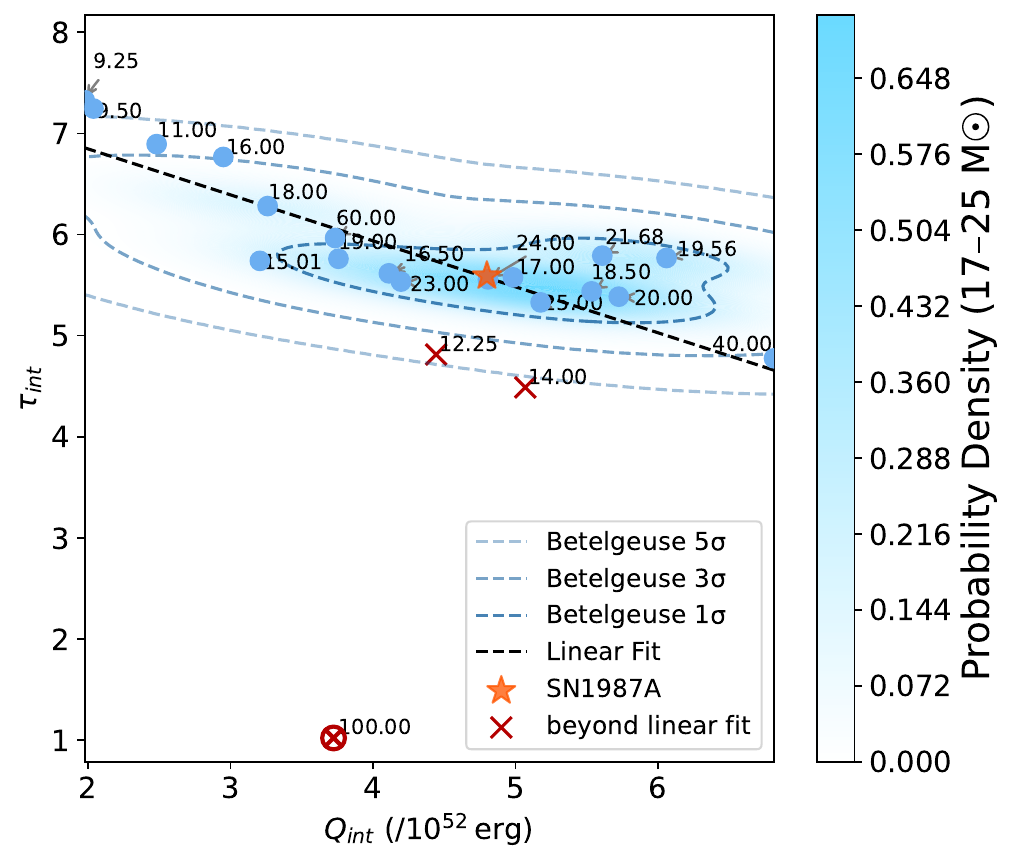}
  \caption{The dashed lines indicate the parameter space corresponding to the possible progenitor mass range of Betelgeuse at different confidence levels after explosion.}\label{fig:trends}
\end{figure}

But the result wasn’t great and didn’t give a strong constraint. This might mean we need more data for better analysis.

\section{Constraining the Progenitor Mass of SN\,1987A}

We constrain the progenitor mass of SN\,1987A by comparing its effective spectral parameter with a grid of 3D CCSN simulations. In the parameter space of \autoref{fig:trends}, the SN\,1987A point lies close to the model with a $24\,M_\odot$ progenitor. This proximity is suggestive but not definitive, given observational uncertainties, intrinsic model scatter, and possible parameter degeneracies. To obtain a more robust constraint, for each simulation we consider the product of two spectral parameters that quantifies the effective area over which the thermal flux---carrying explosion energy $Q$---acts within time $t$ on the diffusion scale $\ell_{\rm diff}$; larger values imply more efficient heat transport. The resulting distributions differ markedly above and below $40\,M_\odot$, indicating a potential transition in explosion behavior. Since spectroscopic analyses place SN\,1987A in the lower-mass regime, we restrict the inference to models with $M_{\star}\!<\!40\,M_\odot$.

\begin{figure}[H]
  \centering
  \includegraphics[width=\linewidth]{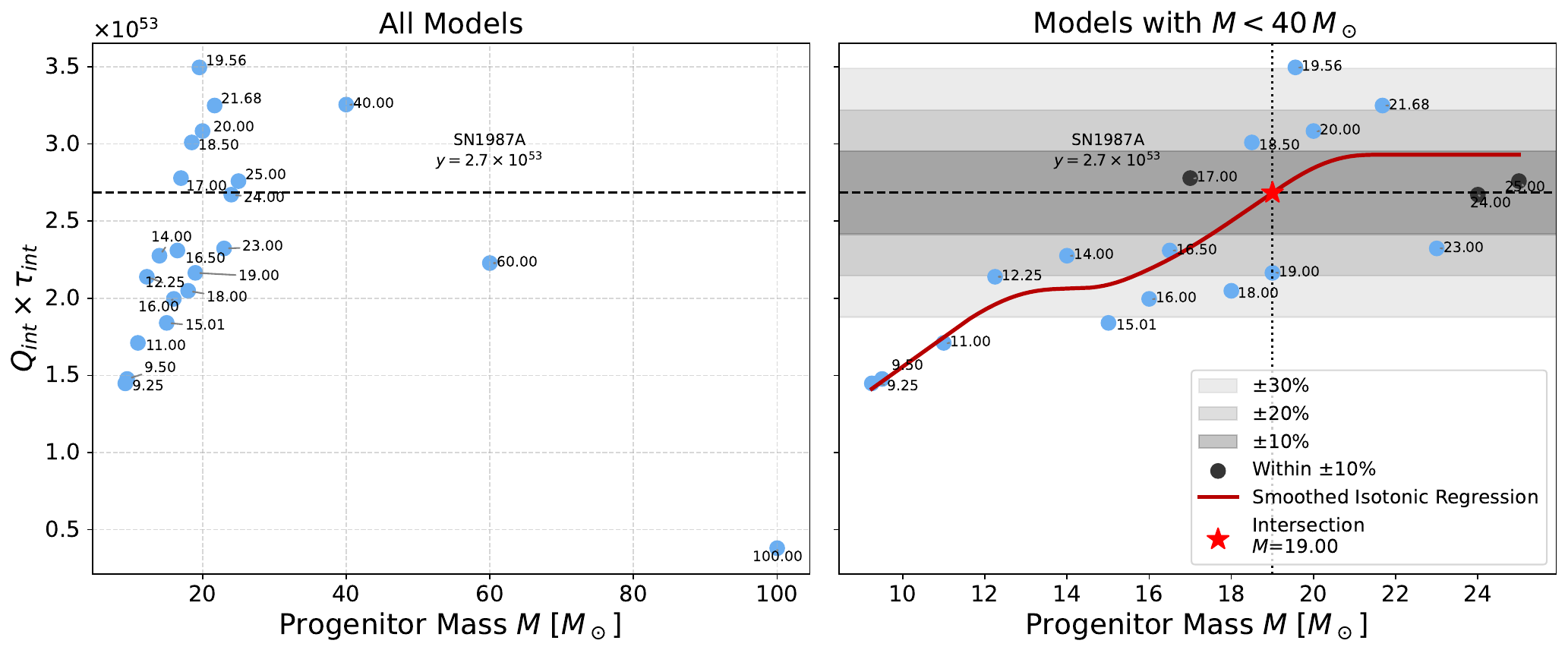}
  \caption{
  \textbf{Left:} Spectral parameter vs.\ progenitor mass; the dashed line marks SN\,1987A's reference value. Models below and above $40\,M_\odot$ occupy distinct regions.
  \textbf{Right:} Smoothed isotonic regression fit to the low-mass models (red), whose intersection with the SN\,1987A reference (red star) gives an estimated progenitor mass of $\sim19\,M_\odot$. Shaded bands denote 10\%, 20\%, and 30\% uncertainties in the parameter.
  }
  \label{supp:Estimated_Mass_SN1987A}
\end{figure}

To place the estimate on the same footing as the simulations, we first convert the parameter $Q$ defined in \autoref{linear relationship} from a neutrino-number-based quantity to neutrino luminosity (erg), obtaining $Q \simeq 5 \times 10^{52}\,\mathrm{erg}$. Applying smoothed isotonic regression \citep{Isotonic} to the low-mass subset yields a best-estimate progenitor mass of $\sim19\,M_\odot$. With a 10\% parameter uncertainty, the inferred range is consistent with stellar-evolution estimates \citep{1989ARA&A..27..629A,2016ApJ...821...38S}; a 30\% uncertainty, however, broadens the interval to encompass nearly the full low-mass model set, underscoring the premium that measurement precision places on progenitor constraints.

We adopt isotonic regression to relate the derived physical parameter to progenitor mass because the simulations exhibit a monotonic but non-linear trend with an inflection near $40\,M_\odot$. While piecewise-linear fits could mimic this behavior, they require an explicit break point and impose a rigid functional form. By contrast, isotonic regression is non-parametric and shape-constrained: it flexibly tracks the monotone trend without presupposing a specific model, thereby preserving the intrinsic structure expected from the underlying physics. For the numerical implementation, we made use of the \texttt{IsotonicRegression} module from the Python package \texttt{scikit-learn} \citep{scikit-learn} together with the \texttt{Statsmodels} package \citep{seabold2010statsmodels}.

\newpage
\section{Parameter Evolution}\label{P:Parameter Evolution}

 To investigate the physical information revealed by the temporal evolution of the model parameters, we examined the evolution curves of $\tau$, $Q$, the gravitational-wave amplitude, and the mean energy parameter.

\subsection{Failed Supernova Explosion}

\begin{figure*}[h]
    \centering 

    \includegraphics[width=0.48\textwidth]{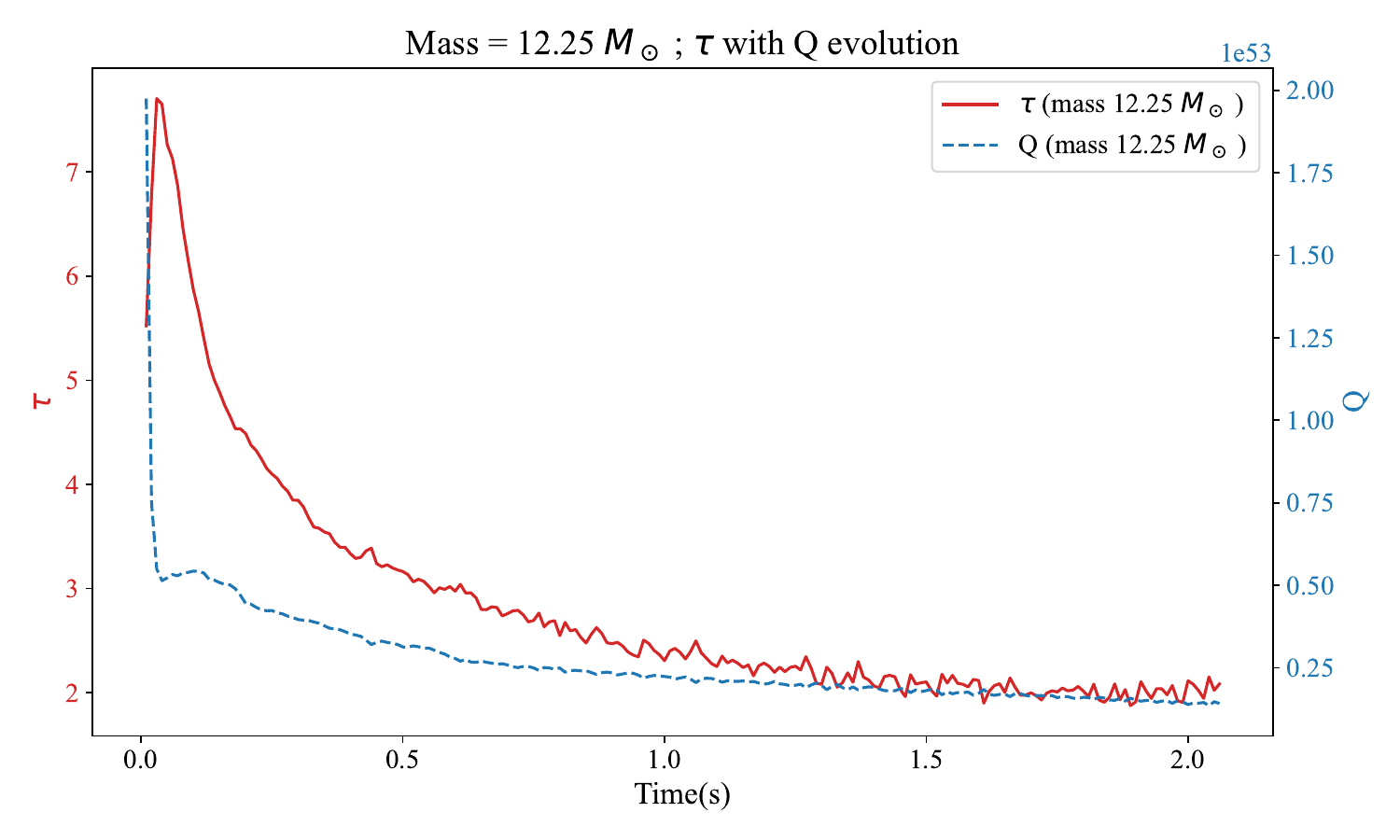}
    \hfill 
    \includegraphics[width=0.48\textwidth]{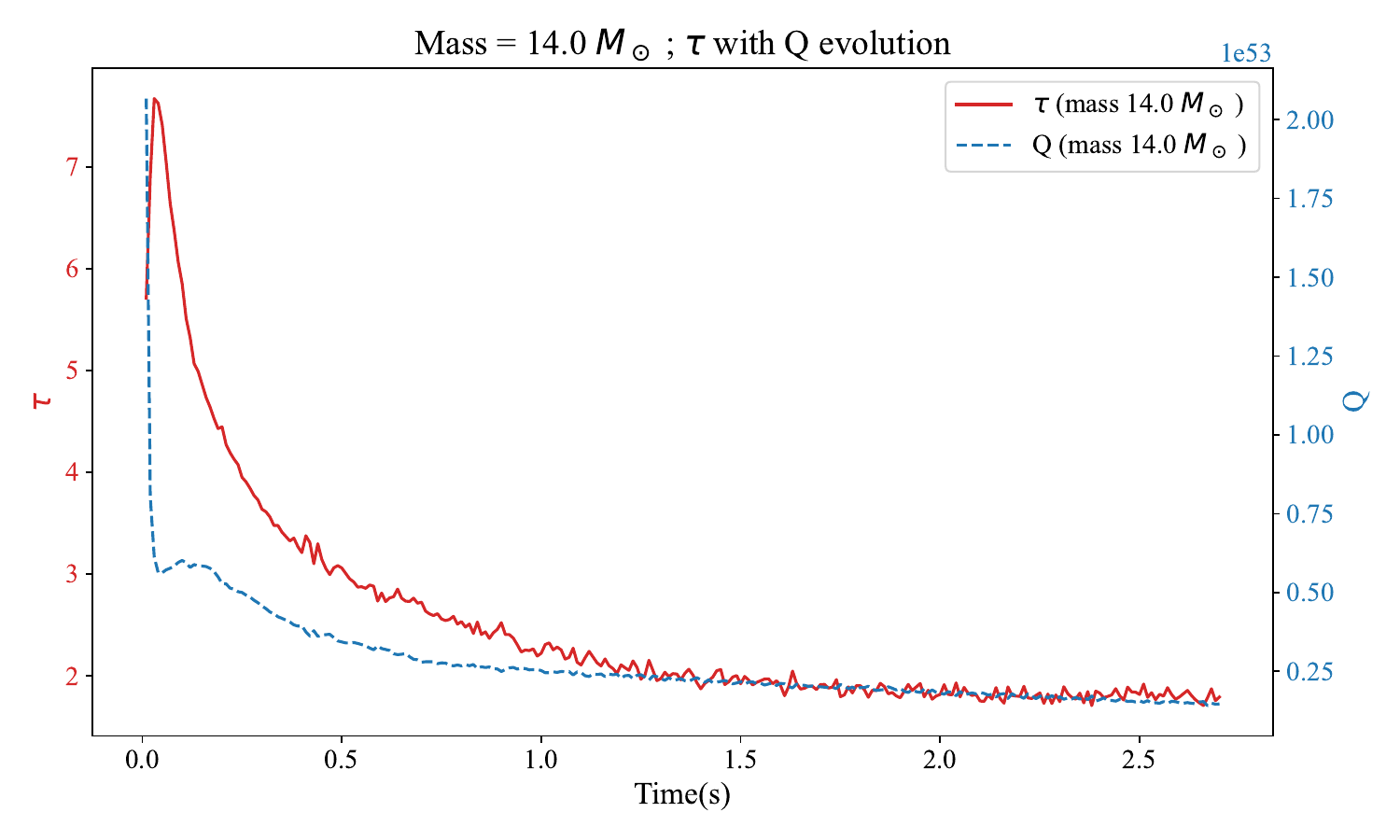}

    \vspace{5mm} 

    \includegraphics[width=0.48\textwidth]{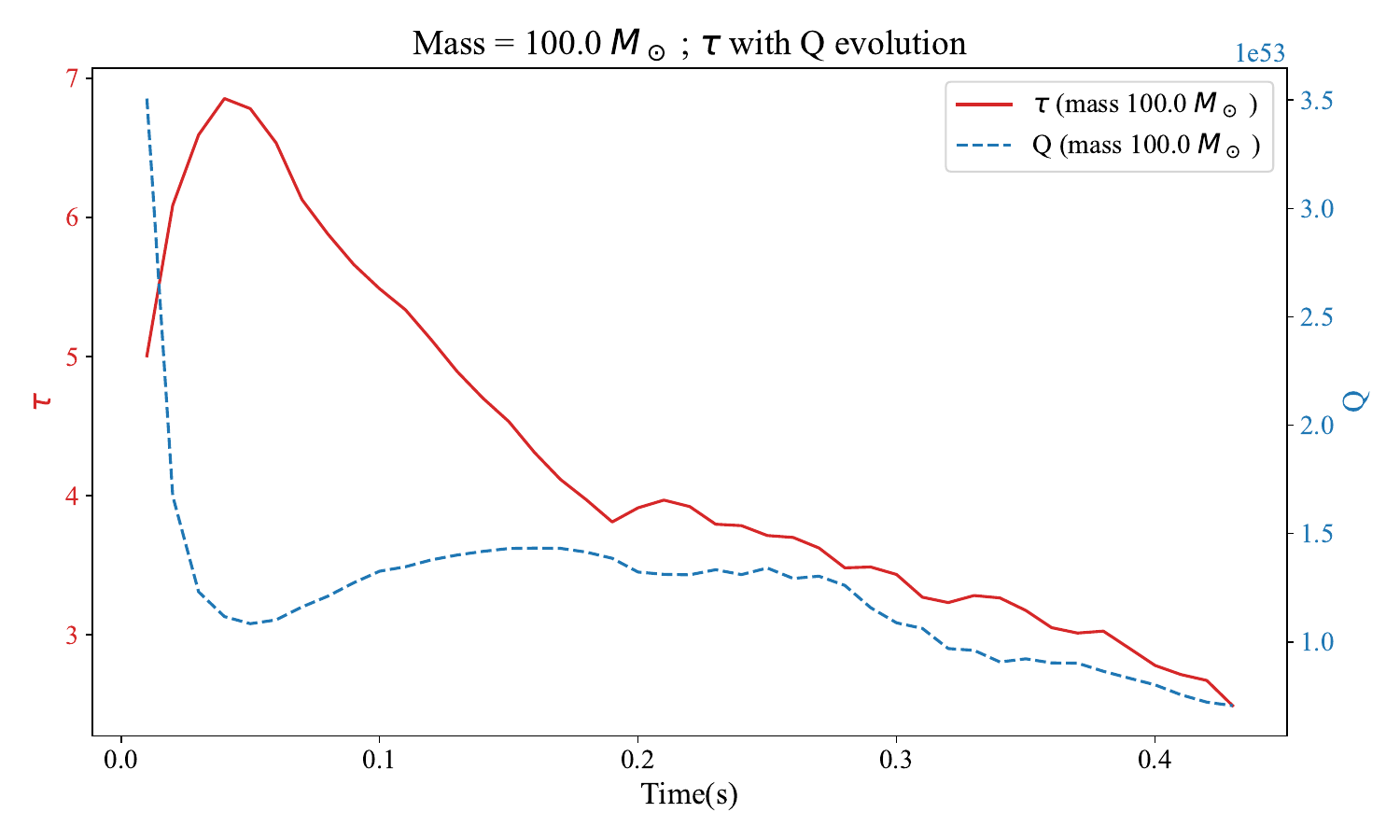}
    \hfill 

    \caption{Parameter evolution diagram of failed supernova explosions.}
    \label{fig:all}
\end{figure*}

\clearpage
\newpage

\subsection{Successfully Exploding Supernova}

\begin{figure*}[h]
    \centering 

    \includegraphics[width=0.48\textwidth]{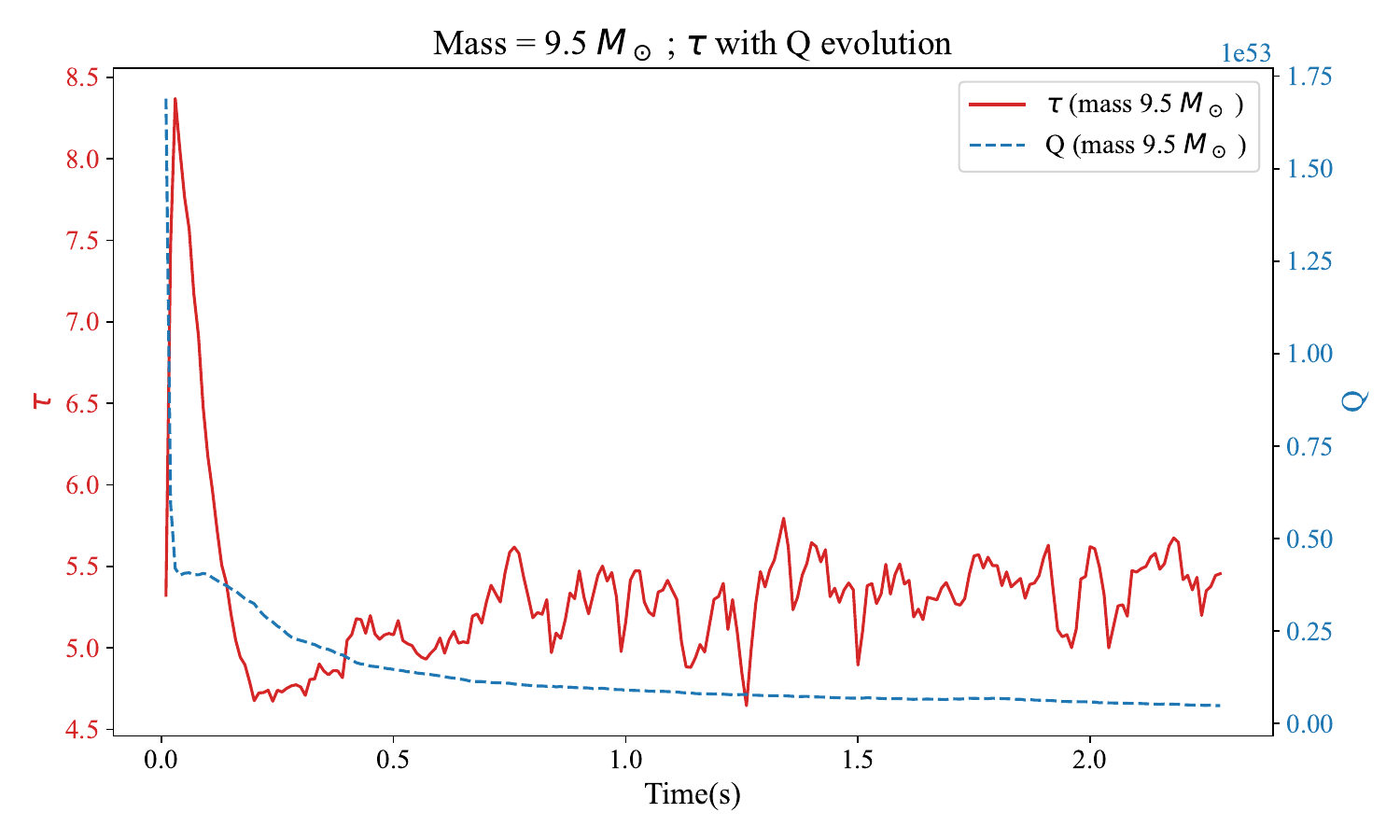}
    \hfill 
    \includegraphics[width=0.48\textwidth]{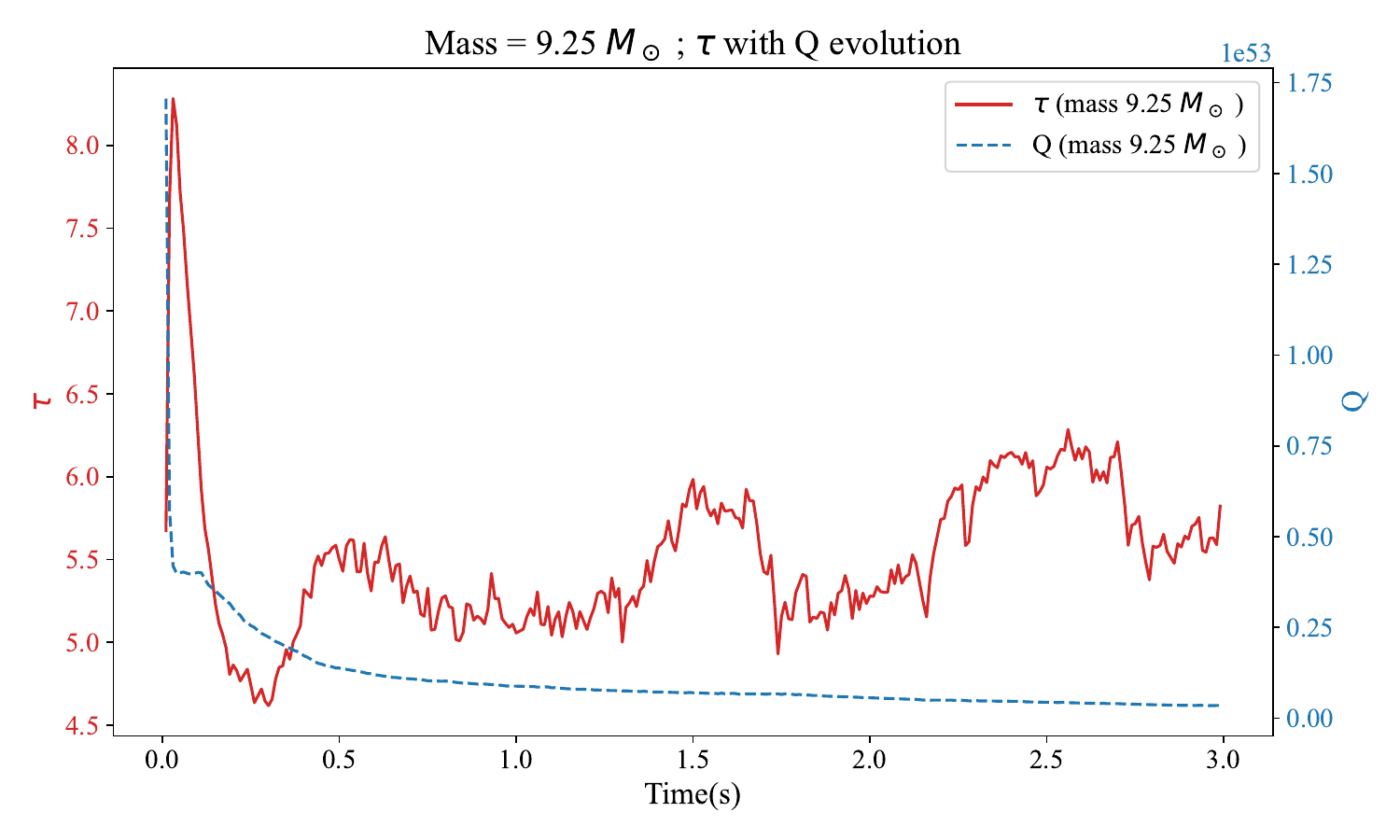}

    \vspace{5mm} 

    \includegraphics[width=0.48\textwidth]{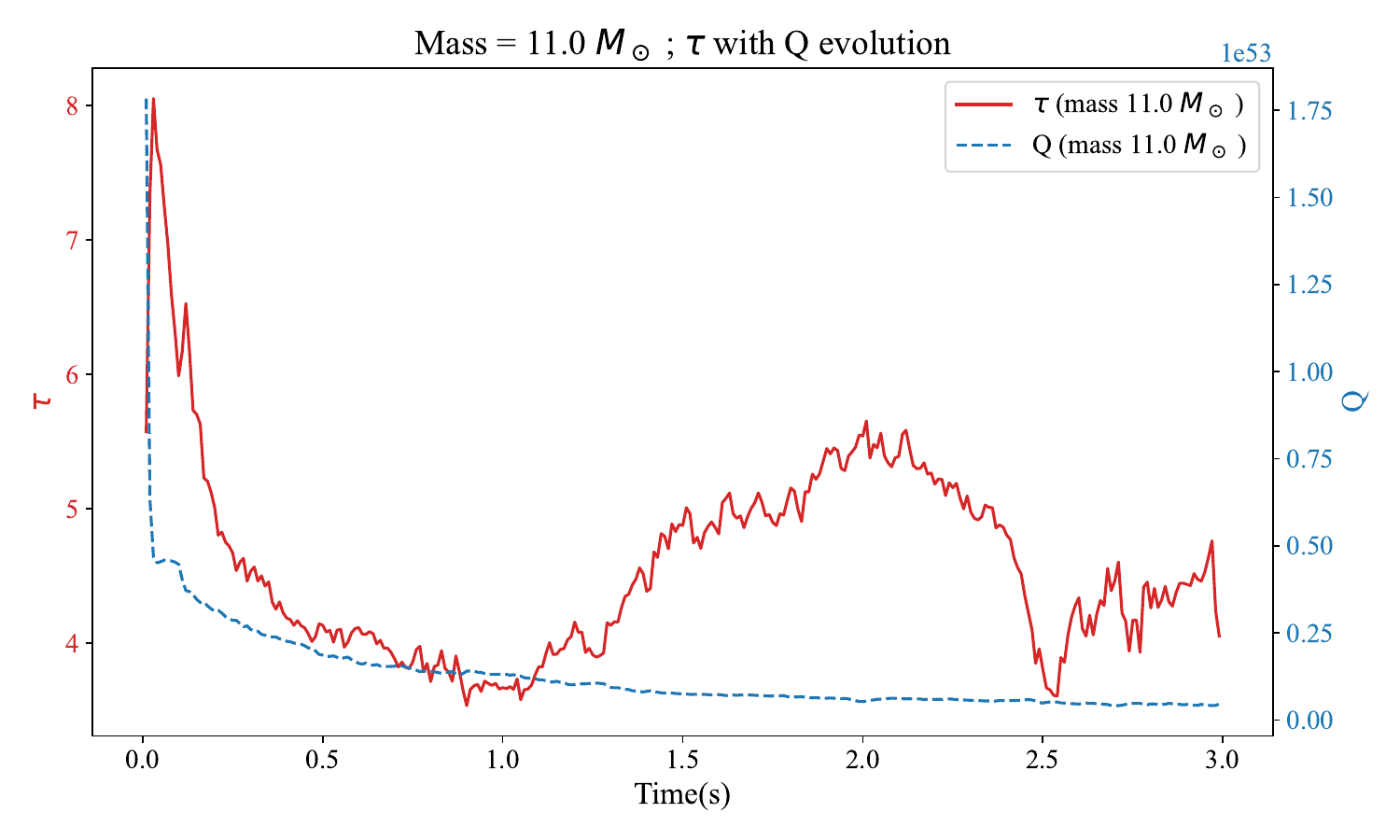}
    \hfill 
\includegraphics[width=0.48\textwidth]{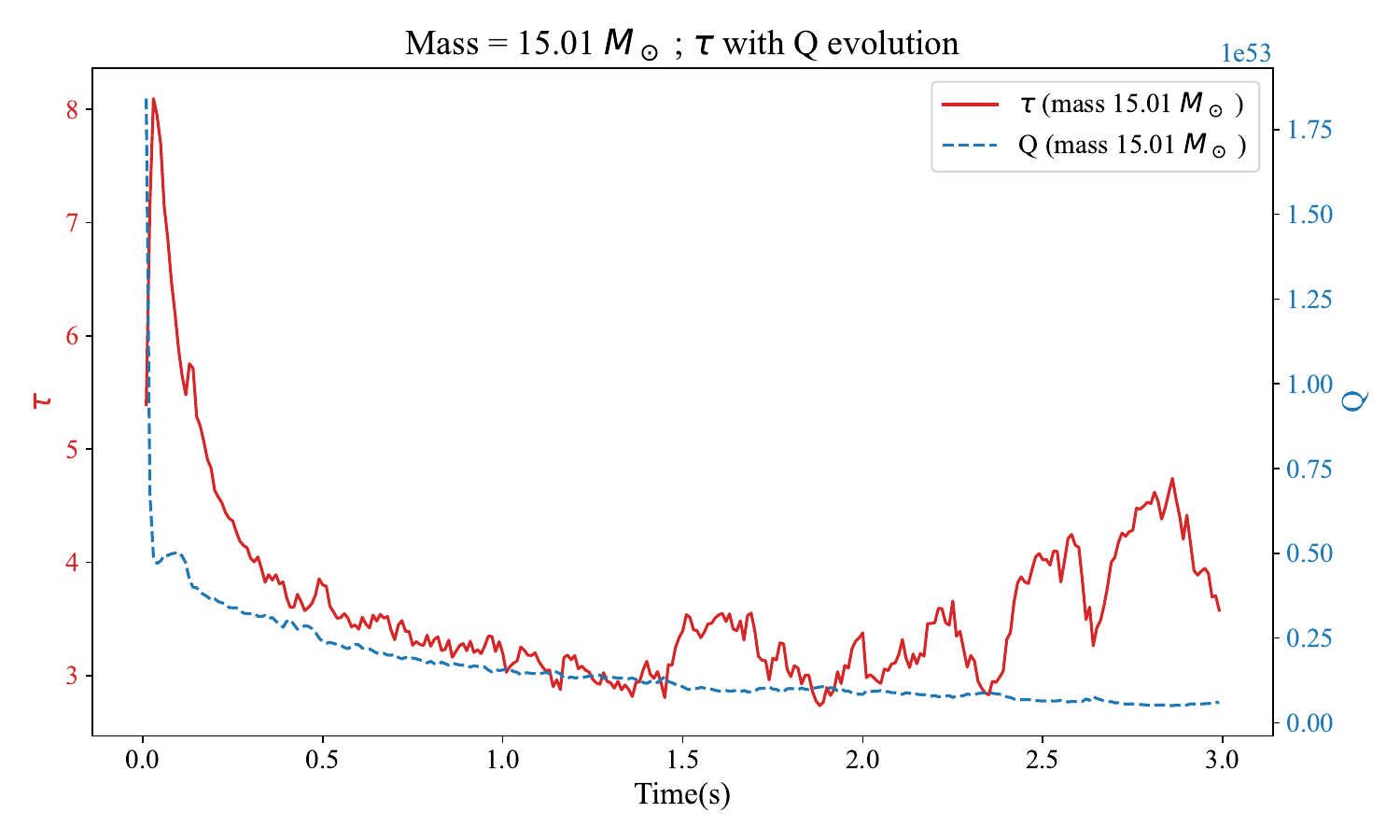}

    \vspace{5mm} 

    \includegraphics[width=0.48\textwidth]{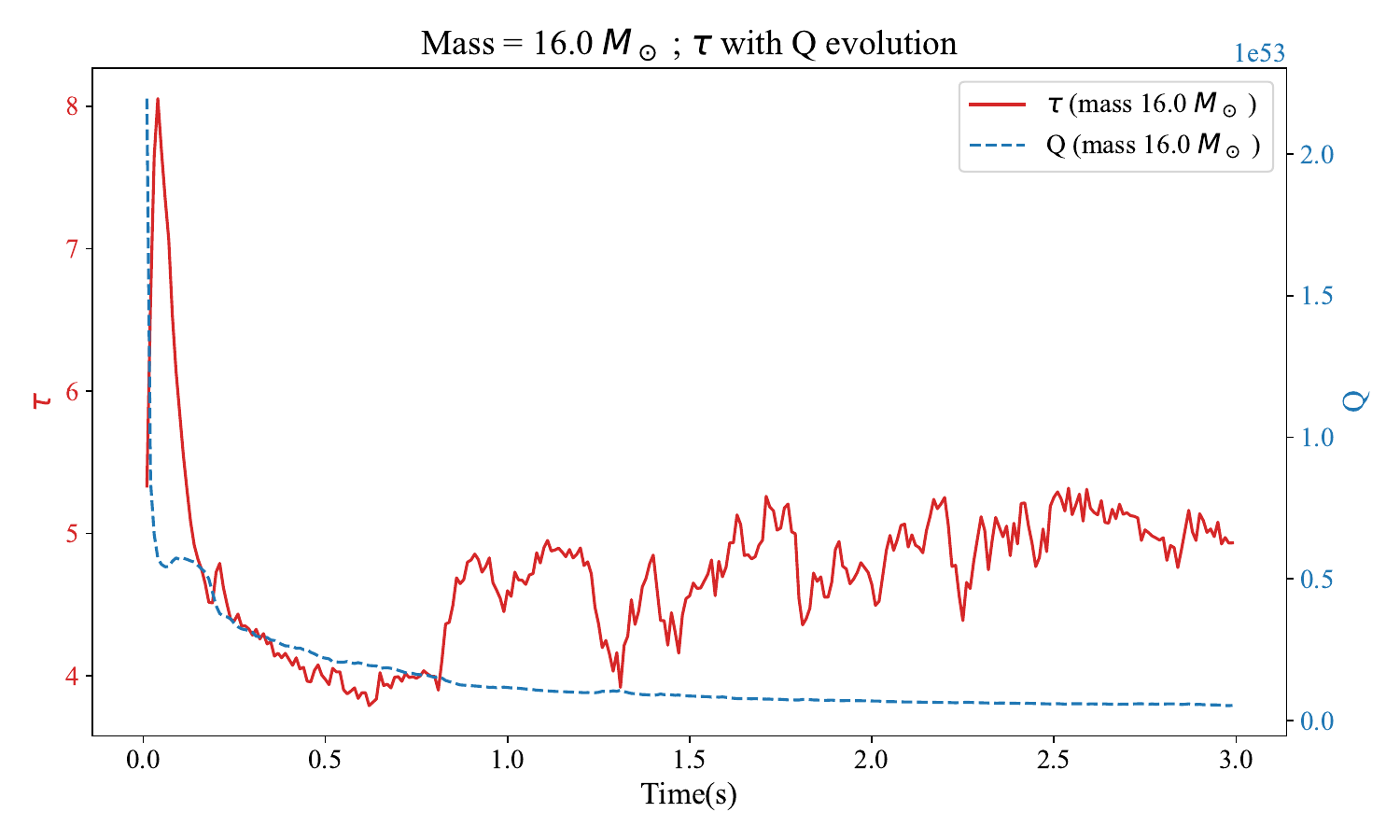}
    \hfill 
\includegraphics[width=0.48\textwidth]{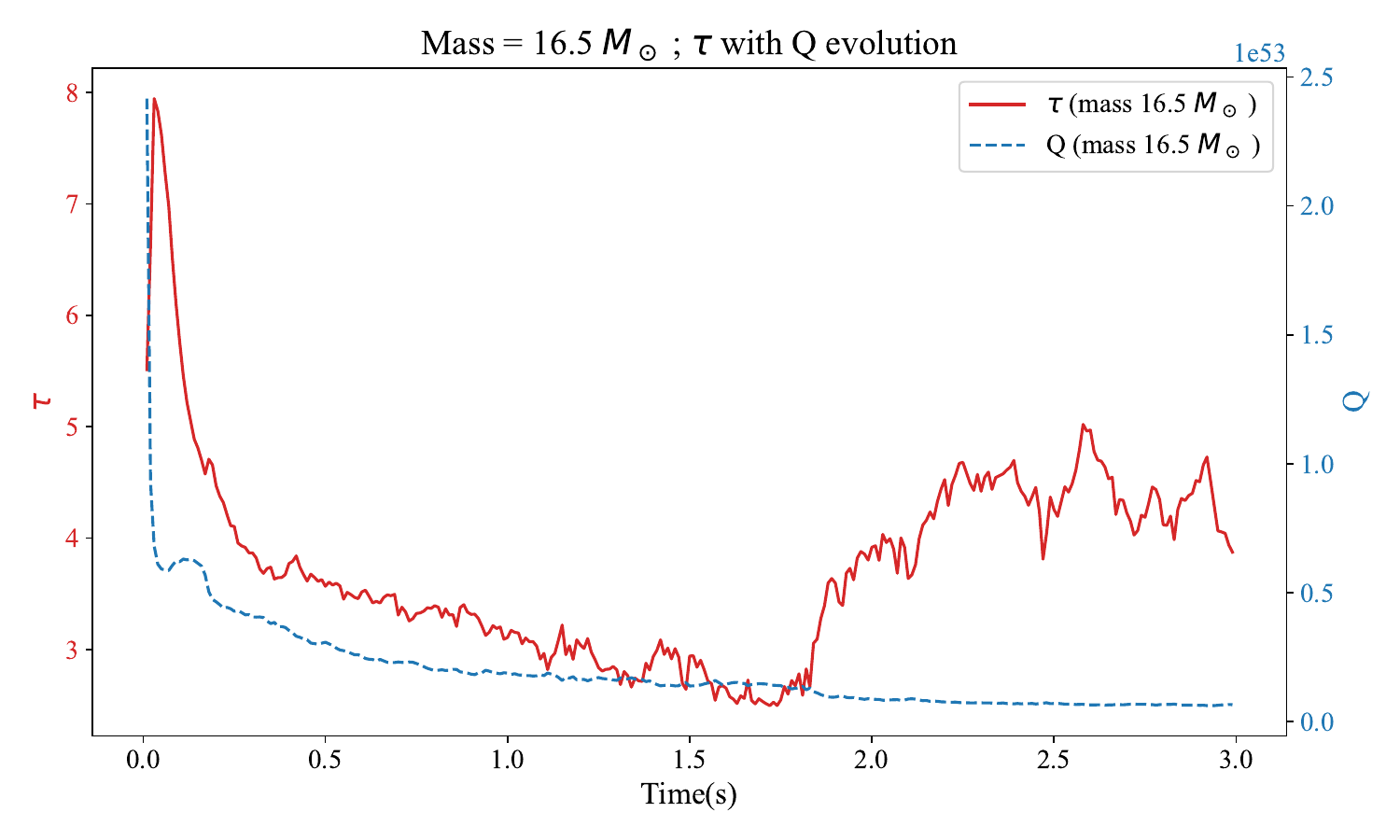}
    \hfill 

    \caption{Evolution diagram of successful supernova explosion parameters.}
    \label{fig:all}
\end{figure*}

\begin{figure*}[h]
    \centering 

    \includegraphics[width=0.48\textwidth]{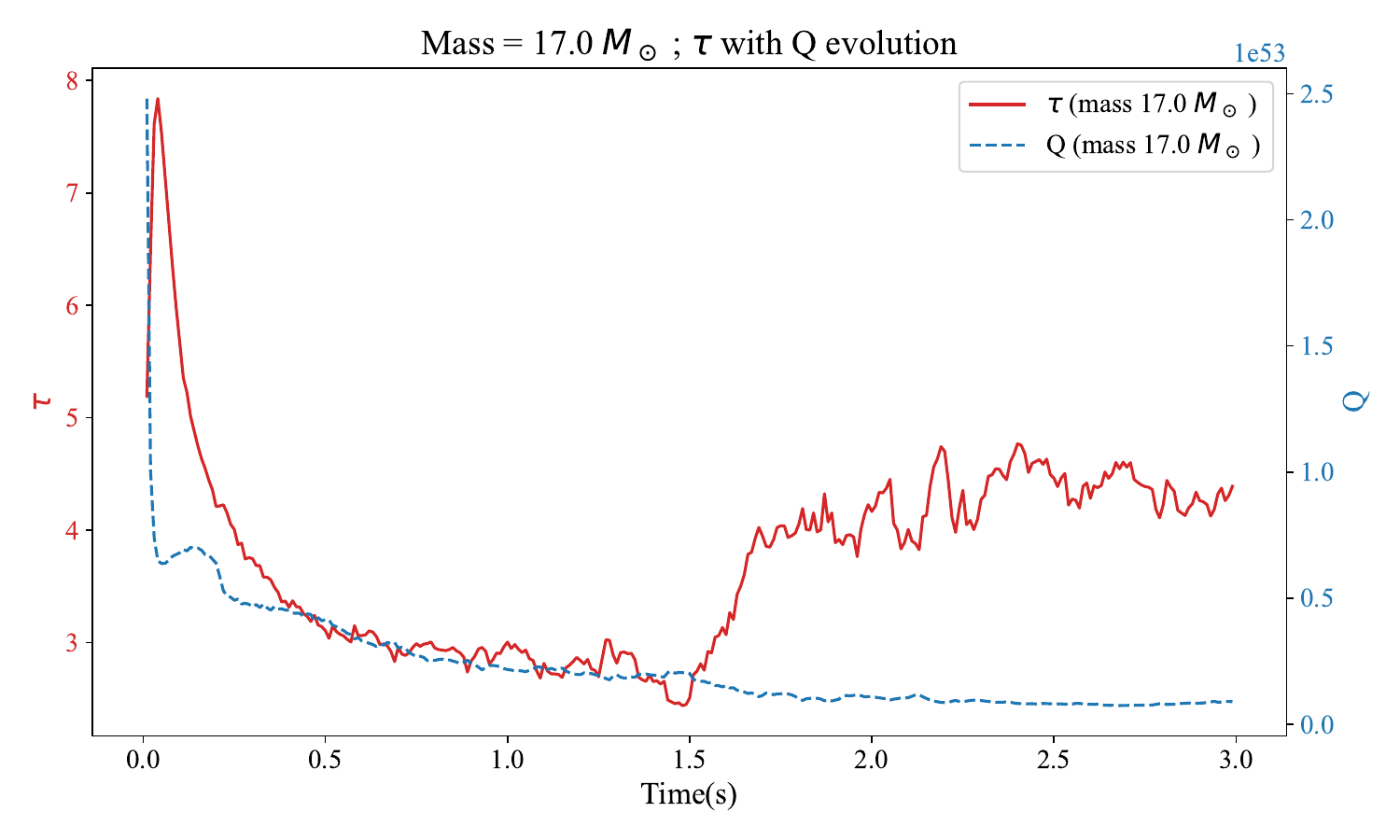}
    \hfill 
    \includegraphics[width=0.48\textwidth]{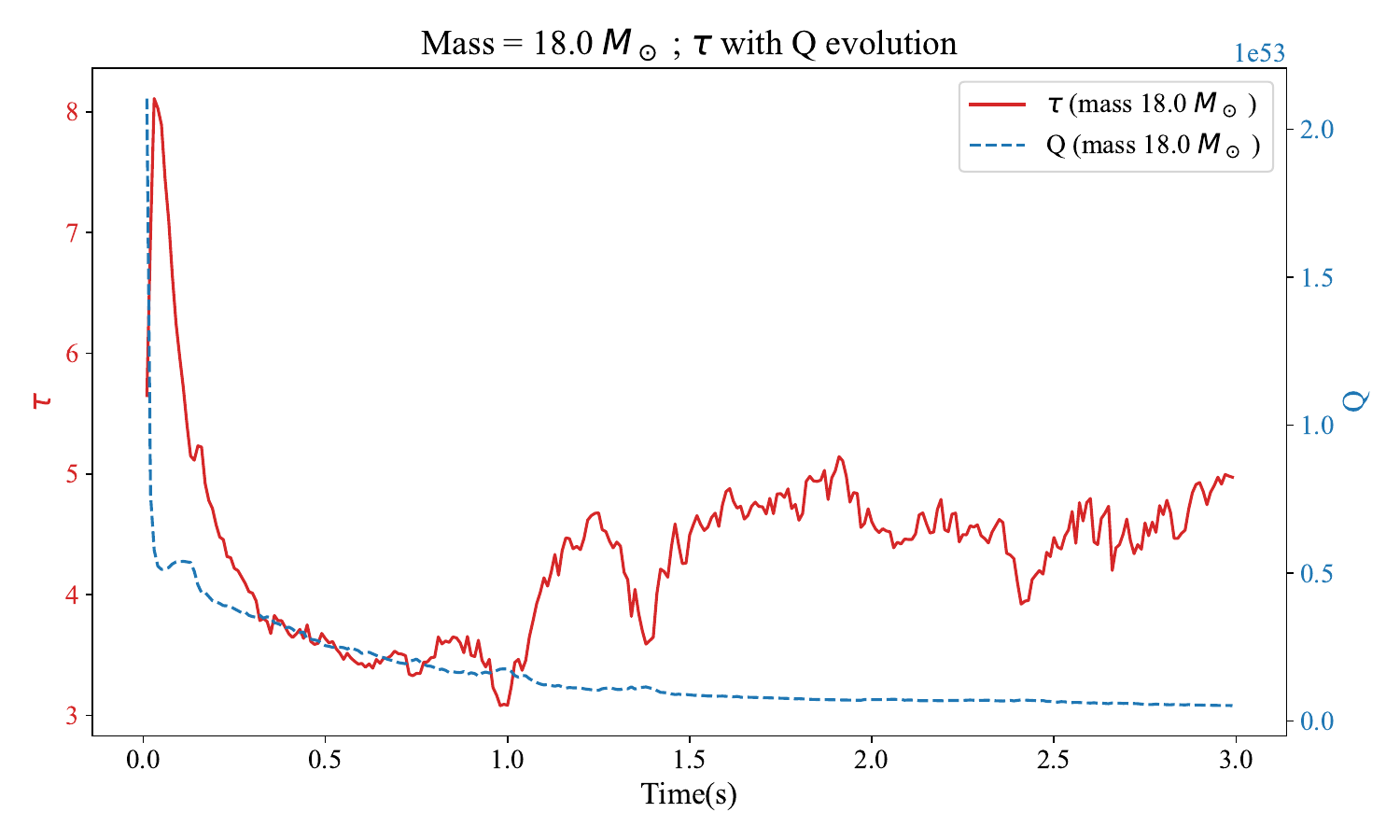}

    \vspace{5mm} 

    \includegraphics[width=0.48\textwidth]{mass_11.0_plot.pdf}
    \hfill 
\includegraphics[width=0.48\textwidth]{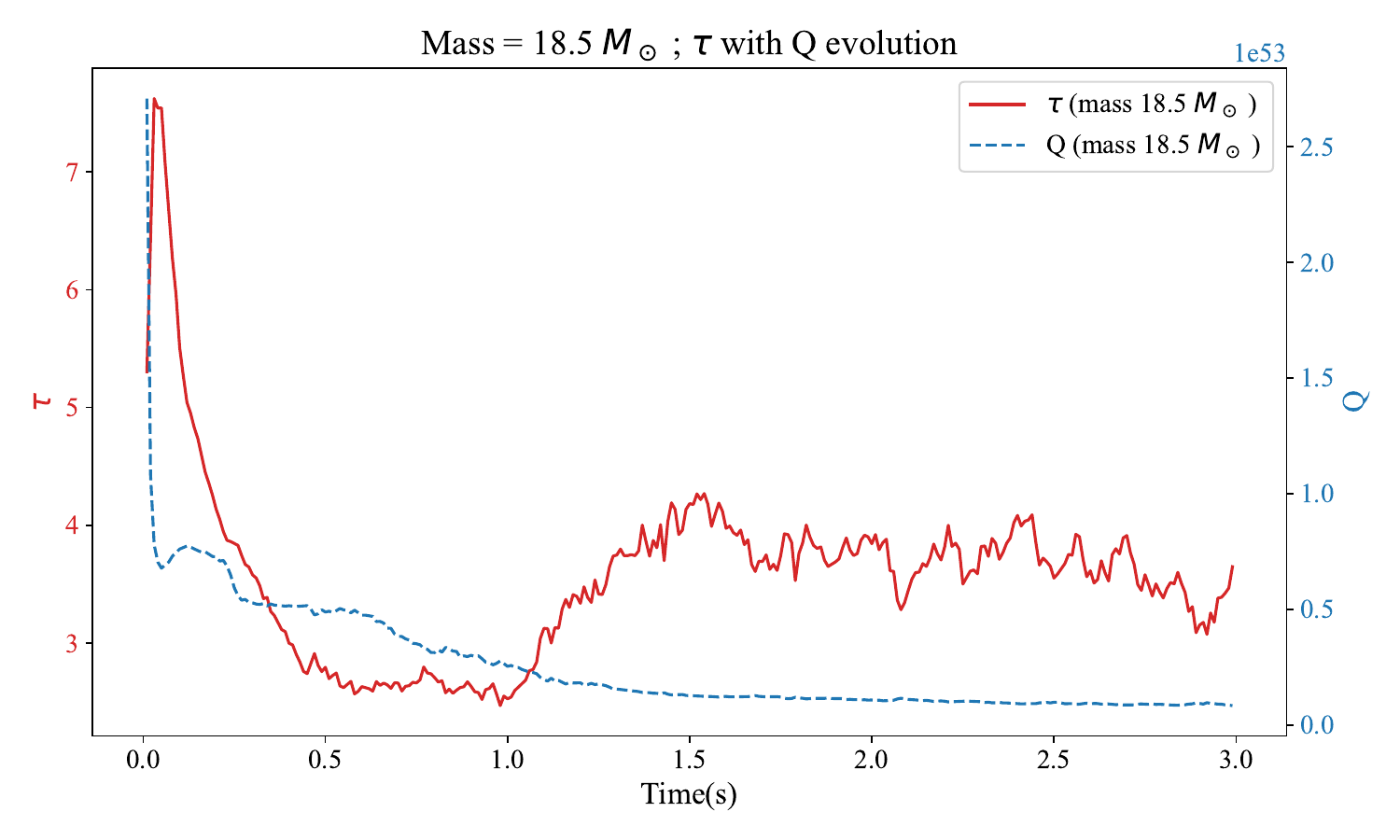}

    \vspace{5mm} 

    \includegraphics[width=0.48\textwidth]{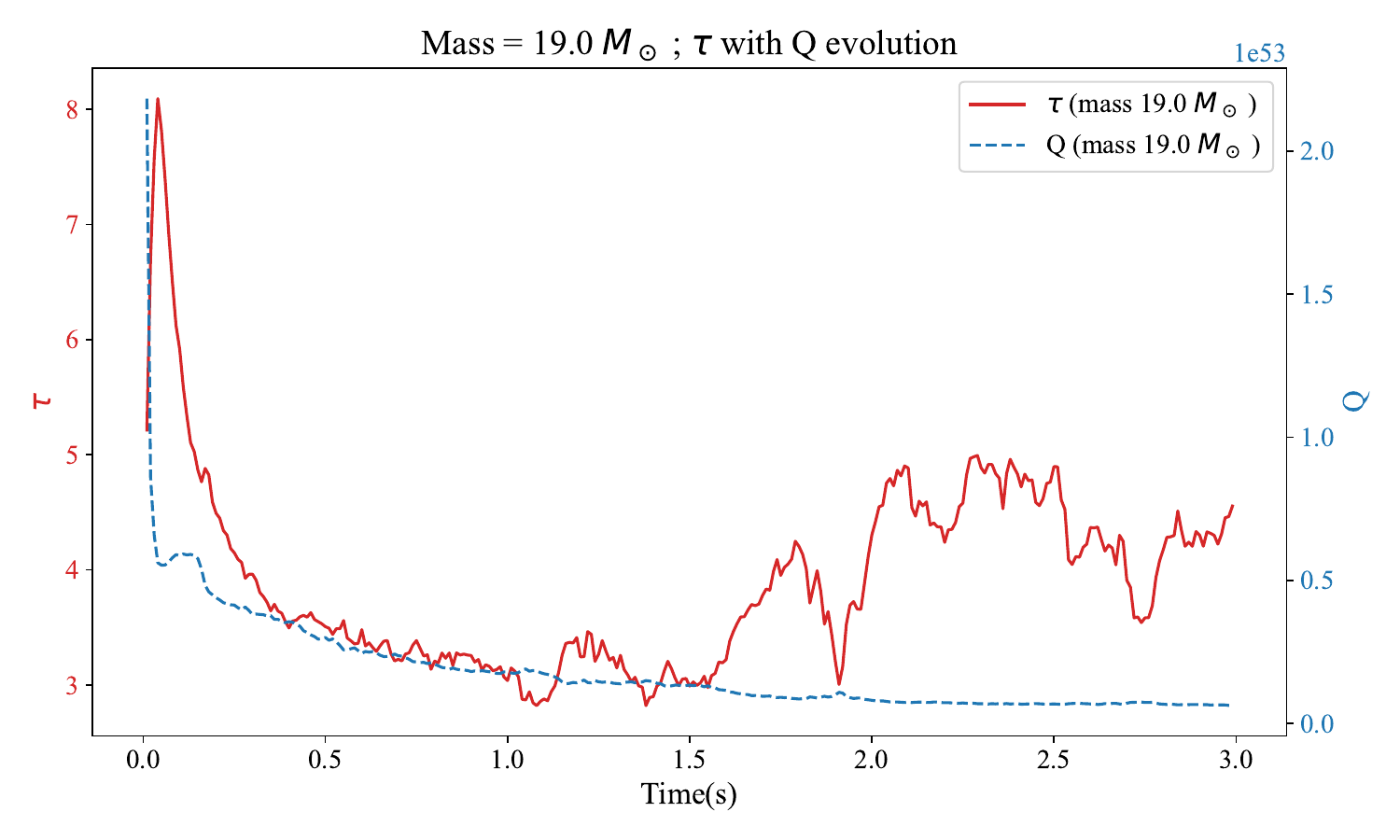}
    \hfill 
\includegraphics[width=0.48\textwidth]{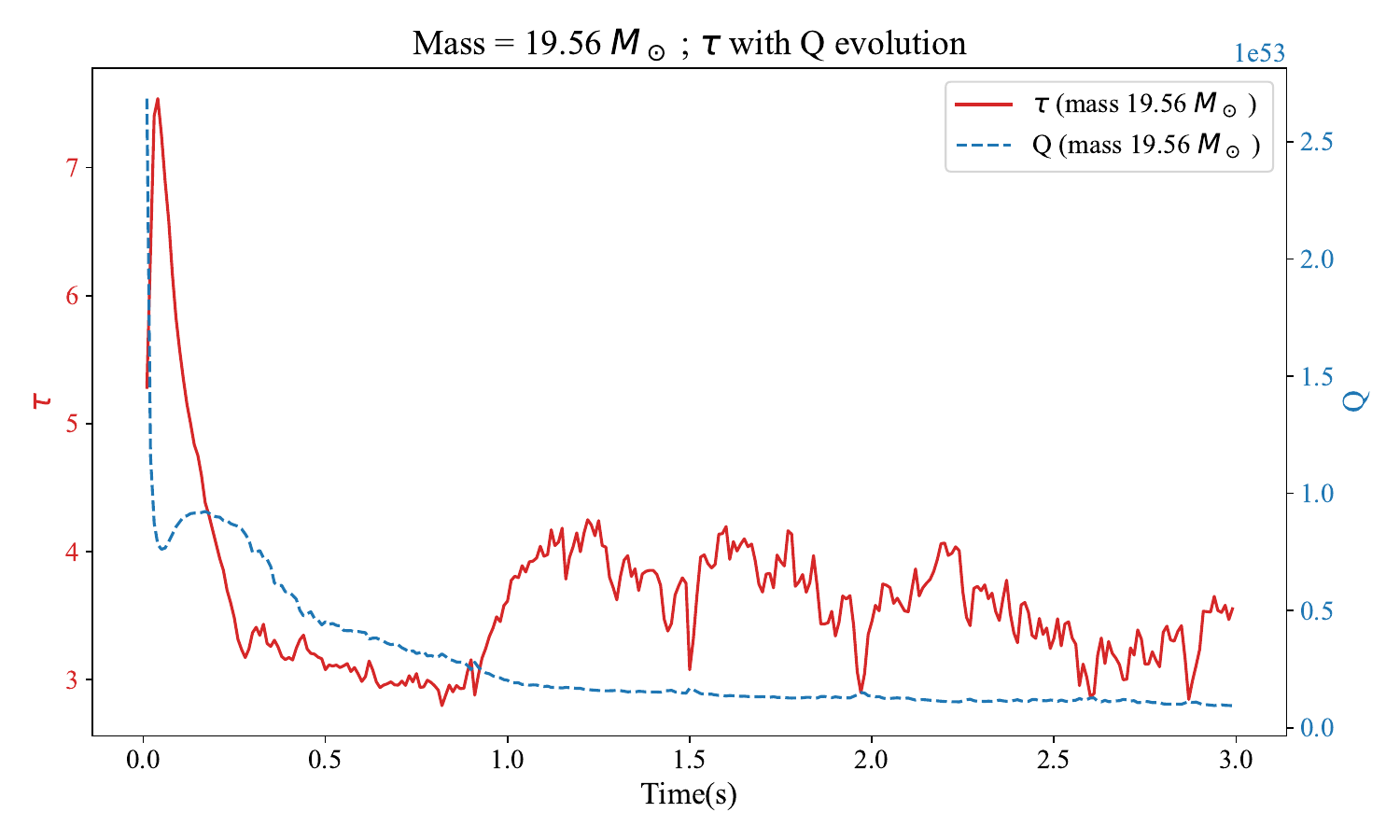}
    \hfill 

    \caption{Evolution diagram of successful supernova explosion parameters.}
    \label{fig:all}
\end{figure*}

\begin{figure*}[h]
    \centering 

    \includegraphics[width=0.48\textwidth]{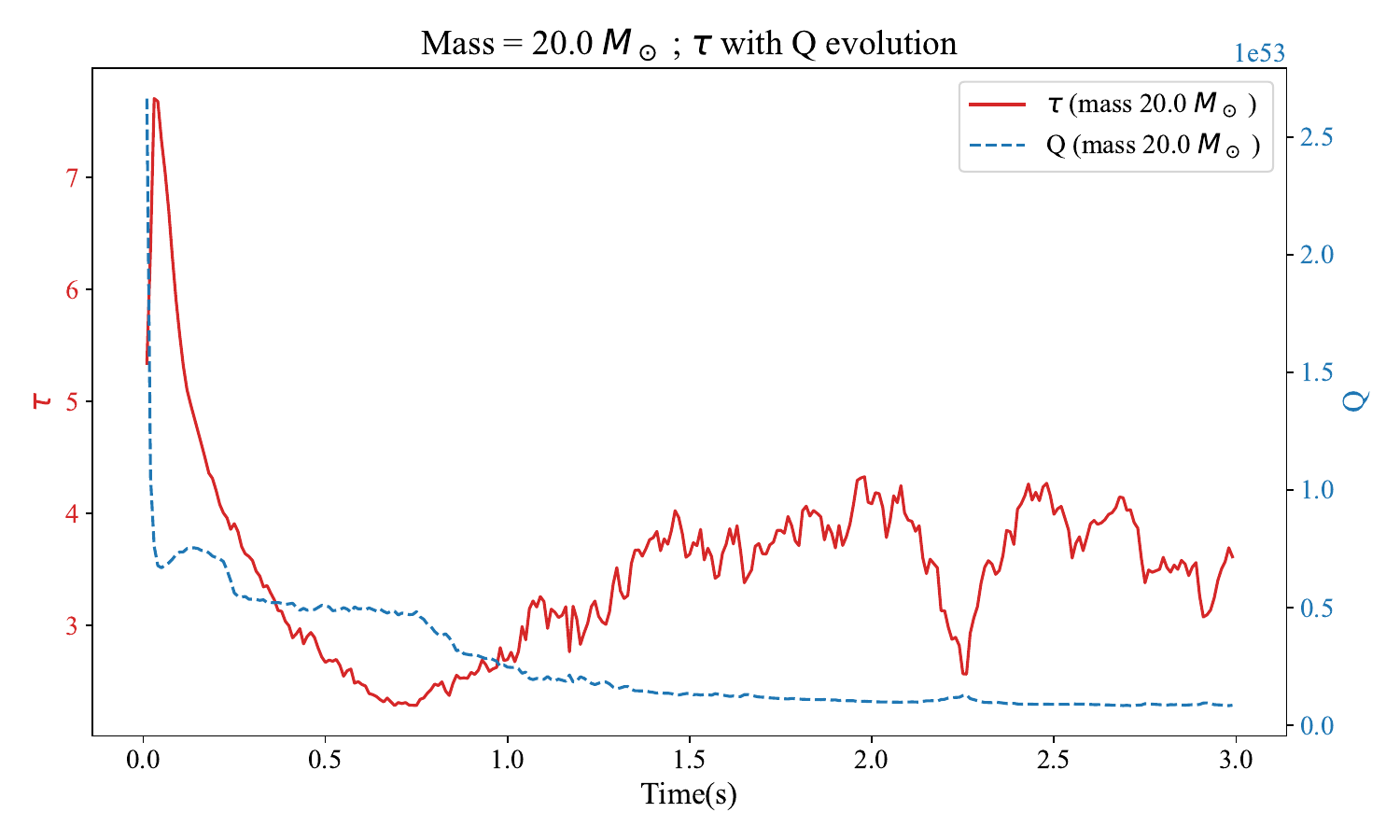}
    \hfill 
    \includegraphics[width=0.48\textwidth]{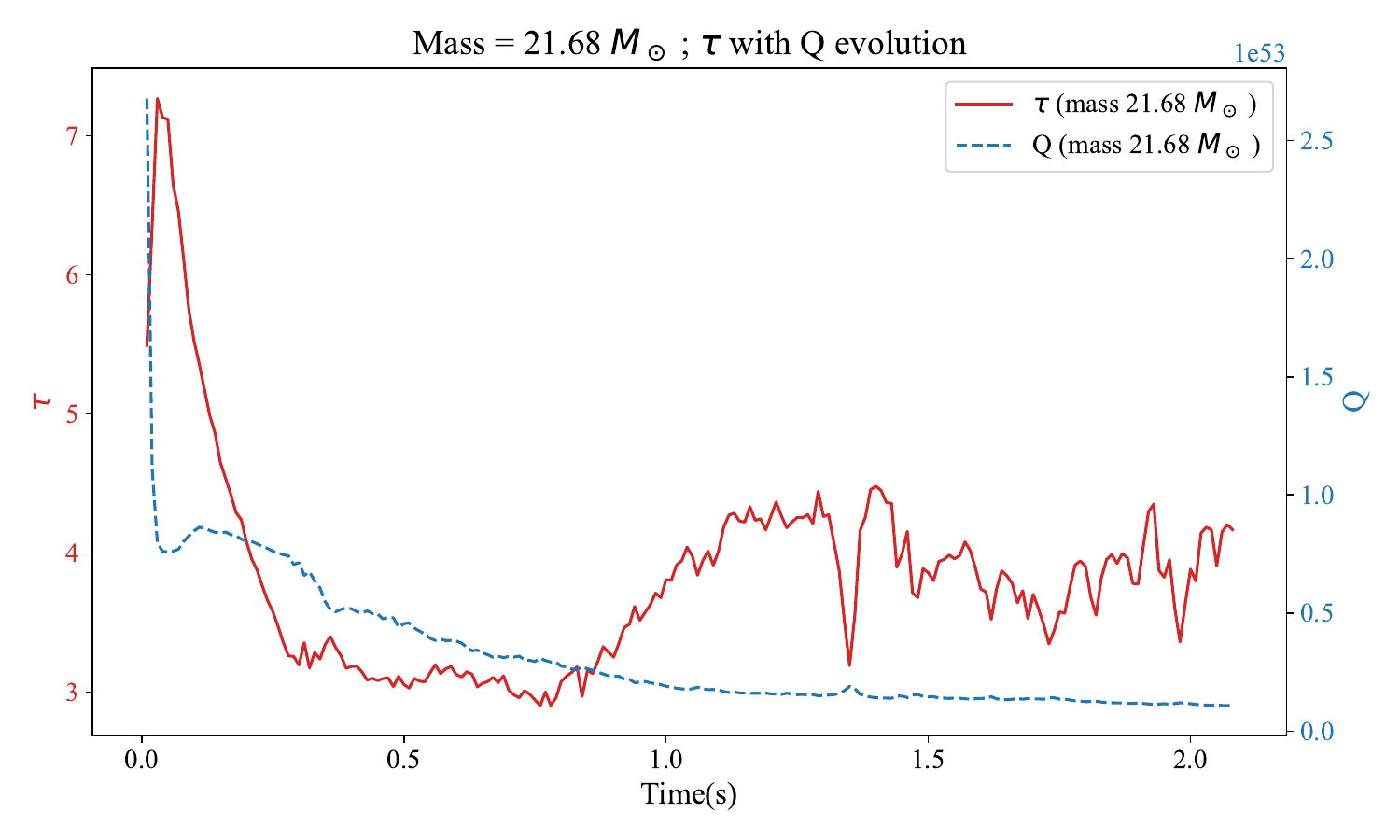}

    \vspace{5mm} 

    \includegraphics[width=0.48\textwidth]{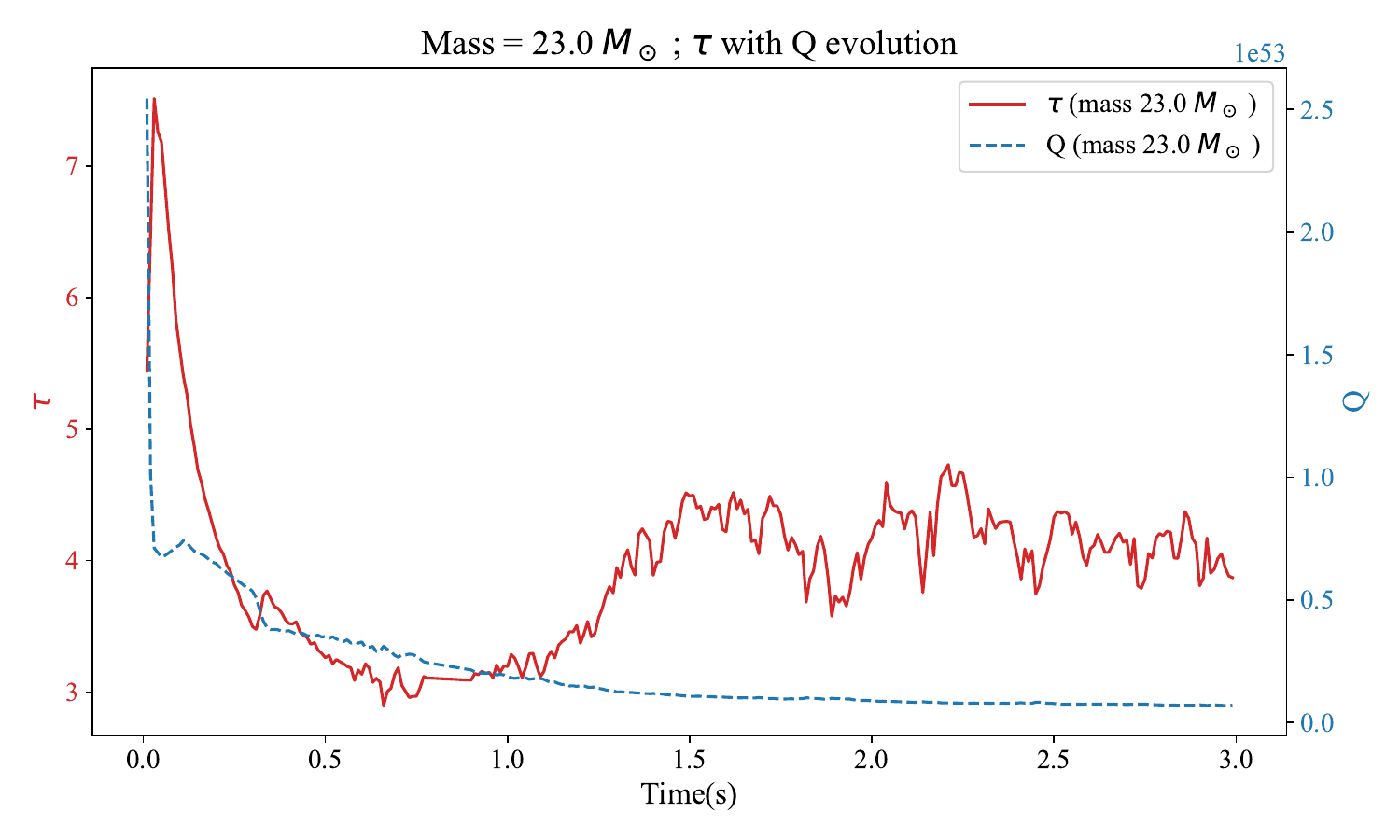}
    \hfill 
\includegraphics[width=0.48\textwidth]{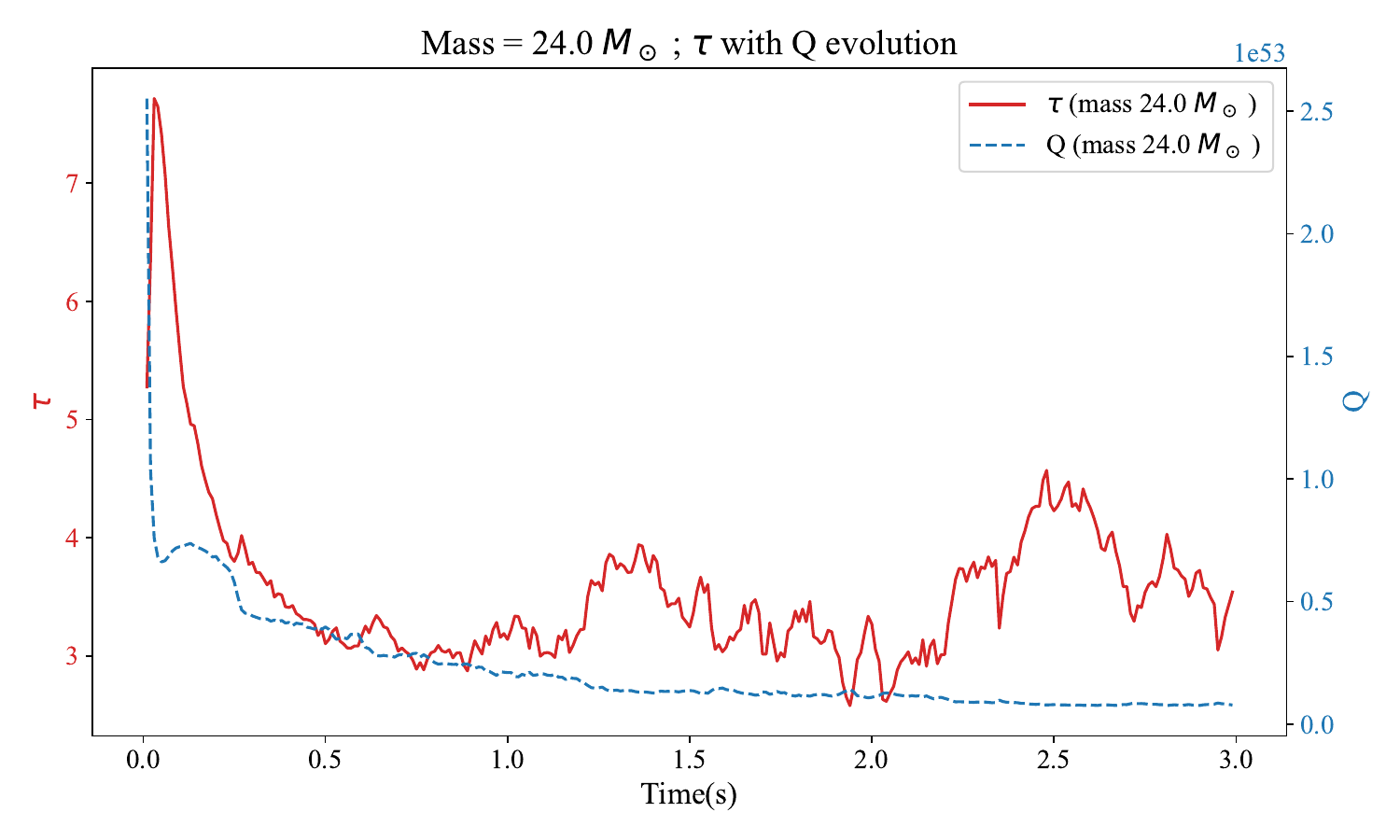}

    \vspace{5mm} 

    \includegraphics[width=0.48\textwidth]{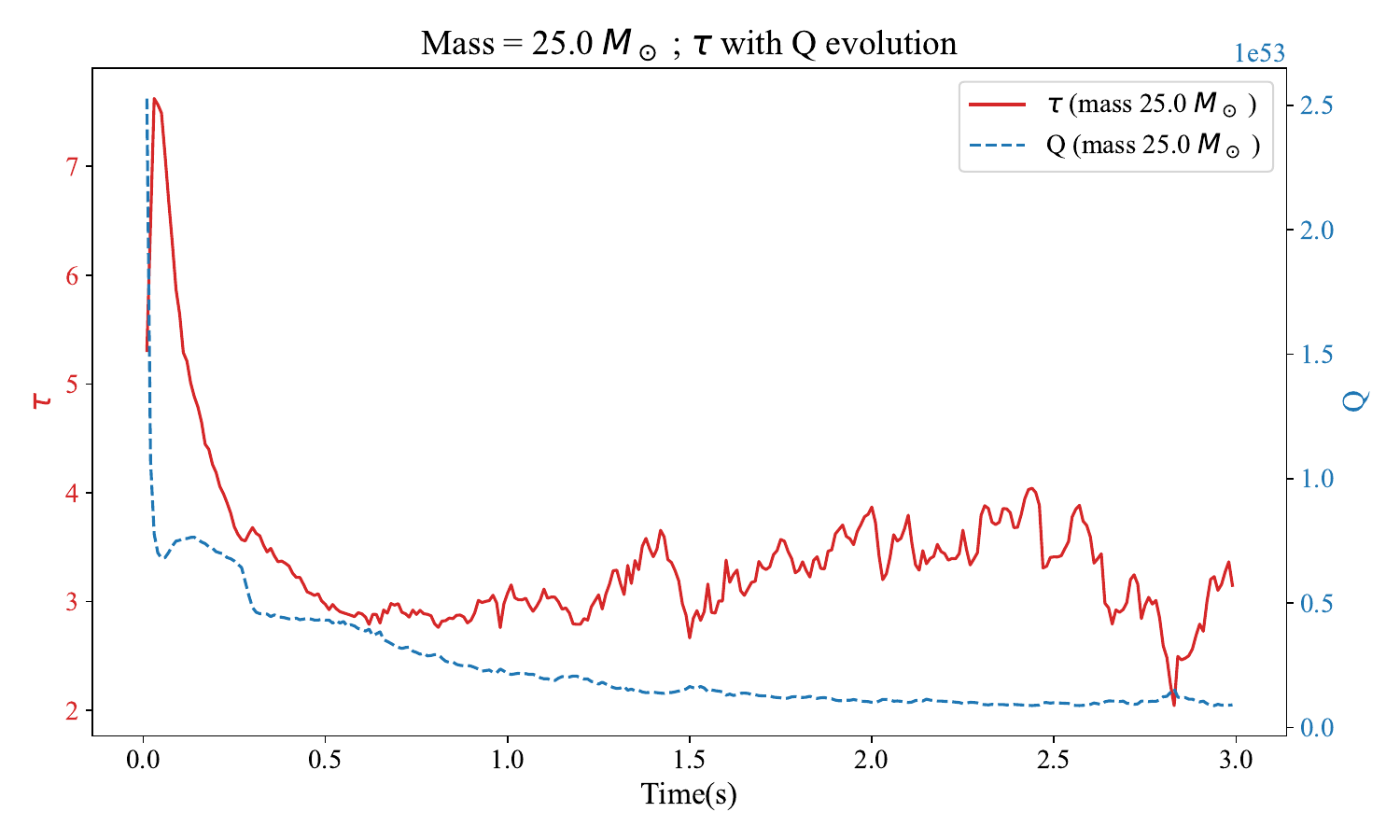}
    \hfill 
\includegraphics[width=0.48\textwidth]{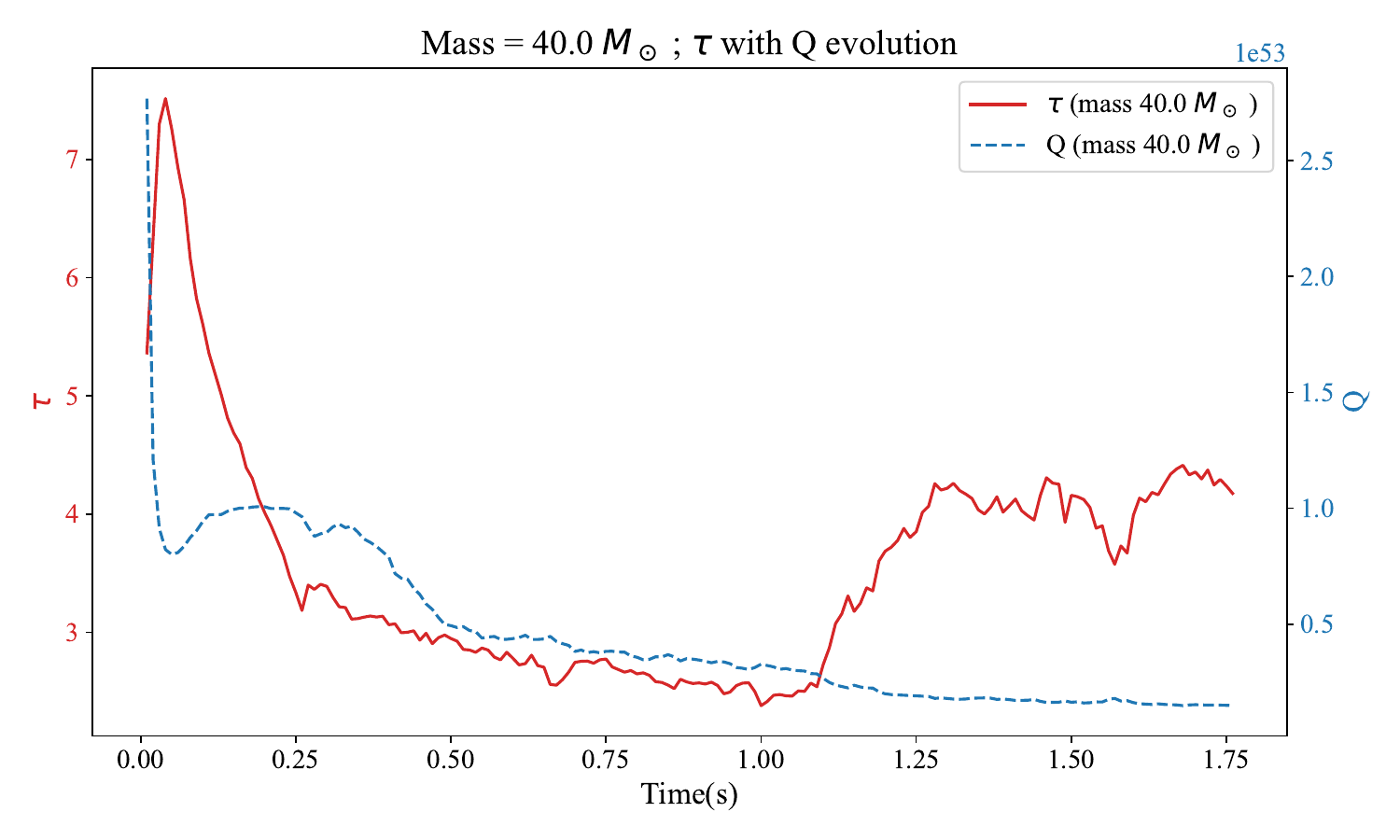}
    \hfill 

\includegraphics[width=0.48\textwidth]{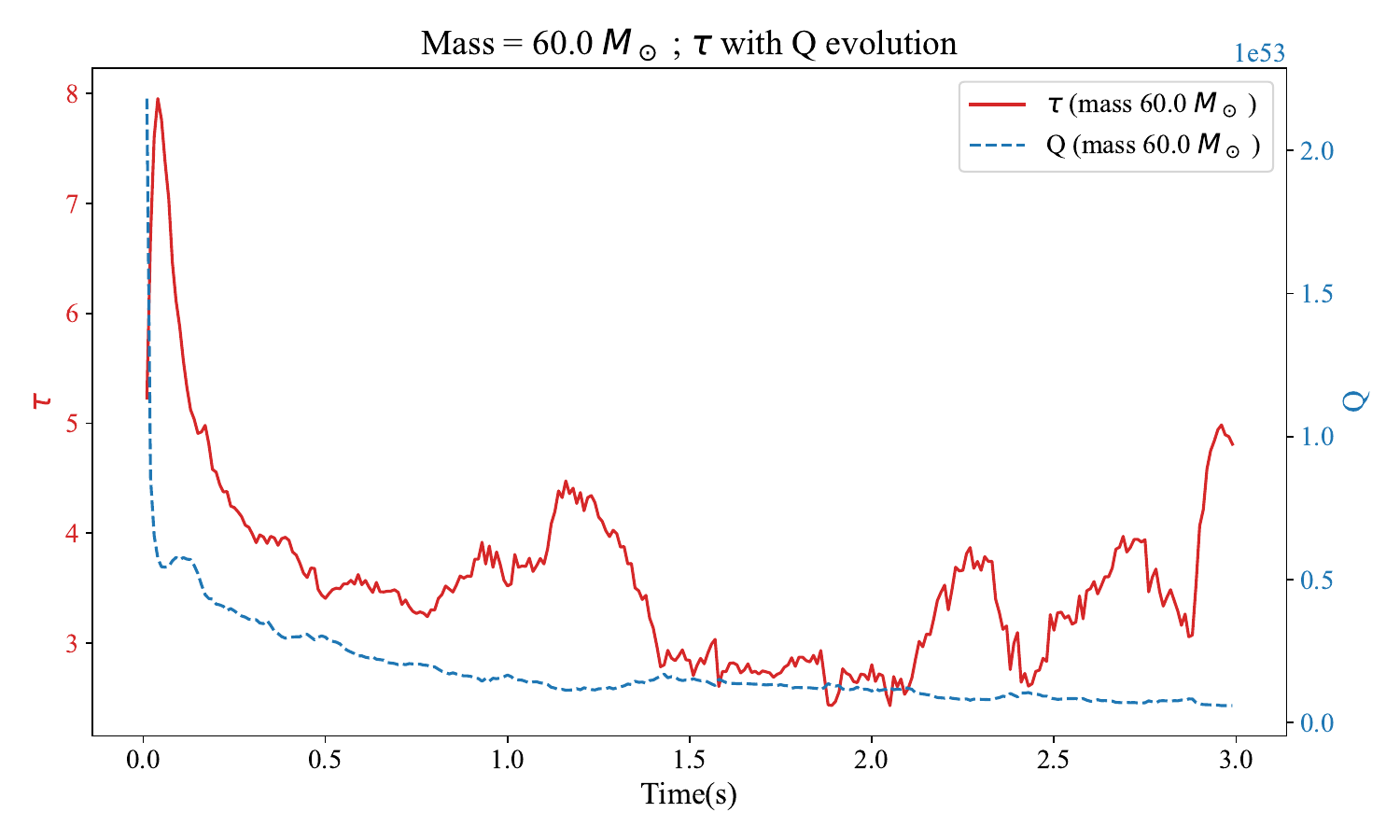}
    \hfill 
    \caption{Evolution diagram of successful supernova explosion parameters.}
    \label{fig:all}
\end{figure*}

\clearpage
\newpage
\subsection{Pattern Similarity Comparison}

\begin{figure*}[h]
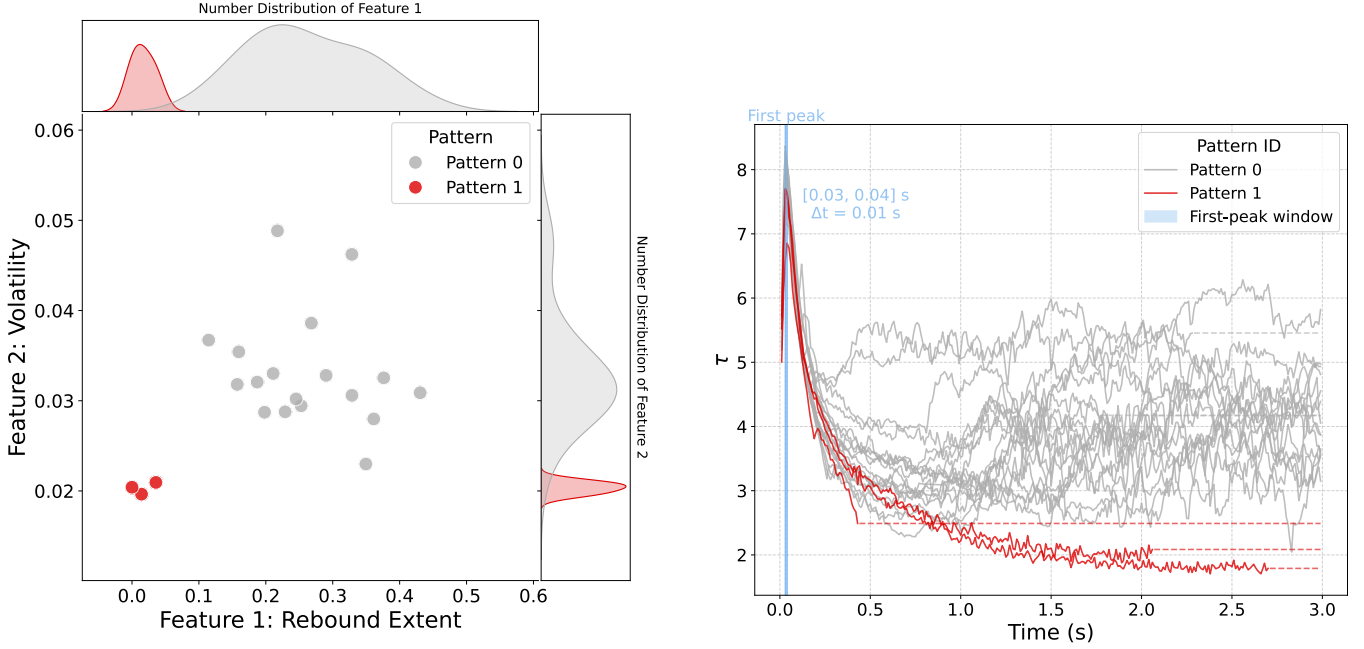

    \centering 

    \includegraphics[width=0.48\textwidth]{feature_space_joint_plot_en.pdf}
    \hfill 
    \includegraphics[width=0.48\textwidth]{tau_pattern_classification_with_peak_window.pdf}

    \vspace{5mm} 

    \caption{Based on unsupervised clustering, we find that the evolution of $\tau$ clearly separates successfully exploding supernovae from failed explosions into two distinct clusters.}
    \label{fig:all}
\end{figure*}
To objectively and quantitatively differentiate the temporal evolution patterns of $\tau$ in our models, we employed an unsupervised clustering method based on feature engineering. The process consists of two main steps: dynamic feature extraction and K-Means clustering.

First, for each model's $\tau(t)$ time series, we normalized its values to a [0, 1] range to eliminate scale effects and focus on the shape of the evolution itself. From the normalized curve, we then extracted three key dynamic features to characterize its evolutionary behavior:

\begin{itemize}
    \item \textbf{Rebound Extent:} Defined as the difference between the final value and the minimum value of the normalized curve. This feature quantifies the degree of recovery after the series reaches a trough. For a monotonically decreasing pattern, this value is expected to be close to zero.
    \item \textbf{Volatility:} Defined as the standard deviation of the first-order differences of the normalized curve. This feature measures the magnitude of oscillations or fluctuations during the evolution.
    \item \textbf{Overall Trend:} Defined as the slope of a linear regression fit to the normalized curve. This feature characterizes the overall direction of the evolution.
\end{itemize}

Based on these three features, each mass model was represented as a three-dimensional feature vector. We then applied the K-Means clustering algorithm to these feature vectors, automatically partitioning the models into two distinct clusters ($k=2$).

The clustering analysis successfully identified two distinct evolutionary patterns, with the following membership:

\begin{itemize}
    \item \textbf{Pattern 0:} This pattern is characterized by a monotonically decreasing or non-rebounding behavior. The models belonging to this cluster (in units of $M_\odot$) are: 9.25, 9.5, 11.0, 15.01, 16.0, 16.5, 17.0, 18.0, 18.5, 19.0, 19.56, 20.0, 21.68, 23.0, 24.0, 25.0, 40.0, and 60.0.
    \item \textbf{Pattern 1:} This pattern exhibits a significant rebound behavior after reaching a trough midway through the evolution. The models belonging to this cluster are: 12.25, 14.0, and 100.0 $M_\odot$.
\end{itemize}

This feature-based classification method provides a quantitative basis for understanding the different mechanisms governing the evolution of $\tau$ across various mass models. This classification result happens to separate successfully exploding supernovae from those that fail to explode.

\begin{figure}[h!]
    \centering
    \begin{subfigure}{\textwidth}
        \centering
        \includegraphics[width=\textwidth]{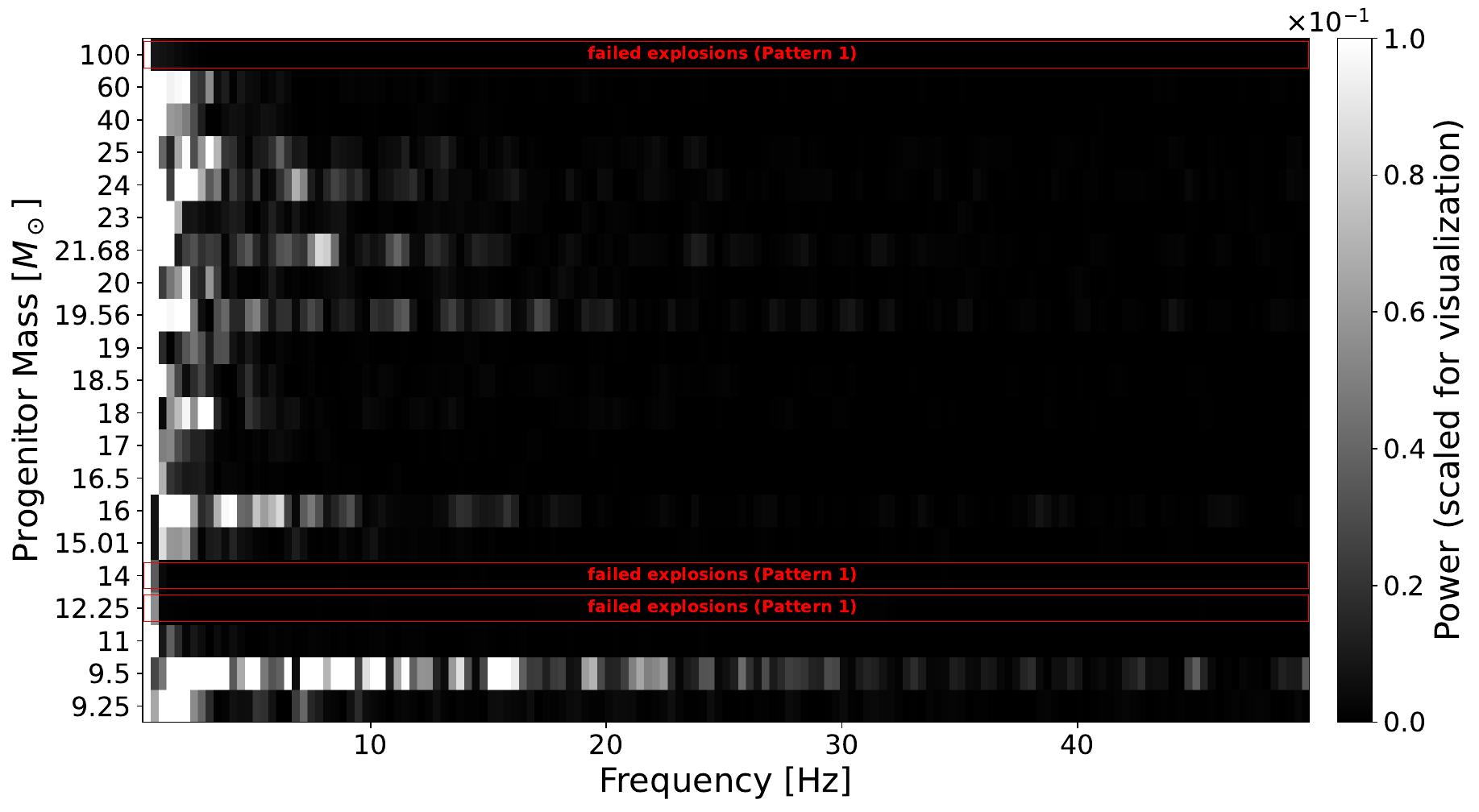}
        \caption{The power spectral density (PSD) of the time series of the neutrino spectral shape parameter, $\tau$, for each simulated progenitor mass. The y-axis lists the progenitor mass, the x-axis represents frequency, and the color intensity corresponds to the spectral power (scaled for visualization). Models resulting in failed explosions are highlighted by red boxes.}     \label{fig:Heartbeat1}
    \end{subfigure}
    \vspace{0.5cm} 
    \begin{subfigure}{\textwidth}
        \centering
        \includegraphics[width=\textwidth]{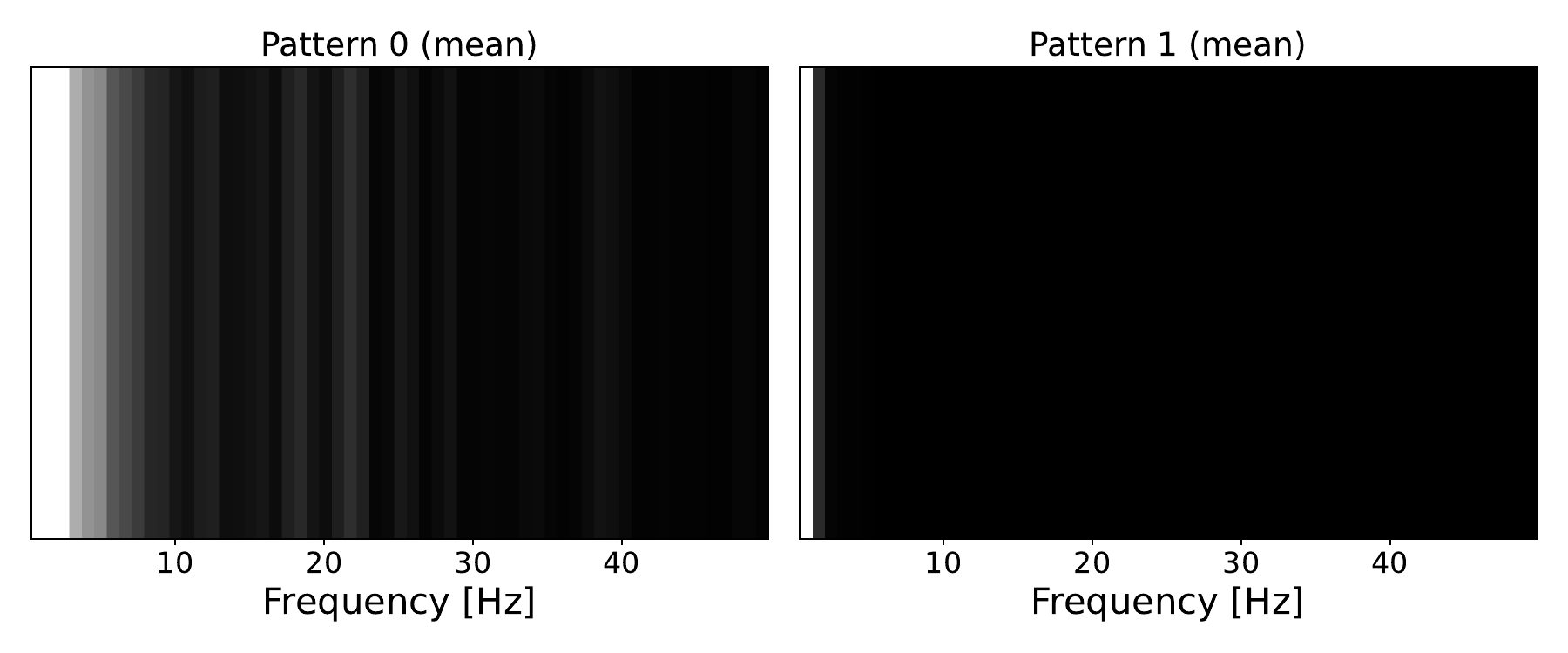}
        \caption{The mean power spectra for two distinct patterns. \textit{Left (Pattern 0):} The average spectrum for successful explosions, characterized by significant and broadly distributed power, reflecting strong hydrodynamic activity (e.g., SASI, convection). \textit{Right (Pattern 1):} The average spectrum for failed explosions, exhibiting a quiescent state with a notable deficit of power.}
    \end{subfigure}
    \caption{The two panels respectively show the frequency–domain behaviour of the neutrino spectral–shape parameter $\tau$ across progenitor models and clustering patterns.}
\label{fig:Heartbeat}
\end{figure}

The frequency-domain characteristics of the evolving neutrino spectral shape parameter, $\tau$, are presented in \autoref{fig:Heartbeat1} and . \autoref{fig:Heartbeat1} displays the power spectral density (PSD) for each progenitor mass, stacked along the y-axis. A striking dichotomy is immediately apparent: models corresponding to successful explosions exhibit rich spectral structures with significant power in the wide-frequency domain. In stark contrast, models that fail to explode appear as quiescent, dark bands with a notable deficit of power, as highlighted by the red boxes. To quantify this visual difference, the lower panels show the mean spectra for these two distinct patterns. Pattern 0, representing the average of the successful explosions, is characterized by significant, broadly distributed power across the entire frequency range shown. Conversely, Pattern 1, the average of the failed events, exhibits a near-zero power floor. This stark visual and quantitative separation reveals a strong correlation between the spectral properties of the $\tau$ parameter's evolution and the final outcome (success or failure) of the core-collapse simulation.

The analysis of our model data is based on the post-explosion evolution of Core-Collapse Supernovae (CCSNe). A crucial distinction must be made between the physics of this observable phase and the physics of the preceding, unobserved explosion mechanism. We posit that the distinct, long-term evolutionary patterns observed in diagnostic parameters, such as the ``Energy Transfer Ability'' ($\tau$), are not independent phenomena. Instead, they can be interpreted as a direct consequence and lasting imprint of the specific physical mechanism that successfully revived the shock moments earlier. The characteristics of the pre-explosion revival process---such as its efficiency, timescale, and hydrodynamical nature---establish the initial conditions that govern the subsequent evolution of the cooling and expanding remnant.

The shock revival mechanism itself is not an instantaneous event. It involves the rapid, non-linear development of hydrodynamical phenomena, including vigorous convection and the Standing Accretion Shock Instability (SASI). Upon the successful relaunch of the shock, the energetic, large-scale fluid structures generated during this critical phase do not immediately dissipate. Rather, they constitute the initial hydrodynamical state for the post-explosion evolution that our data captures. In this view, the patterns observed in the post-explosion $\tau$ parameter serve as a fossil record of these initial conditions. For successful explosions, the observed pattern featuring a secondary rebound and sustained oscillations suggests an initial state endowed with significant, large-scale turbulent energy. The post-explosion dynamics we observe can thus be understood as the continued evolution and damping of the very instabilities that were integral to the successful revival event. Conversely, the monotonically decaying pattern associated with failed events reflects a system that never underwent this violent hydrodynamical transformation.

In summary, while our data is strictly from the post-explosion phase, the observed dynamics are not decoupled from the preceding explosion mechanism. We argue that the distinct evolutionary patterns in $\tau$ can be rigorously interpreted as the causal aftermath of the physical processes that occurred during shock revival. Therefore, the quantitative analysis of these post-explosion patterns serves as a valuable diagnostic tool, offering indirect but powerful insights into the necessary conditions and nature of the unobserved, pre-explosion mechanism itself. The framework of Self-Organized Criticality (SOC) may provide a theoretical basis for understanding how the system arrives at this critical juncture, where the subsequent evolution carries the definitive imprint of a successful or failed event.

From the parameter evolution plots, it can be seen that when the energy changes sharply, the energy transport rate is positively correlated with the trend of energy variation. In addition, there is a common feature among successfully exploding supernovae: after the energy variation becomes steady, the energy transport rate recovers to a higher, dynamically fluctuating level. In contrast, failed supernova explosions lack this recovery process. We propose two complementary interpretations that can be unified into a single theoretical picture:

The first interpretation is based on a transition of the dominant physical process. We posit that the initial, violent energy release is governed by a prompt and transient core process directly associated with the shock revival itself, which dictates the initial peak in the energy transport rate. As this primary process wanes and its influence subsides, the system does not become quiescent. Instead, other more persistent physical processes—such as fully developed, large-scale turbulence or convection—take over to become the new dominant driver of energy transport. The observed "restorative rebound," therefore, signifies the system's successful transition from a transient "burst mode" to a turbulence-dominated, sustainable "dynamic equilibrium mode."

The second interpretation frames this phenomenon within the theory of Self-Organized Criticality (SOC). From this perspective, a successful explosion can be viewed as a large-scale "avalanche" triggered when the system crosses a critical threshold. The initial, violent energy release corresponds to this primary avalanche itself. Subsequently, it is an intrinsic and defining property of a critical system that it will automatically evolve back towards the edge of criticality in preparation for subsequent energy release events (of potentially smaller scales). Therefore, the observed "restorative rebound and subsequent fluctuations" are not driven by a new, distinct mechanism, but are a direct manifestation of the system spontaneously returning to and maintaining its critical state after undergoing a major energy reconfiguration.

\subsection{Co-evolution of $\tau$ and GW amplitude.}
\begin{figure*}[h]
    \centering 

    \includegraphics[width=0.48\textwidth]{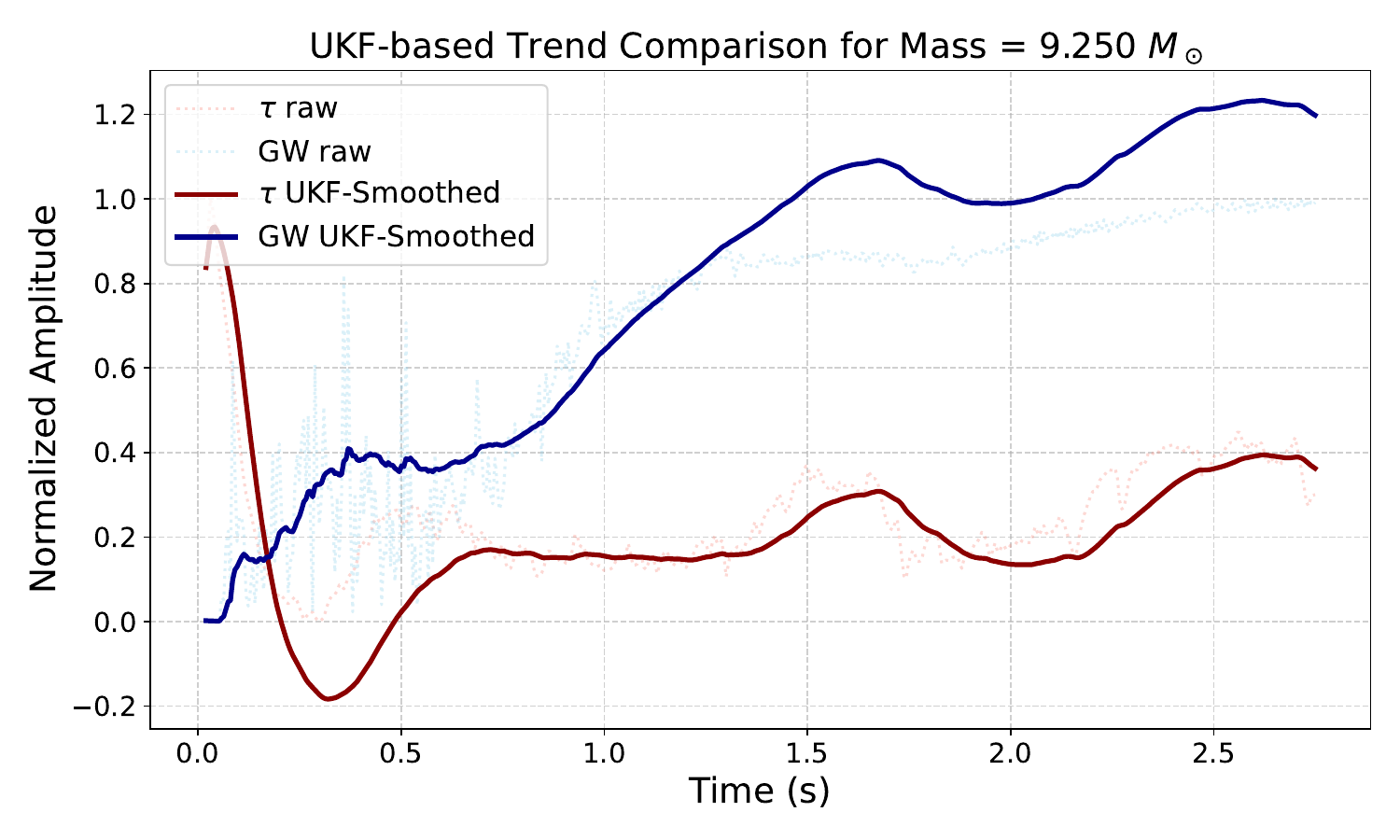}
    \hfill 
    \includegraphics[width=0.48\textwidth]{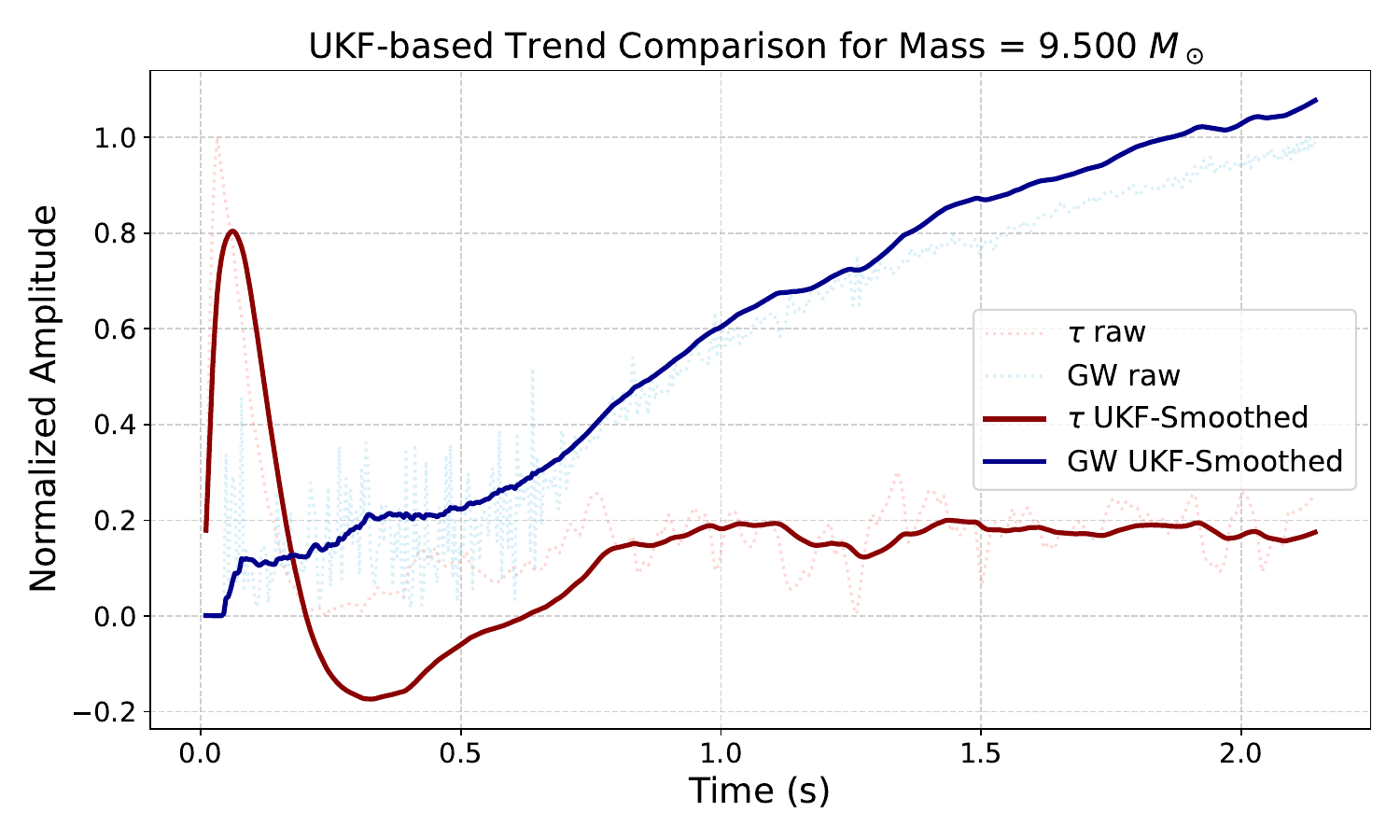}

    \vspace{5mm} 

    \includegraphics[width=0.48\textwidth]{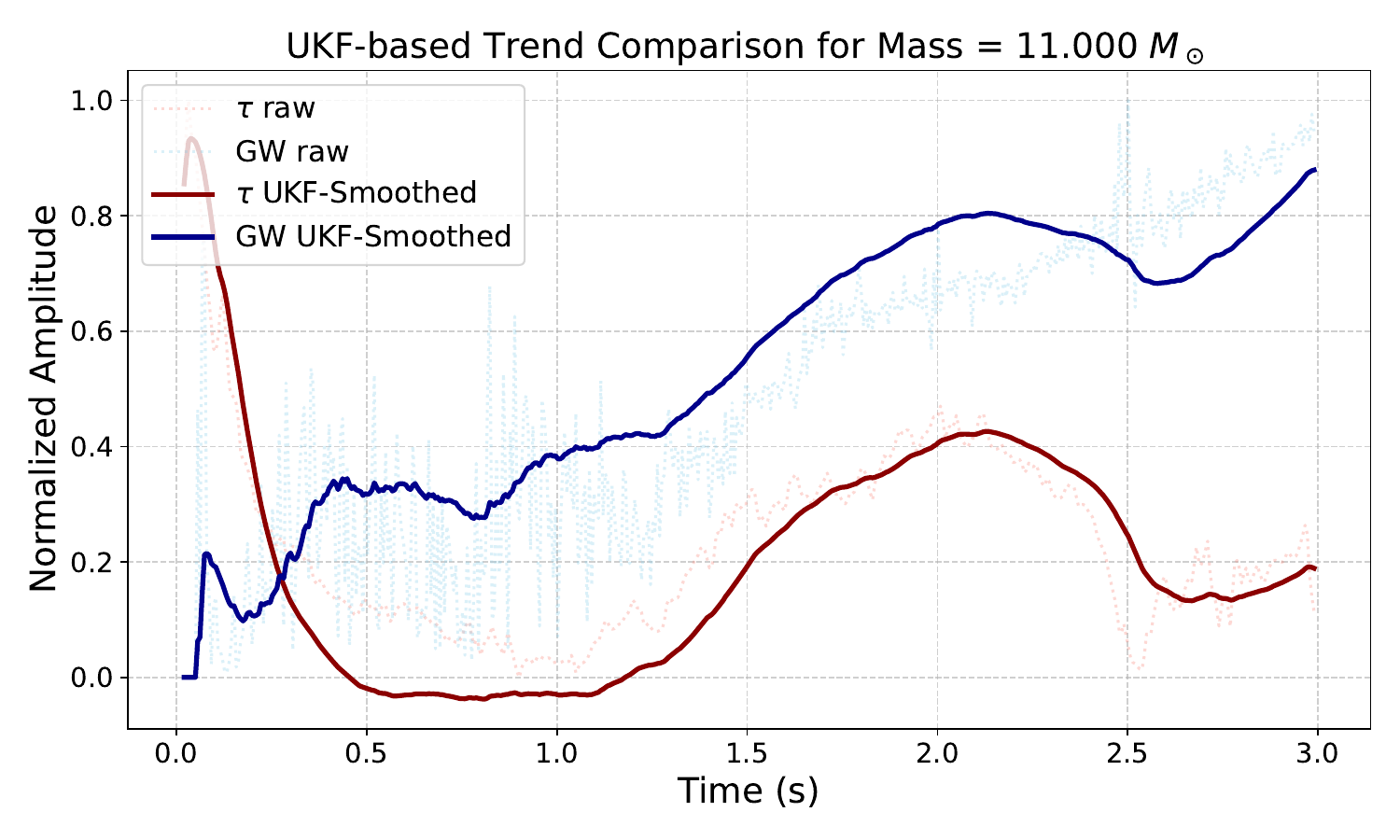}
    \hfill 
\includegraphics[width=0.48\textwidth]{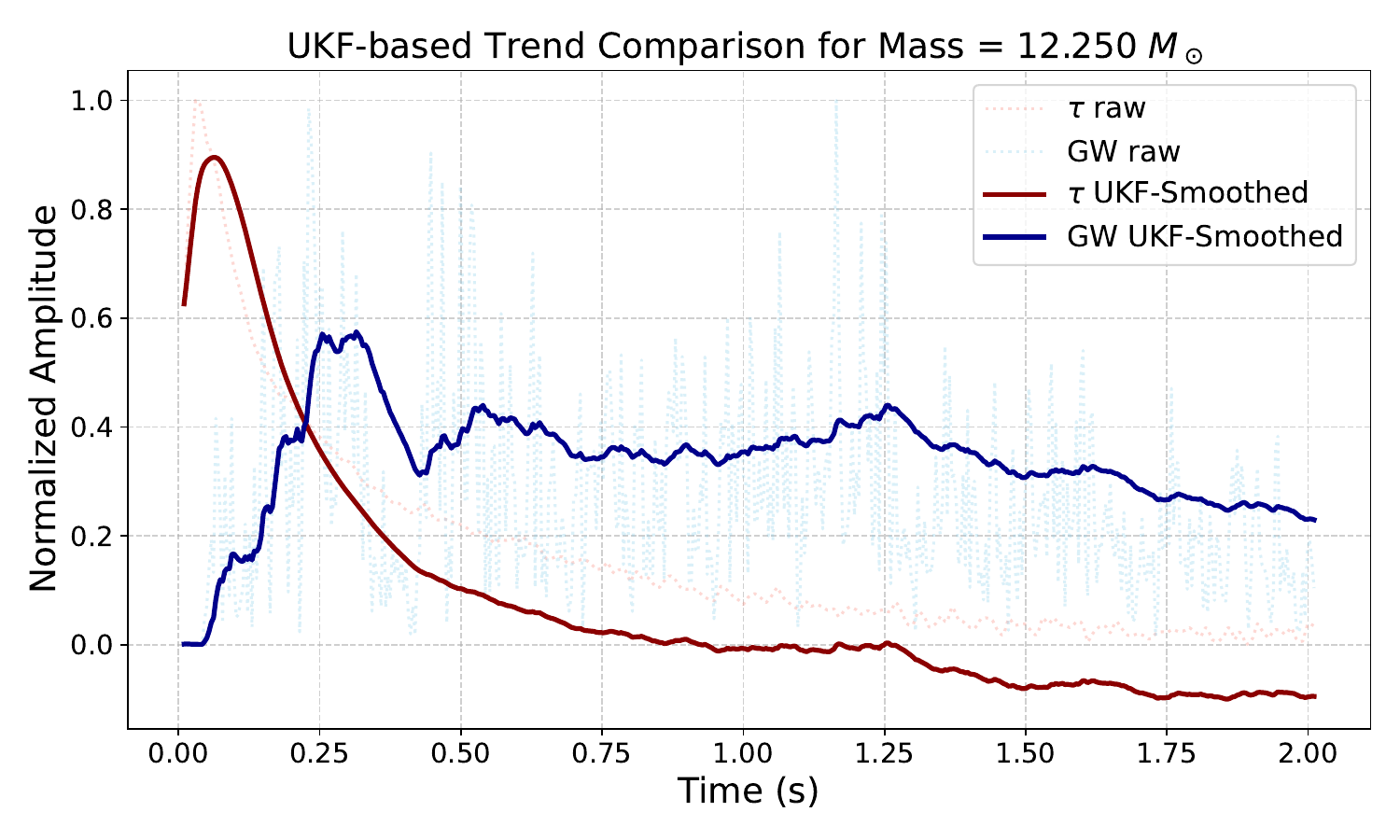}

    \vspace{5mm} 

    \includegraphics[width=0.48\textwidth]{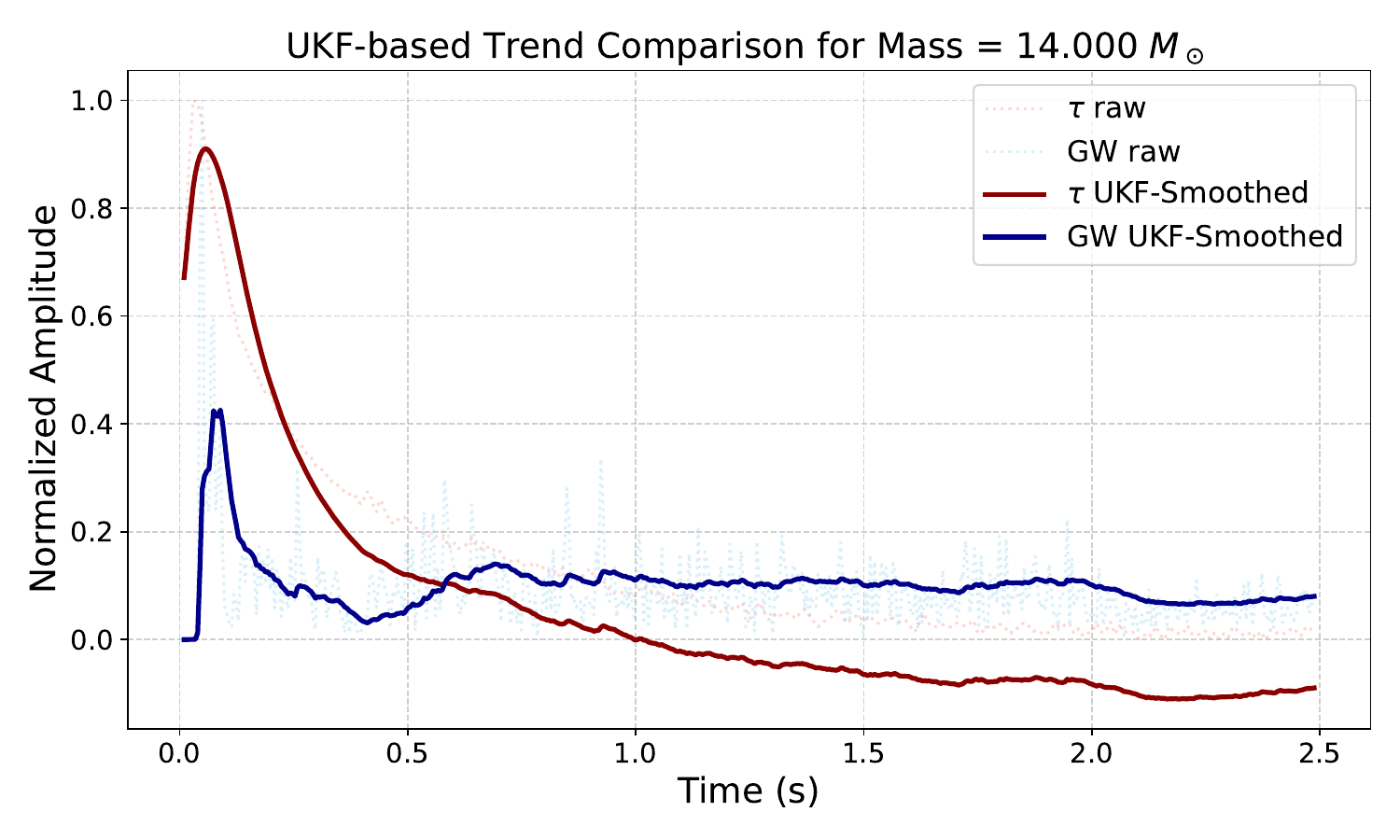}
    \hfill 
\includegraphics[width=0.48\textwidth]{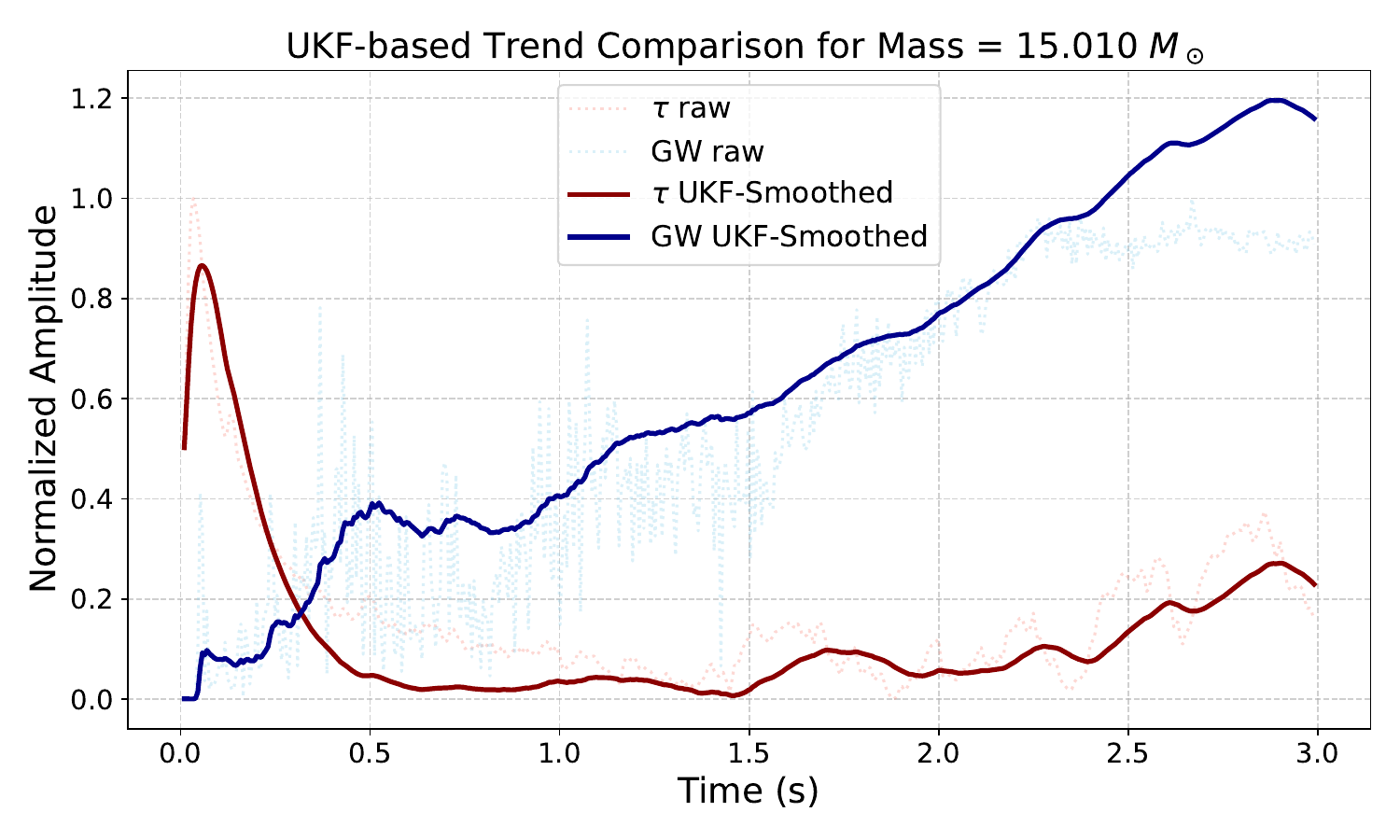}
    \hfill 

    \caption{Evolution diagrams of gravitational wave amplitude and $\tau$ under different progenitor star mass models.}
    \label{fig:all}
\end{figure*}

\begin{figure*}[h]
    \centering 

    \includegraphics[width=0.48\textwidth]{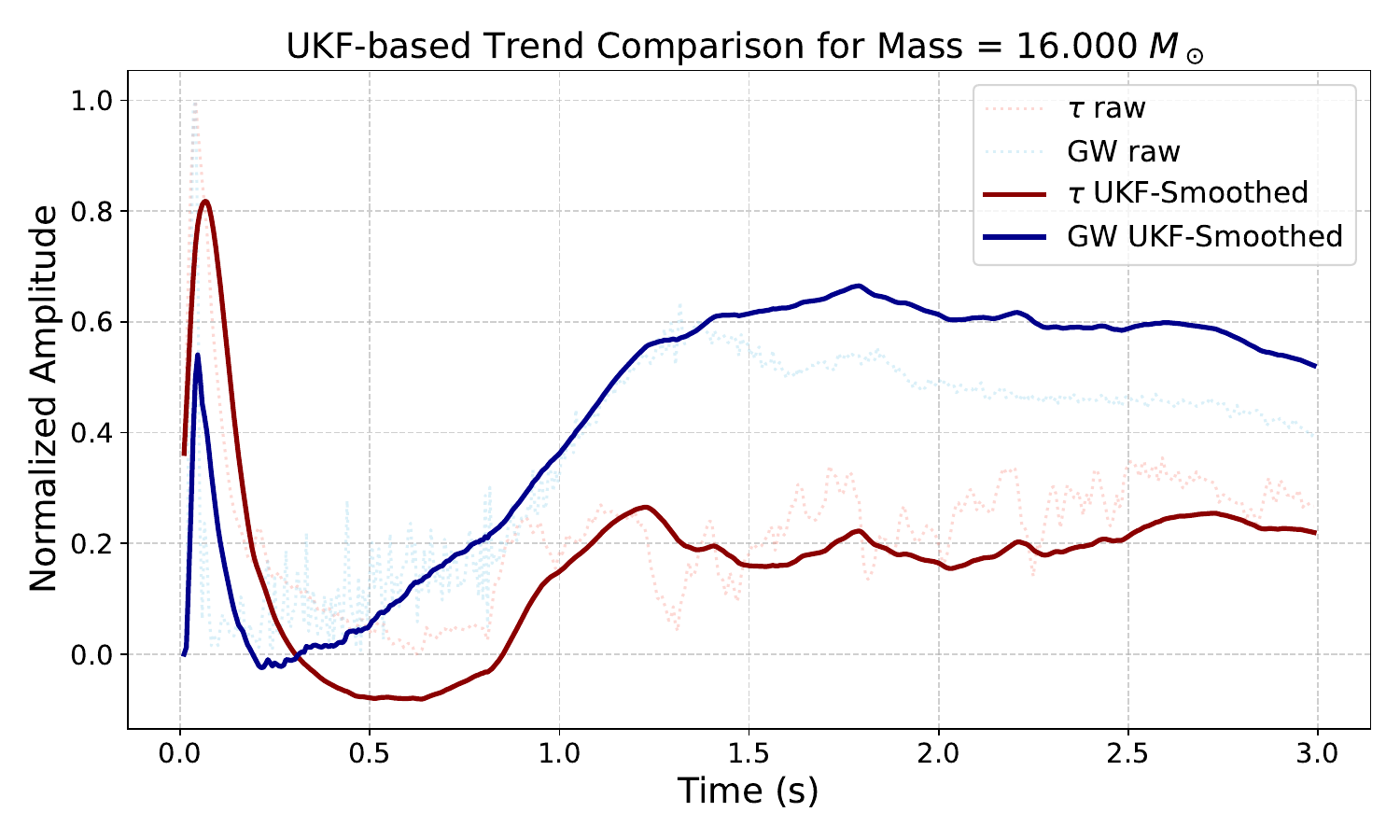}
    \hfill 
    \includegraphics[width=0.48\textwidth]{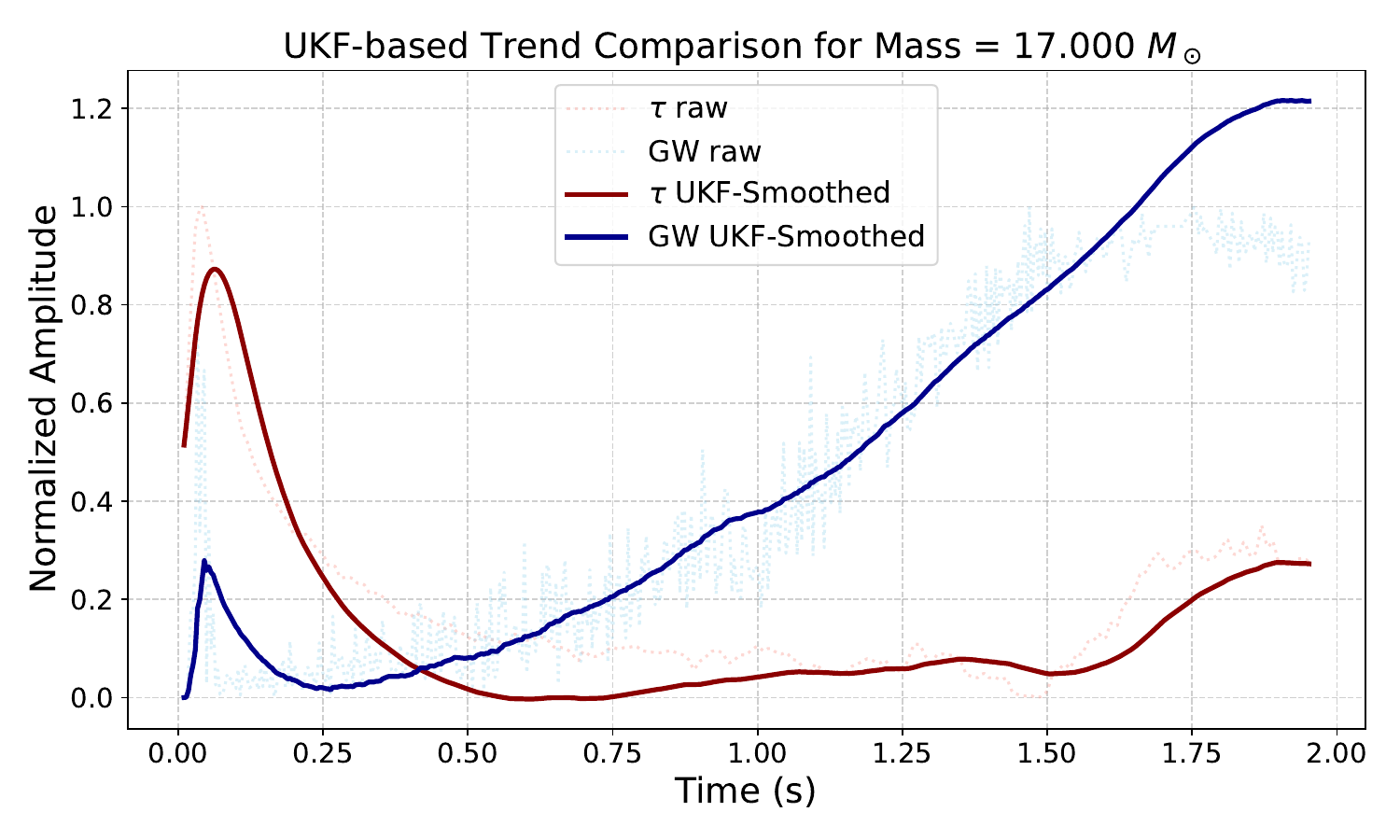}

    \vspace{5mm} 

    \includegraphics[width=0.48\textwidth]{GWCo-evolution_18.000.pdf}
    \hfill 
\includegraphics[width=0.48\textwidth]{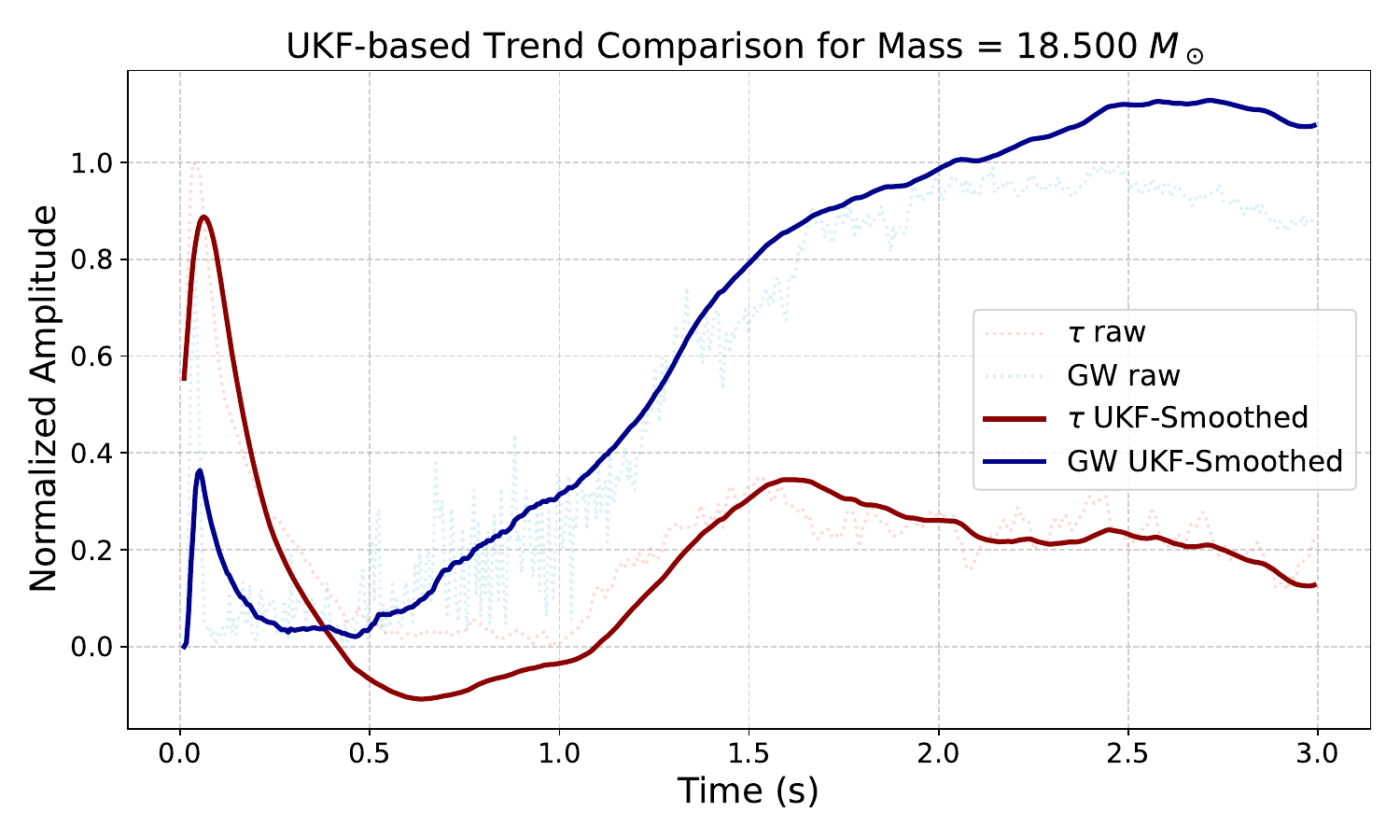}

    \vspace{5mm} 

    \includegraphics[width=0.48\textwidth]{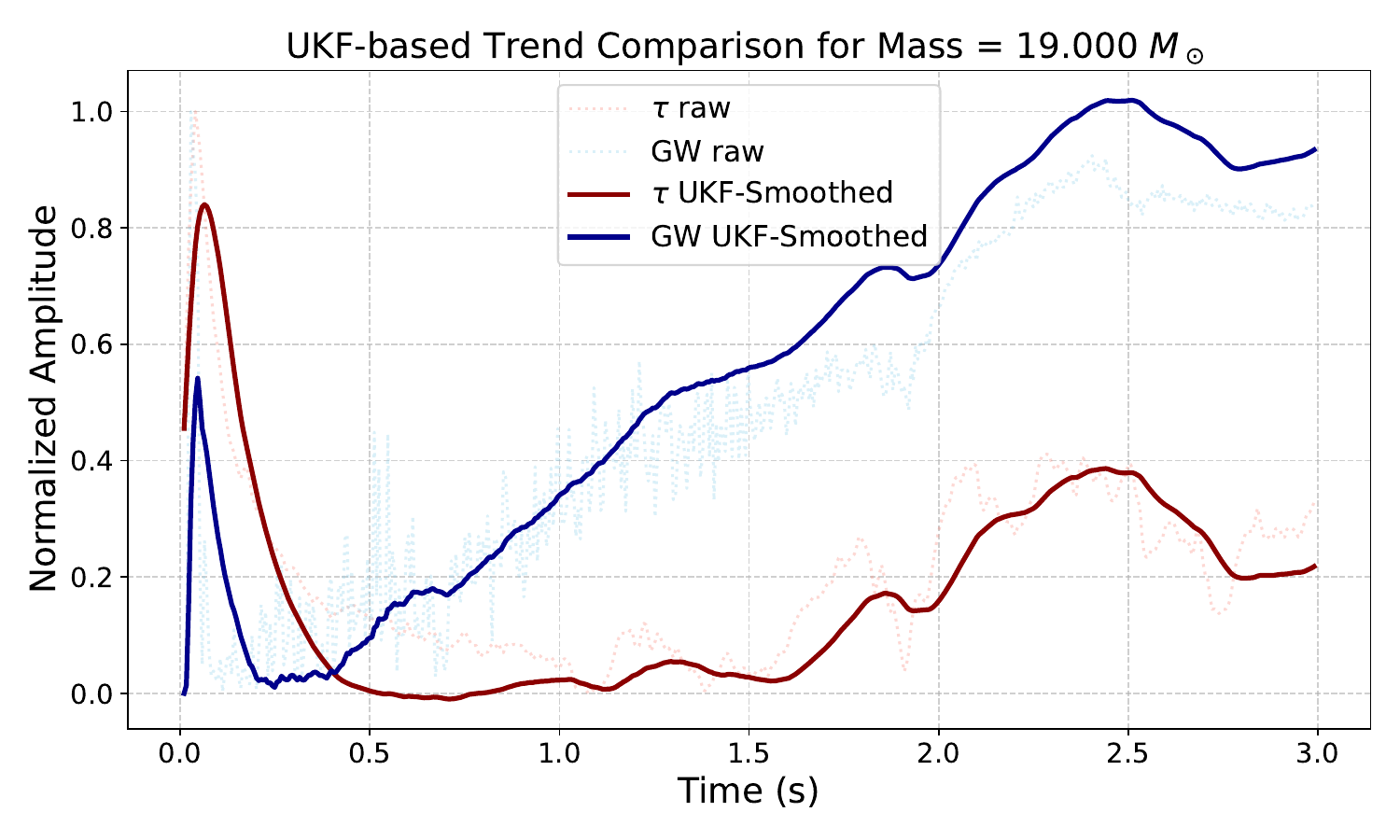}
    \hfill 
\includegraphics[width=0.48\textwidth]{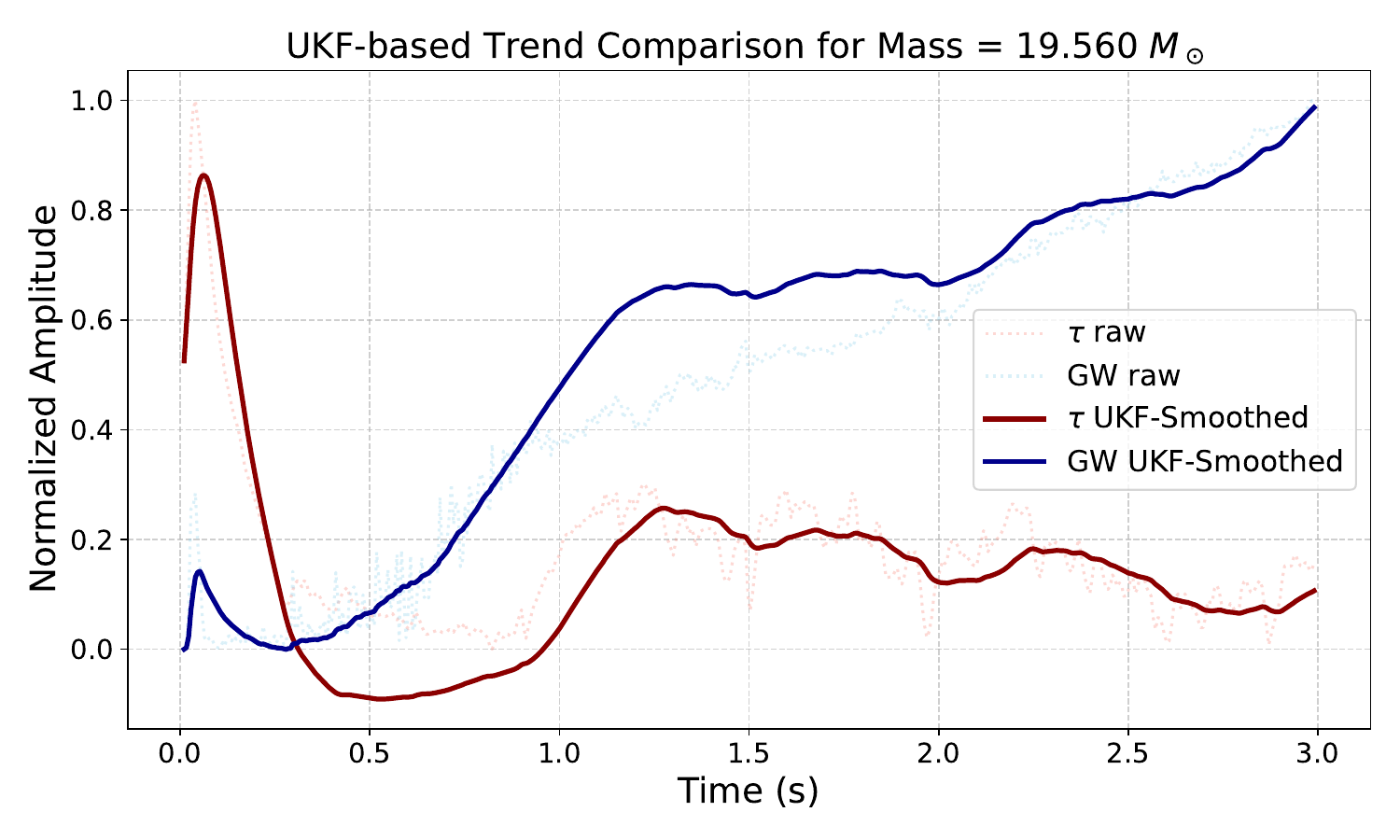}
    \hfill 

    \caption{Evolution diagrams of gravitational wave amplitude and $\tau$ under different progenitor star mass models.}
    \label{fig:all}
\end{figure*}

\begin{figure*}[h]
    \centering 

    \includegraphics[width=0.48\textwidth]{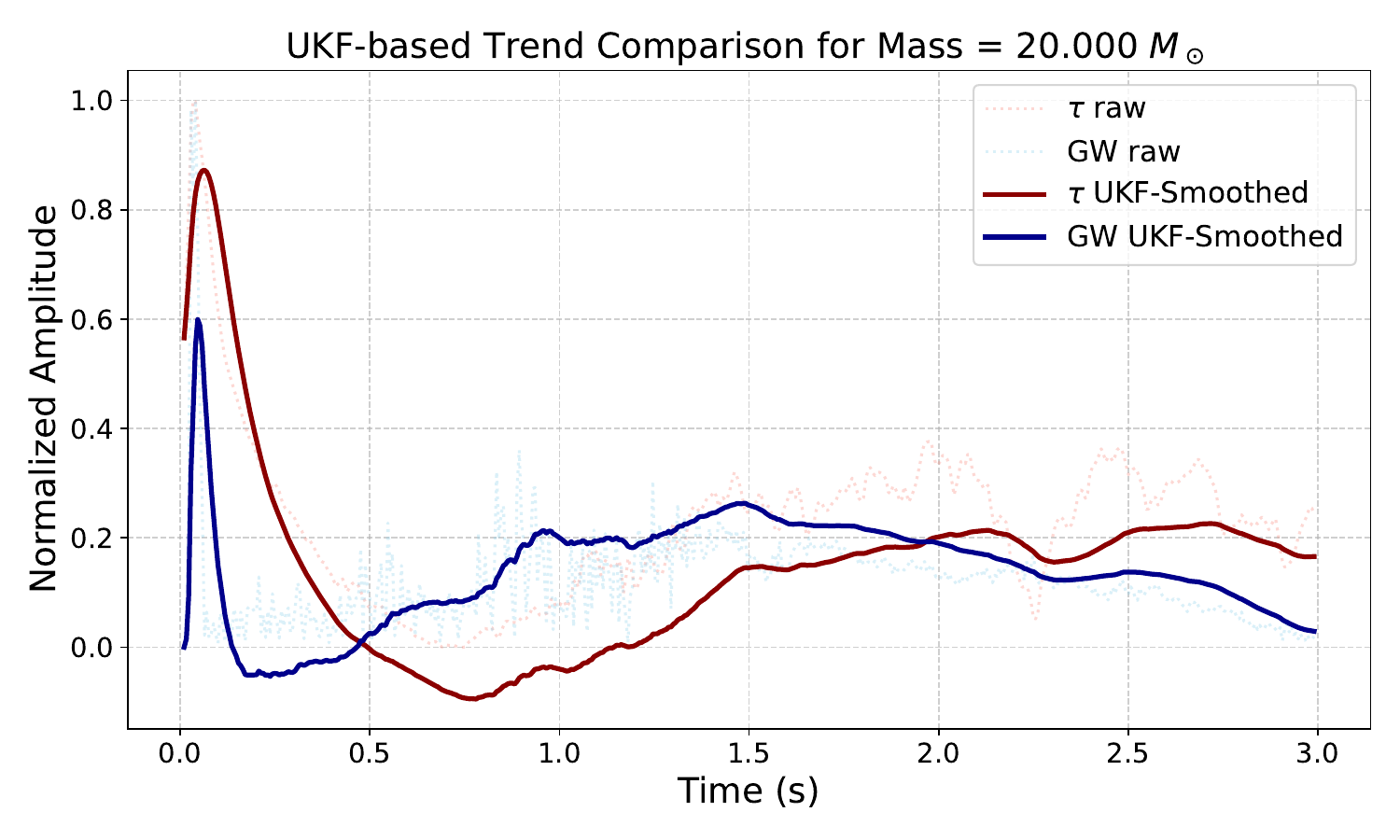}
    \hfill 
    \includegraphics[width=0.48\textwidth]{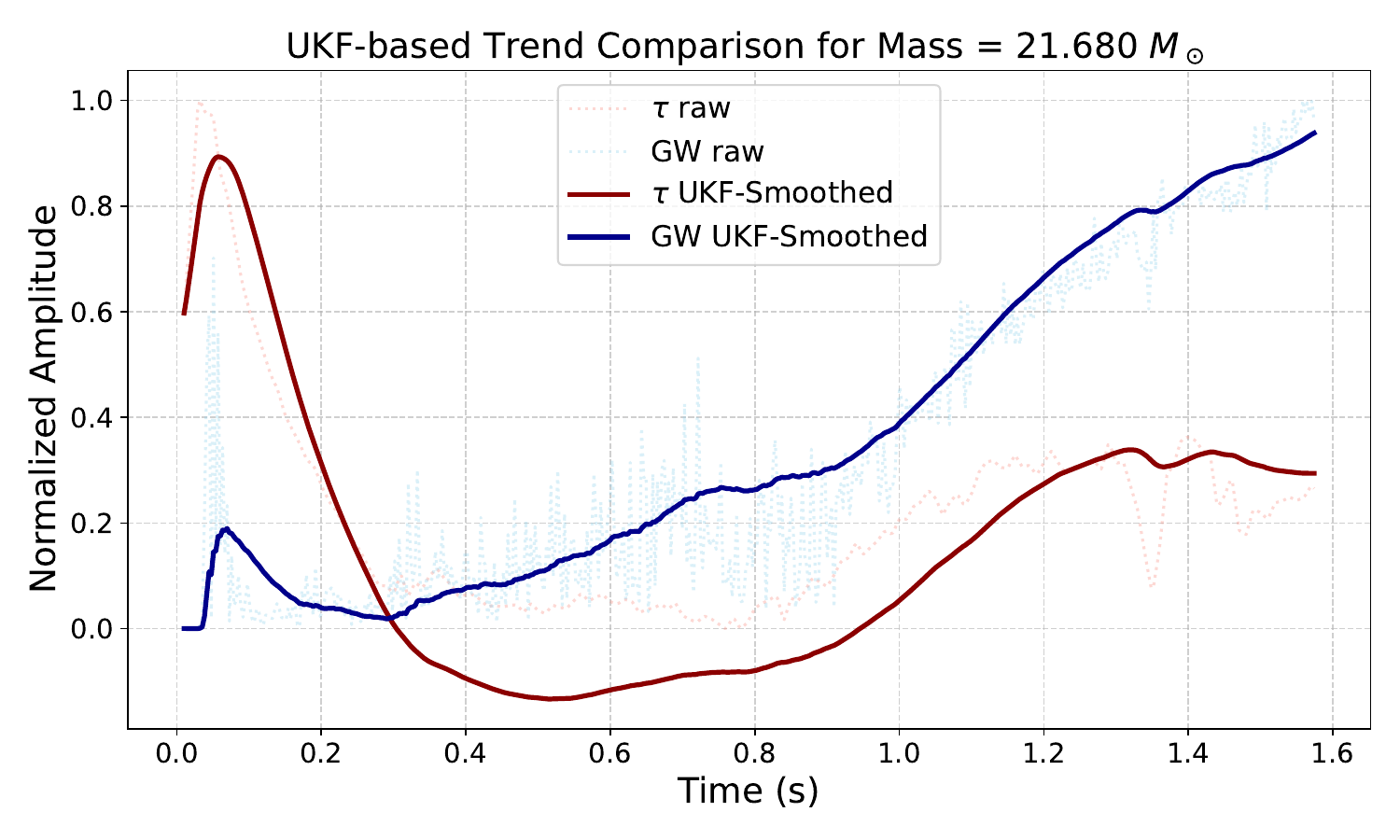}

    \vspace{5mm} 

    \includegraphics[width=0.48\textwidth]{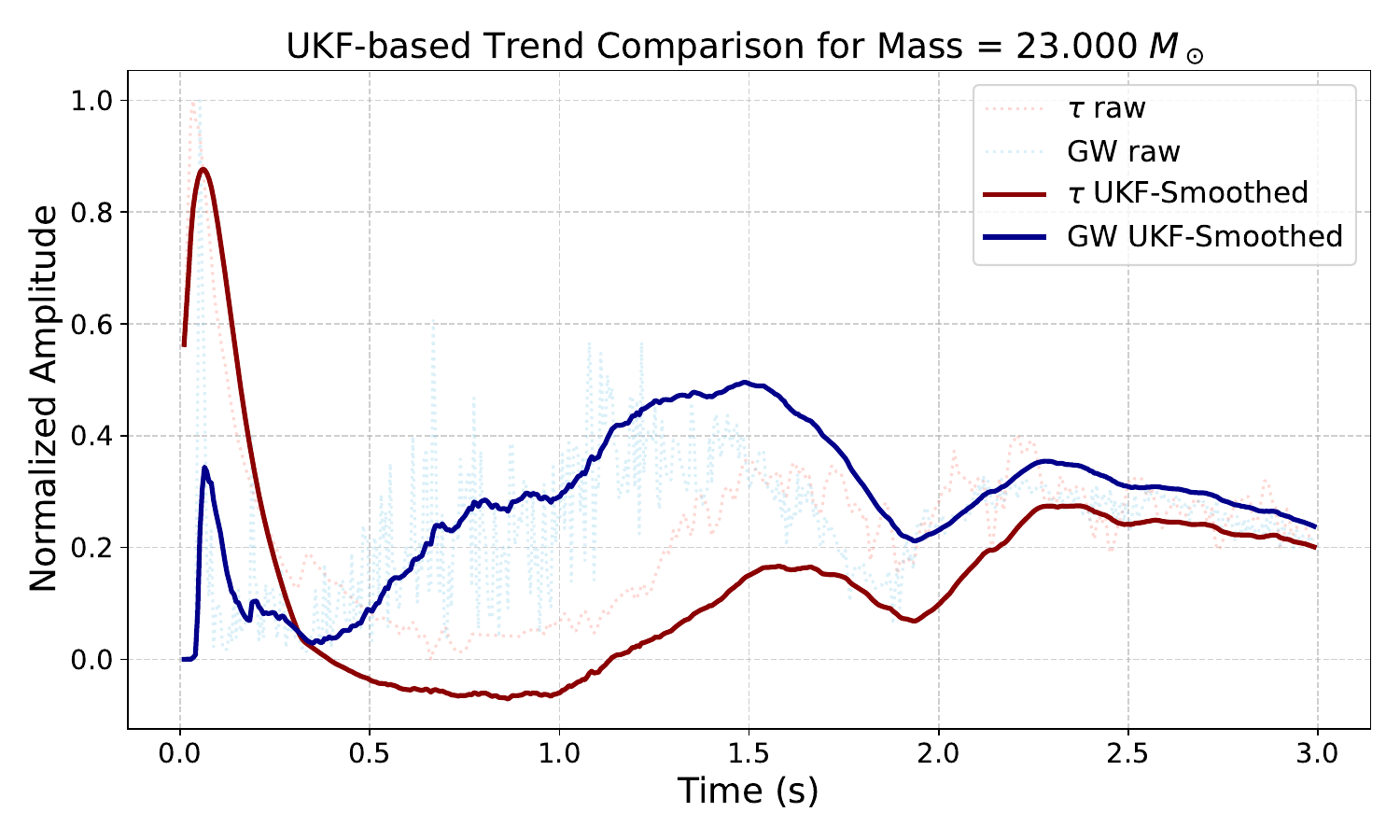}
    \hfill 
\includegraphics[width=0.48\textwidth]{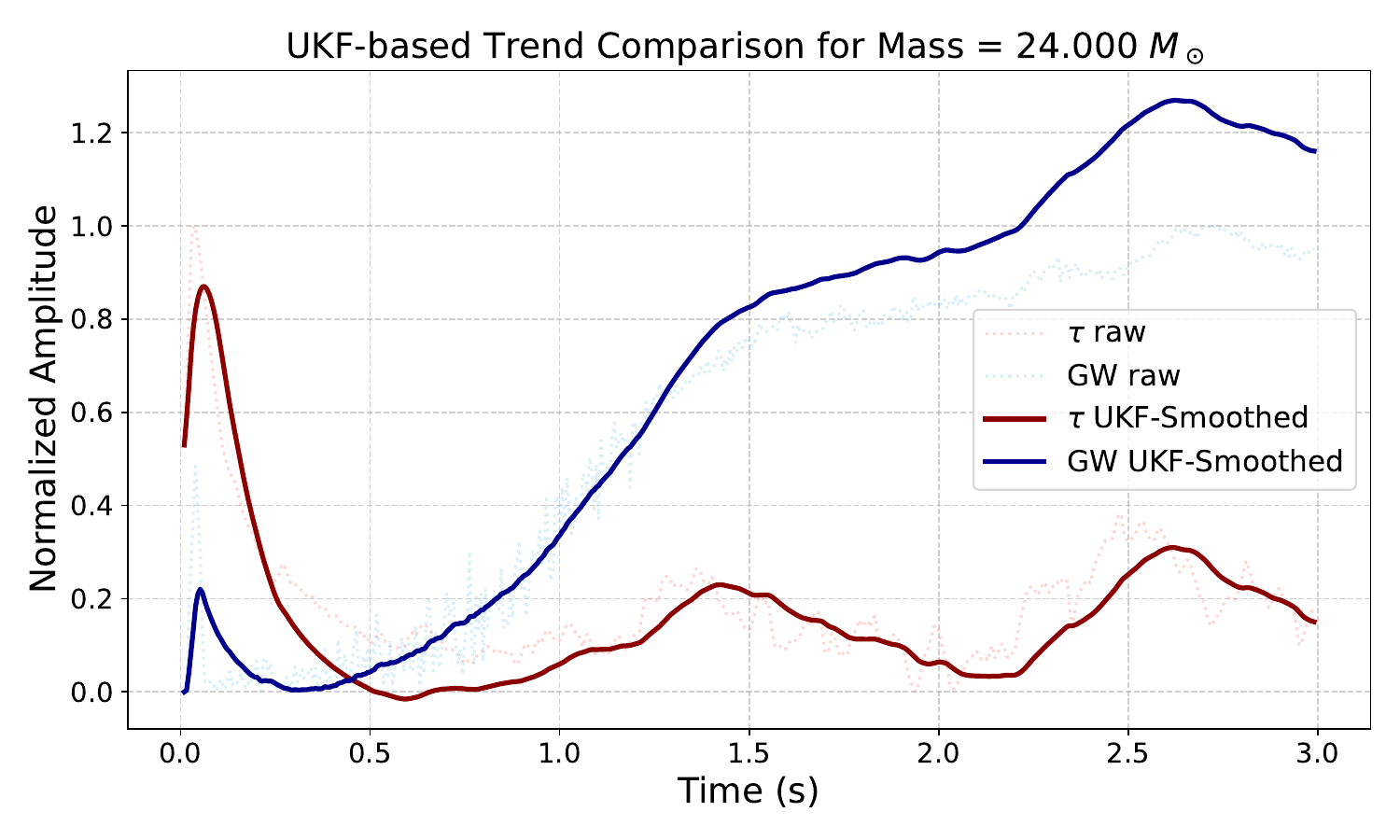}

    \vspace{5mm} 

    \includegraphics[width=0.48\textwidth]{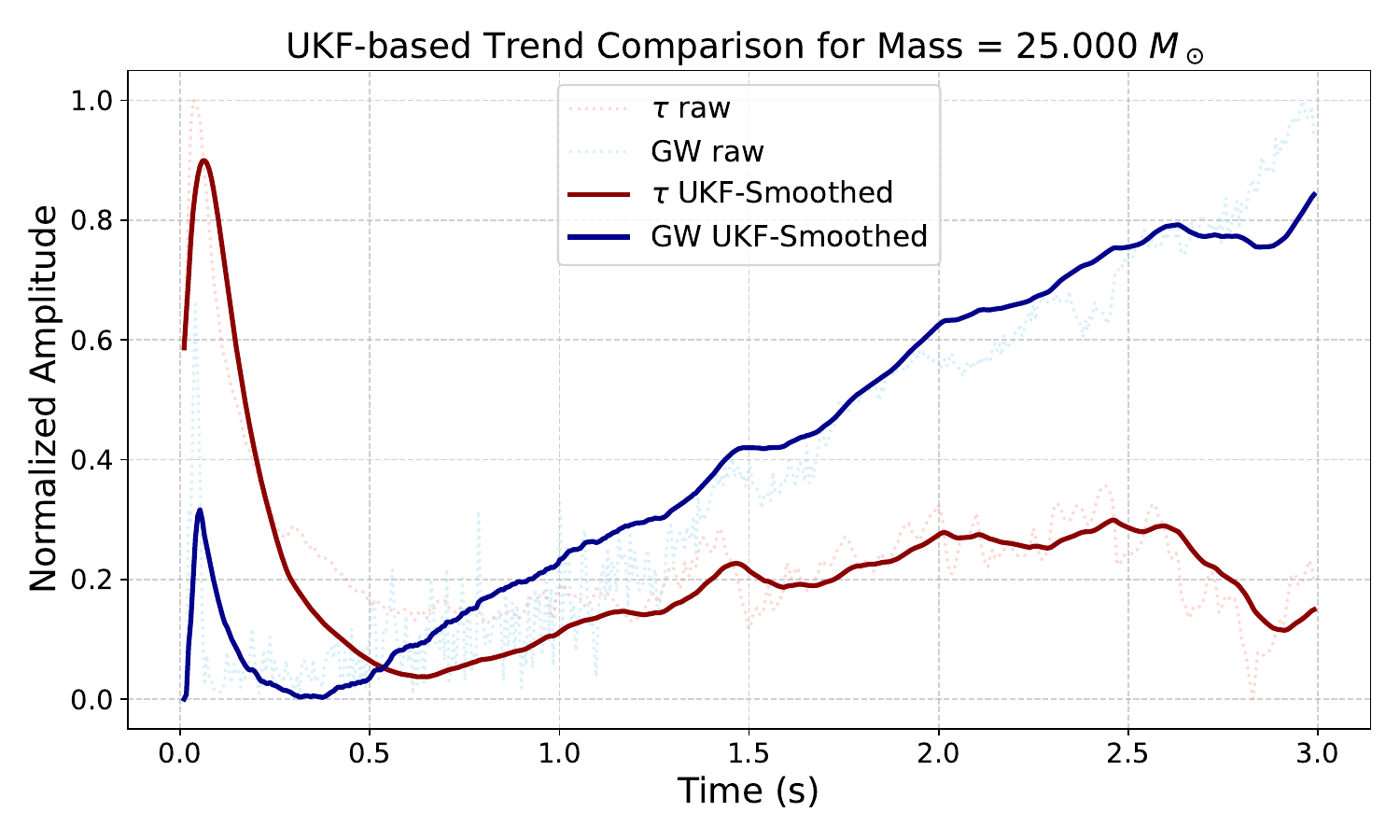}
    \hfill 
\includegraphics[width=0.48\textwidth]{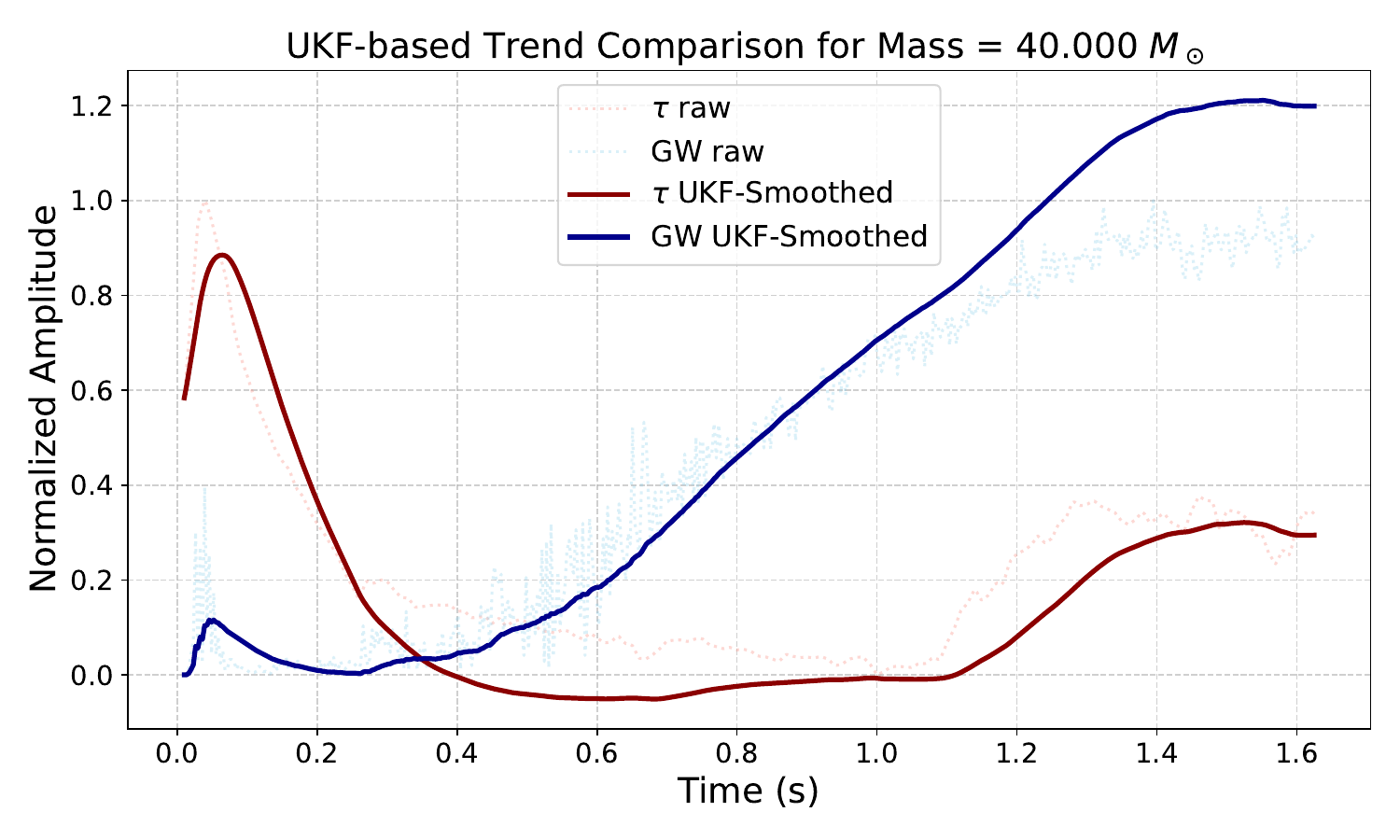}
    \hfill 

    \includegraphics[width=0.48\textwidth]{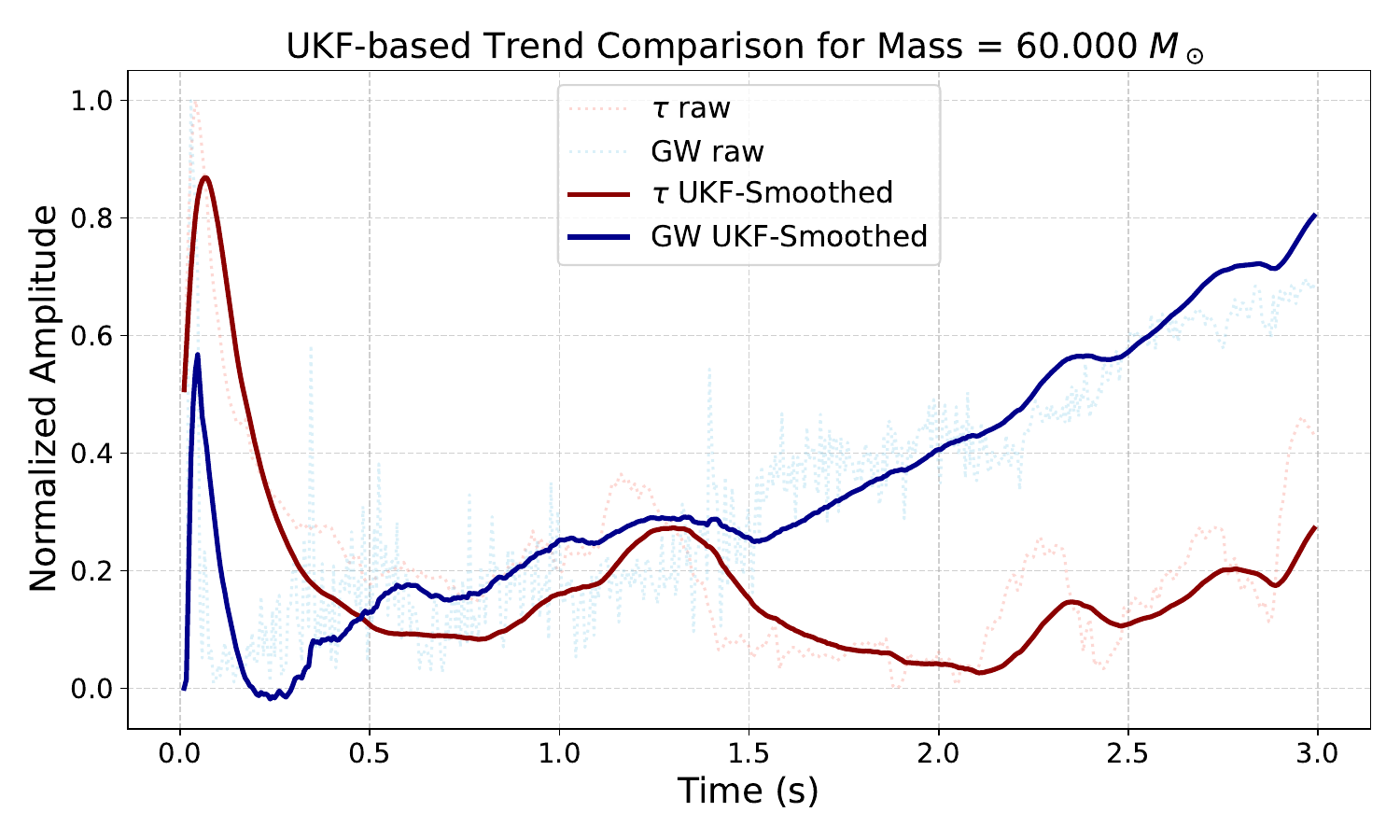}
    \caption{Evolution diagrams of gravitational wave amplitude and $\tau$ under different progenitor star mass models.}
    \label{fig:all}
\end{figure*}

Through in-depth analysis of our model data, we identify a key physical phenomenon: the diagnostic “energy transfer rate” ($\tau$) exhibits a pronounced, synergistic temporal evolution with the total gravitational-wave strain amplitude, defined as

\begin{equation}
\text{amp} = \sqrt{h_{+}^{2} + h_{\times}^{2}}.
\end{equation}



In core-collapse supernovae, we identify a clear co-evolutionary relationship between the diagnostic parameter $\tau$ and the total gravitational-wave (GW) amplitude. To characterize this dynamic relationship, we adopt a state-space approach based on the Unscented Kalman Filter (UKF). Given the highly nonlinear nature of the physical processes involved in supernova explosions (e.g., fluid dynamics), the UKF is capable of tracking the underlying dynamic states of fluctuating time series more accurately than traditional linear models, making it an ideal choice for such problems. Its key principle is to bypass direct approximation of the nonlinear functions and instead employ a deterministic sampling strategy: a set of representative sample points, known as \textbf{sigma points}, is selected around the current state to fully capture the characteristics of its probability distribution (e.g., mean and covariance). These points are then directly propagated through the state-transition model, thereby yielding a more accurate estimation of the new state and its uncertainty. To implement this, we construct a state-space model consisting of a state vector, an observation vector, a state-transition model, and an observation model, allowing us to simultaneously track the temporal evolution of both signals and estimate their dynamic correlation.

At any discrete time step $k$, the system state vector $\vect{x}_k$ is a four-dimensional vector defined as:
\begin{equation}
    \vect{x}_k = \begin{bmatrix}
        x_{k,0} \\ x_{k,1} \\ x_{k,2} \\ x_{k,3}
    \end{bmatrix} = \begin{bmatrix}
        \mathrm{amp}_k \\ v_{\mathrm{amp},k} \\ \tau_k \\ v_{\tau,k}
    \end{bmatrix}
\end{equation}
Here, $\mathrm{amp}_k$ and $\tau_k$ are the values of the two signals at time $k$, while $v_{\mathrm{amp},k}$ and $v_{\tau,k}$ are their respective first time derivatives (i.e., velocities).

We directly measure the values of the two signals; hence, the observation vector $\vect{z}_k$ is a two-dimensional vector:
\begin{equation}
    \vect{z}_k = \begin{bmatrix}
        \mathrm{amp}_{\text{measured},k} \\ \tau_{\text{measured},k}
    \end{bmatrix}
\end{equation}

We adopt a discrete-time constant velocity (CV) model to describe state evolution. This model assumes that within a small time step $\Delta t$, the signal velocity remains constant, and the value updates according to that velocity. The state transition can be expressed as:
\begin{equation}
    \vect{x}_k = f(\vect{x}_{k-1}) + \vect{w}_{k-1} = \mat{F} \vect{x}_{k-1} + \vect{w}_{k-1}
\end{equation}
where $\vect{w}_{k-1}$ is the process noise and $\mat{F}$ is the state-transition matrix:
\begin{equation}
    \mat{F} = \begin{bmatrix}
        1 & \Delta t & 0 & 0 \\
        0 & 1 & 0 & 0 \\
        0 & 0 & 1 & \Delta t \\
        0 & 0 & 0 & 1
    \end{bmatrix}
\end{equation}

The observation model relates the state vector to the actual measurements. In this model, we directly observe the 0th and 2nd elements of the state vector (i.e., the values of $\mathrm{amp}$ and $\tau$):
\begin{equation}
    \vect{z}_k = h(\vect{x}_k) + \vect{v}_k = \mat{H} \vect{x}_k + \vect{v}_k
\end{equation}
where $\vect{v}_k$ is the observation noise, and $\mat{H}$ is the observation matrix:
\begin{equation}
    \mat{H} = \begin{bmatrix}
        1 & 0 & 0 & 0 \\
        0 & 0 & 1 & 0
    \end{bmatrix}
\end{equation}

The UKF estimates the state via a ``predict–update'' loop. Assuming that at step $k-1$ we have the posterior state estimate $\vect{x}_{k-1}$ and its covariance $\mat{P}_{k-1}$:

predict:
\begin{enumerate}
    \item \textbf{Generate Sigma Points}: From $\vect{x}_{k-1}$ and $\mat{P}_{k-1}$, generate $2n+1$ sigma points $\bm{\mathcal{X}}_{k-1}$ (with state dimension $n=4$) whose weighted mean and covariance match the state distribution exactly.
    \item \textbf{Propagate Sigma Points}: Pass each sigma point through the state-transition model:
    \begin{equation}
        \bm{\mathcal{X}}_{k|k-1}^{(i)} = f(\bm{\mathcal{X}}_{k-1}^{(i)}) = \mat{F} \bm{\mathcal{X}}_{k-1}^{(i)}, \quad i=0, \dots, 2n
    \end{equation}
    \item \textbf{Compute Predicted Mean and Covariance}:
    \begin{align}
        \vect{x}_{k|k-1} &= \sum_{i=0}^{2n} W_m^{(i)} \bm{\mathcal{X}}_{k|k-1}^{(i)} \\
        \mat{P}_{k|k-1} &= \sum_{i=0}^{2n} W_c^{(i)} (\bm{\mathcal{X}}_{k|k-1}^{(i)} - \vect{x}_{k|k-1})(\bm{\mathcal{X}}_{k|k-1}^{(i)} - \vect{x}_{k|k-1})^T + \mat{Q}_{k-1}
    \end{align}
    where $W_m^{(i)}$ and $W_c^{(i)}$ are the weights for the mean and covariance, and $\mat{Q}_{k-1}$ is the process-noise covariance.
\end{enumerate}

update:
\begin{enumerate}
    \item \textbf{Predict Observations}: Pass $\bm{\mathcal{X}}_{k|k-1}$ through the observation model:
    \begin{equation}
        \bm{\mathcal{Z}}_{k|k-1}^{(i)} = h(\bm{\mathcal{X}}_{k|k-1}^{(i)}) = \mat{H} \bm{\mathcal{X}}_{k|k-1}^{(i)}
    \end{equation}
    \item \textbf{Compute Predicted Observation Mean and Covariance}:
    \begin{align}
        \vect{z}_{k|k-1} &= \sum_{i=0}^{2n} W_m^{(i)} \bm{\mathcal{Z}}_{k|k-1}^{(i)} \\
        \mat{P}_{zz} &= \sum_{i=0}^{2n} W_c^{(i)} (\bm{\mathcal{Z}}_{k|k-1}^{(i)} - \vect{z}_{k|k-1})(\bm{\mathcal{Z}}_{k|k-1}^{(i)} - \vect{z}_{k|k-1})^T + \mat{R}_k
    \end{align}
    where $\mat{R}_k$ is the observation-noise covariance.
    \item \textbf{Compute Cross-Covariance}:
    \begin{equation}
        \mat{P}_{xz} = \sum_{i=0}^{2n} W_c^{(i)} (\bm{\mathcal{X}}_{k|k-1}^{(i)} - \vect{x}_{k|k-1})(\bm{\mathcal{Z}}_{k|k-1}^{(i)} - \vect{z}_{k|k-1})^T
    \end{equation}
    \item \textbf{Kalman Gain}:
    \begin{equation}
        \mat{K}_k = \mat{P}_{xz} \mat{P}_{zz}^{-1}
    \end{equation}
    \item \textbf{State Update}:
    \begin{align}
        \vect{x}_k &= \vect{x}_{k|k-1} + \mat{K}_k(\vect{z}_k - \vect{z}_{k|k-1}) \\
        \mat{P}_k &= \mat{P}_{k|k-1} - \mat{K}_k \mat{P}_{zz} \mat{K}_k^T
    \end{align}
\end{enumerate}
This predict–update cycle is iterated for every time step in the series.

To enable the model to infer potential coupling between the two signals, we construct a ``coupled'' process-noise matrix $\mat{Q}$. We assume that the random acceleration noise of the two signals is correlated, with correlation coefficient $\rho$. Based on a continuous white-noise acceleration model, the discretized $\mat{Q}$ is:
\begin{equation}
    \mat{Q} = \begin{bmatrix}
        \sigma_1^2 \frac{\Delta t^3}{3} & \sigma_1^2 \frac{\Delta t^2}{2} & \rho\sigma_1\sigma_2 \frac{\Delta t^3}{3} & \rho\sigma_1\sigma_2 \frac{\Delta t^2}{2} \\
        \sigma_1^2 \frac{\Delta t^2}{2} & \sigma_1^2 \Delta t & \rho\sigma_1\sigma_2 \frac{\Delta t^2}{2} & \rho\sigma_1\sigma_2 \Delta t \\
        \rho\sigma_1\sigma_2 \frac{\Delta t^3}{3} & \rho\sigma_1\sigma_2 \frac{\Delta t^2}{2} & \sigma_2^2 \frac{\Delta t^3}{3} & \sigma_2^2 \frac{\Delta t^2}{2} \\
        \rho\sigma_1\sigma_2 \frac{\Delta t^2}{2} & \rho\sigma_1\sigma_2 \Delta t & \sigma_2^2 \frac{\Delta t^2}{2} & \sigma_2^2 \Delta t
    \end{bmatrix}
\end{equation}
Here, $\sigma_1^2$ and $\sigma_2^2$ denote the process-noise variances for $\mathrm{amp}$ and $\tau$, respectively, and $\rho$ is the process-noise correlation coefficient. Since both parameters characterize the same CCSN system and share the same process noise, we set $\rho = 1$. As evident from the evolution curves of the various models, the UKF is still able to closely track the observational data, which supports the validity of this assumption. The off-diagonal blocks in $\mat{Q}$ couple the two subsystems at the model level.

After processing all $N$ data points, the final state covariance matrix $\mathbf{P}_N$ is obtained. The UKF derivative correlation is defined as the correlation coefficient between the velocity components $v_{\mathrm{amp}}$ (state index 1) and $v_{\tau}$ (state index 3):
\begin{equation}
    \rho'_{\text{UKF}} = \frac{ \mathbf{P}_{N}(1, 3) }{ \sqrt{ \mathbf{P}_{N}(1, 1) \cdot \mathbf{P}_{N}(3, 3) } }
\end{equation}
where $\mathbf{P}_{N}(i, j)$ denotes the $(i,j)$ element of $\mathbf{P}_N$. This value quantifies the linear correlation between the underlying rates of change of the two signals after noise suppression.

\begin{figure*}[h]
    \centering 

    \includegraphics[width=0.8\textwidth]{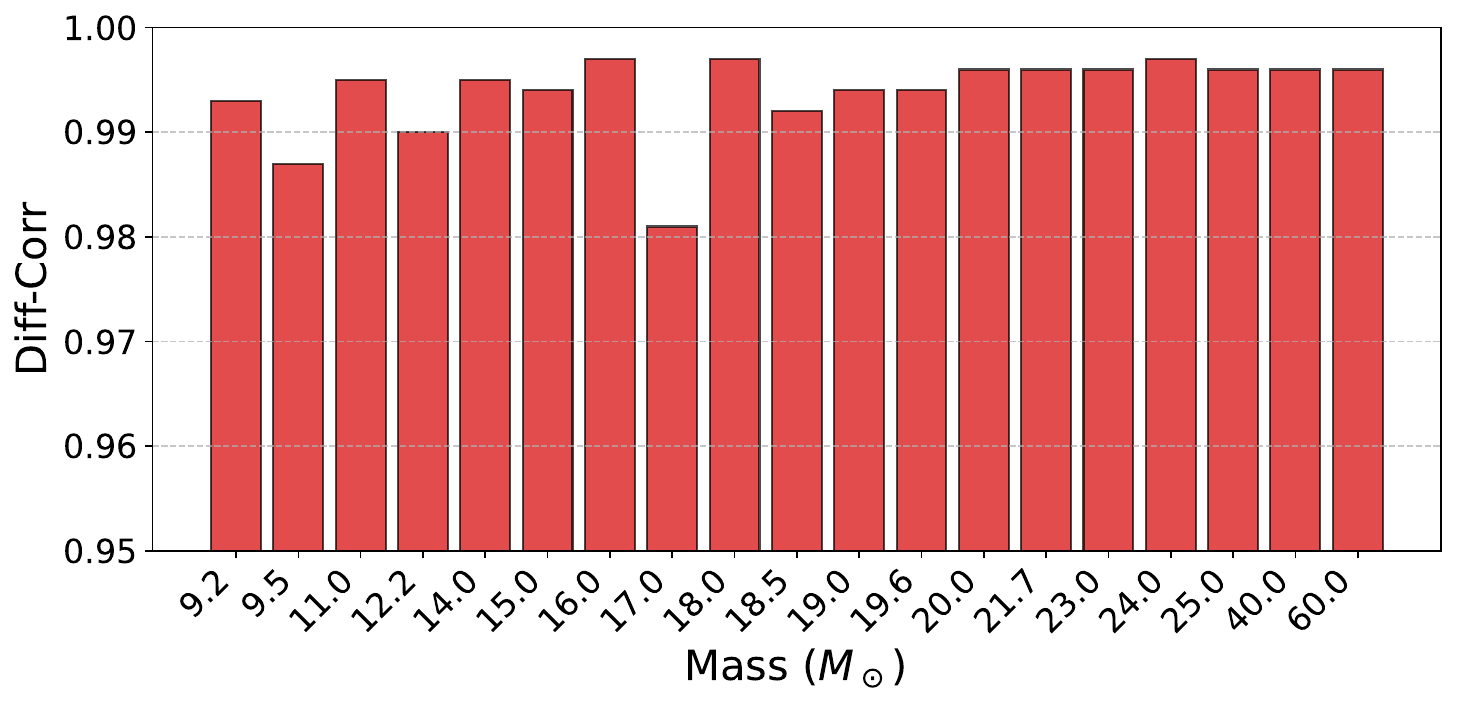}
    \caption{This figure illustrates the diff-corr parameters for $\tau$ and gravitational wave amplitude across various models. Notably, all of the models exhibit extremely high correlation.}
    \label{fig:diff_corr}
\end{figure*}

In the context of core-collapse supernovae, a high diff-corr value (e.g., $>0.9$) suggests that changes in the energy transfer rate are tightly coupled with the corresponding variations in gravitational-wave emission. In contrast, lower values imply reduced dynamical correlation or even trend reversal between the internal energy transport processes and the gravitational-wave signal. Our analysis reveals that, across all progenitor models studied, the evolution rates of these two quantities exhibit a significant and consistent strong positive correlation, with $\rho'_{\mathrm{UKF}}$ exceeding $0.98$ in all cases (see \autoref{fig:diff_corr}). This result indicates that the UKF’s dynamic modeling capability has uncovered an almost perfect underlying synchrony between the evolution rates of the two physical processes. Such near-perfect synchrony suggests that, regardless of whether the explosion ultimately succeeds or fails, the growth and decay rates of $\tau$ and $\mathrm{amp}$ are tightly coupled. This behavior is likely driven by large-scale, non-spherical hydrodynamic instabilities (e.g., convection, SASI) in the post-shock region, in which $\tau$ tracks the energy efficiency of the central engine while $\mathrm{amp}$ captures the asymmetry of its geometry. A detailed frequency-resolved analysis shows that above 100 Hz this correlation becomes insignificant. This result may indicate that gravitational waves at different frequencies originate with richer physical details.

Notably, this co-evolution follows two distinctly different patterns depending on the supernova's outcome. In successful explosions, $\tau$ and amp undergo an initial rapid rise and decline, followed by a pronounced recovery leading to sustained, dynamic fluctuations at a higher level. In contrast, failed explosions display a single-peaked behavior followed by a monotonic decay of both parameters until they vanish.

We propose a two-tiered physical interpretation:

1. Shared physical engine  
   The strong, universal correlation between $\tau$ and amp points to a common driving mechanism: large-scale, non-spherical hydrodynamic instabilities in the supernova core—such as convection and the Standing Accretion Shock Instability (SASI). Here, $\tau$ quantifies the engine’s energetic efficiency, while amp directly measures its geometric asymmetry. Their synchronous evolution is thus a natural consequence of these shared origins.

2. Diagnostic explosion outcome  
    In successful explosions, the “recovery plus sustained fluctuations” pattern implies that the turbulent engine not only reached the critical power needed to revive the shock, but also established and maintained a new, self-sustaining dynamic balance. This reflects a positive feedback loop in energy deposition overcoming continued accretion, sustaining the outward shock via a “boiling” turbulent state—hallmarks of explosion success. In failed explosions, the “monotonic decay” pattern indicates that while the engine was initially triggered (producing the initial peak), it failed to sustain energy feedback. With energy depleted and accretion persisting, the engine “quenched,” leading to irreversible decay in both $\tau$ and amp.

In summary, the co-evolution of $\tau$ and amp underscores the central role of non-spherical fluid instabilities in explosion dynamics. Whether this evolution leads to a sustained dynamic equilibrium or to monotonic demise directly reflects the engine's capacity to self-sustain. This finding offers a powerful multi-messenger diagnostic: real-time observation of supernova neutrino spectra—closely tied to the system’s energy distribution—may provide direct inference of the hidden energy transfer process during collapse.

\clearpage
\newpage

\section{Inference on the Conditions for Successful Supernova Explosions}\label{PF}

From the perspective of early energy release, failed supernova explosions appear indistinguishable from successful ones. Therefore, we argue that the subsequent divergence in the evolution of $\tau$ is not caused by the initial energy outburst itself, but rather originates from the internal dynamical structure of the progenitor star prior to explosion. 

A potential explanation for the late-time rise and oscillation of the $\tau$ parameter is the effect of the shock wave: the first peak arises as the shock passes through the neutrinosphere, and as it propagates into the outer layers, a successful supernova explosion drives the outer material into a turbulent state with enhanced energy transport efficiency. In contrast, failed supernovae lack sufficient energy, causing the shock to dissipate and thus preventing any improvement in energy transport efficiency.

By analyzing the evolution plot, a natural interpretation is that the shock receives an energy boost after passing through the neutrinosphere, as the timescale of neutrino delayed heating is on the order of several hundred milliseconds — consistent with what is shown in the figure. It is observed that, for successfully exploding supernovae, the gravitational wave amplitude begins to increase typically between 0.2 and 0.5 seconds. During this period, the shock has propagated several hundred kilometers and has moved beyond the neutrinosphere. This time window likely corresponds to the onset of the neutrino delayed heating mechanism, which becomes effective and drives the secondary heating of the shock. It can also be noted that in the later phase, when $\tau$ begins to fluctuate (about at 0.5$\sim$1s), the gravitational wave amplitude consistently shows an increasing trend. We interpret this as a consequence of asymmetric matter flows such as turbulence, which not only enhance the rate of heat transport but also generate and amplify gravitational wave emission. 

We summarize the supernova explosion picture inferred from the neutrino energy spectra as follows: the explosion begins with an initial shock rebound, accompanied by the first peak in gravitational wave emission. This is followed by neutrino delayed heating, which provides secondary energy injection to the shock; the resulting acceleration of the shock reverses the decline in gravitational wave amplitude and leads to its subsequent increase. As the revived shock propagates outward, the “gain region” surrounding the proto-neutron star becomes filled with vigorous turbulence, convection, and non-axisymmetric motions such as the Standing Accretion Shock Instability (SASI), which further amplify the gravitational wave emission.

\begin{figure}[h!]
    \centering
    
    \includegraphics[width=0.7\textwidth]{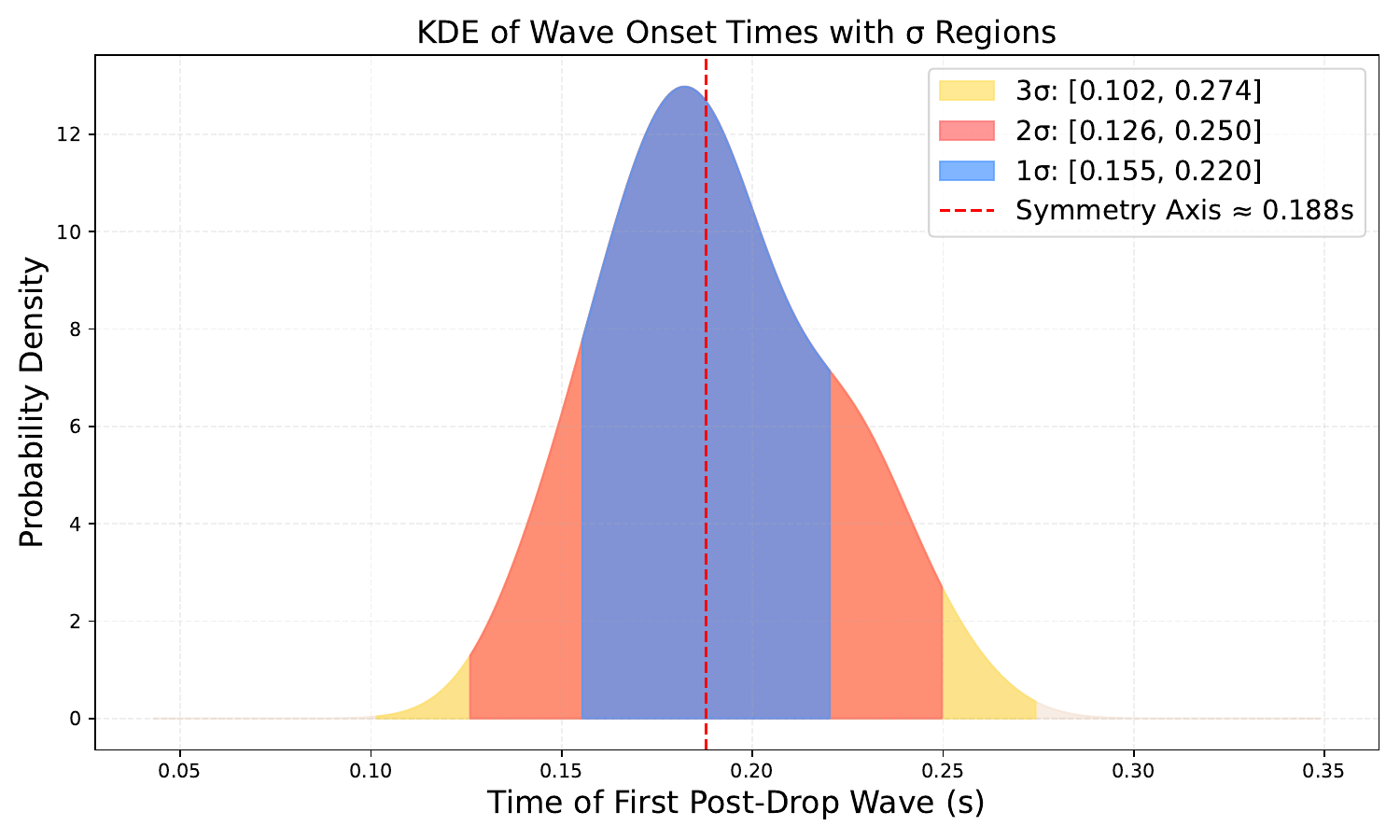}
    
    \vspace{5mm} 
    
    \includegraphics[width=0.7\textwidth]{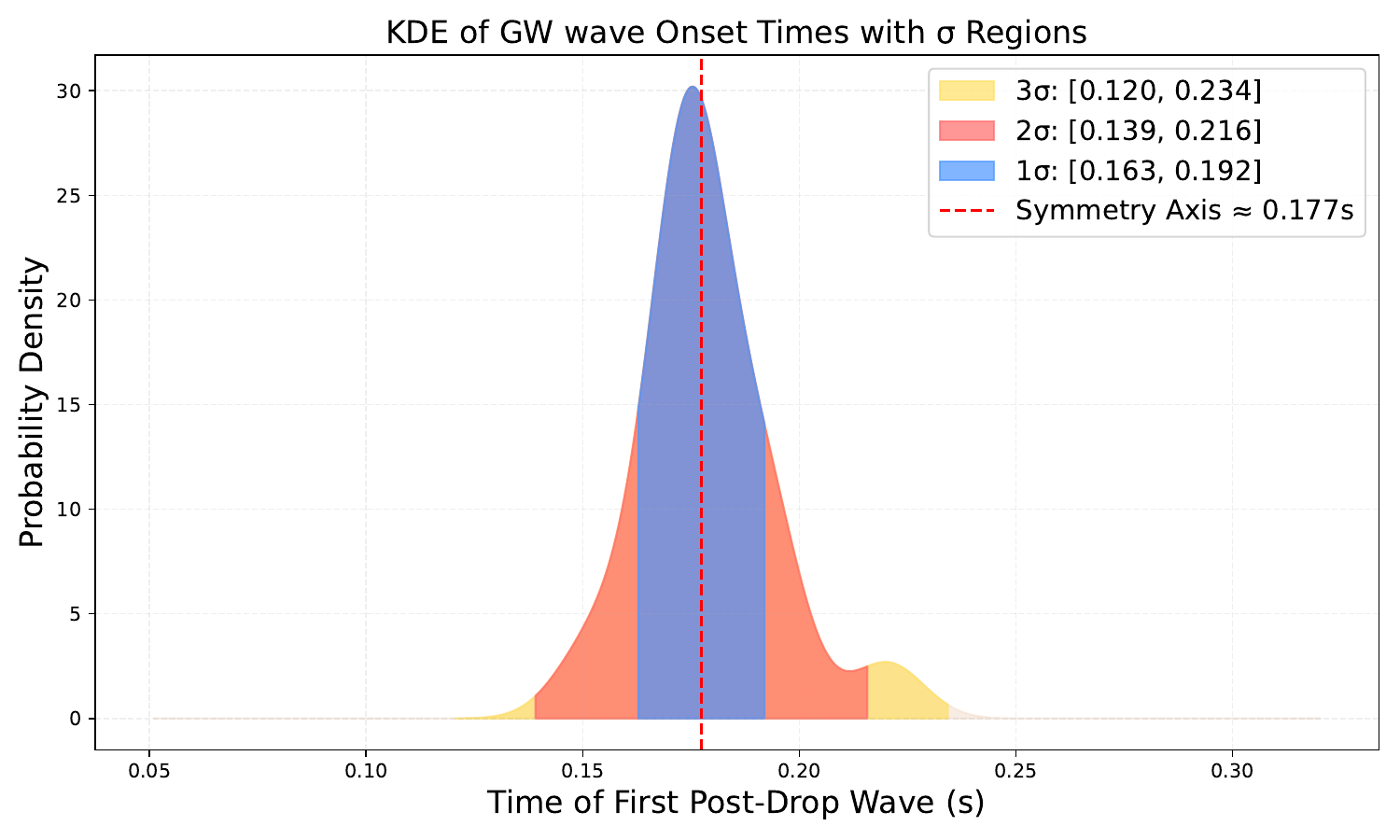}
    
    \caption{
        \raggedright Time distributions of key events in supernova explosion models, derived from a set of simulations.
        \textbf{Top panel:} Probability density distribution for the time of the first sudden drop in mean neutrino energy post-bounce. The distribution, obtained via Kernel Density Estimation (KDE), peaks at approximately 0.188\,s.
        \textbf{Bottom panel:} Probability density distribution for the time at which the gravitational wave (GW) amplitude transitions from a decreasing to an increasing trend, peaking at approximately 0.177\,s.
        In both plots, the colored regions represent the 1$\sigma$, 2$\sigma$, and 3$\sigma$ confidence intervals. The close temporal agreement between these two events strongly suggests a common physical origin, likely the onset of the neutrino-delayed heating mechanism, which energizes the shock revival and drives the enhancement of the GW signal.
        \label{fig:kde_onset_times} 
    }
    \label{fig:Times}
\end{figure}

Moreover, if the neutrino delayed heating mechanism is at work, the neutrino spectrum is expected to soften during the corresponding time period. Therefore, we examined the data and plotted the time evolution of the neutrino mean energy as well as the energy corresponding to the peak of the spectral flux at each moment. The evolutionary trend reveals that, for all successfully exploding supernovae, both the neutrino mean energy and the energy corresponding to the spectral flux peak exhibit a characteristic pattern during the early post-bounce phase: an initial phase of spectral hardening with a decreasing rate of increase, followed by a plateau where the energies remain approximately constant, and finally a phase of fluctuations with an overall softening tendency. Notably, during the early hardening phase—around 0.18 seconds post-bounce, which coincides with the moment when the gravitational wave amplitude transitions from decreasing to increasing(0.177s, as shown in \autoref{fig:Times}) —both energy indicators exhibit a sudden drop, after which their evolution becomes unstable and oscillatory. This feature may signify the onset of the neutrino delayed heating mechanism playing a role in powering the explosion. The amplification of the gravitational wave amplitude is believed to stem from the inherently non-spherical nature of the shock revival process, which is characterized by strong turbulence and pronounced asymmetry. 

Between approximately 1.0 and 1.5 seconds, both the neutrino mean energy and the energy corresponding to the spectral flux peak begin to decline and enter a phase of dynamic fluctuations. This behavior indicates the onset of cooling at the surface of the proto-neutron star. Notably, this period also coincides with an enhancement in $\tau$, suggesting that in addition to radiative losses, other channels of energy transport may be contributing to the temperature decrease.

The above conclusions derived from the neutrino energy spectra are consistent with the description in \citep{Janka:2006fh,Choi:2025igp}, demonstrating the effectiveness of our parametrization of the neutrino spectral form. This suggests that the shape parameters of supernova neutrino spectra, derived from our toy model, can serve as one of the indicators in multi-messenger astronomy. As they are more focused on describing the physical processes associated with neutrino emission and the thermodynamic conditions during the explosion, these parameters may become powerful tools for extracting information about the explosion mechanism and nucleosynthesis.

\begin{figure*}[h]
    \centering 

    \includegraphics[width=0.48\textwidth]{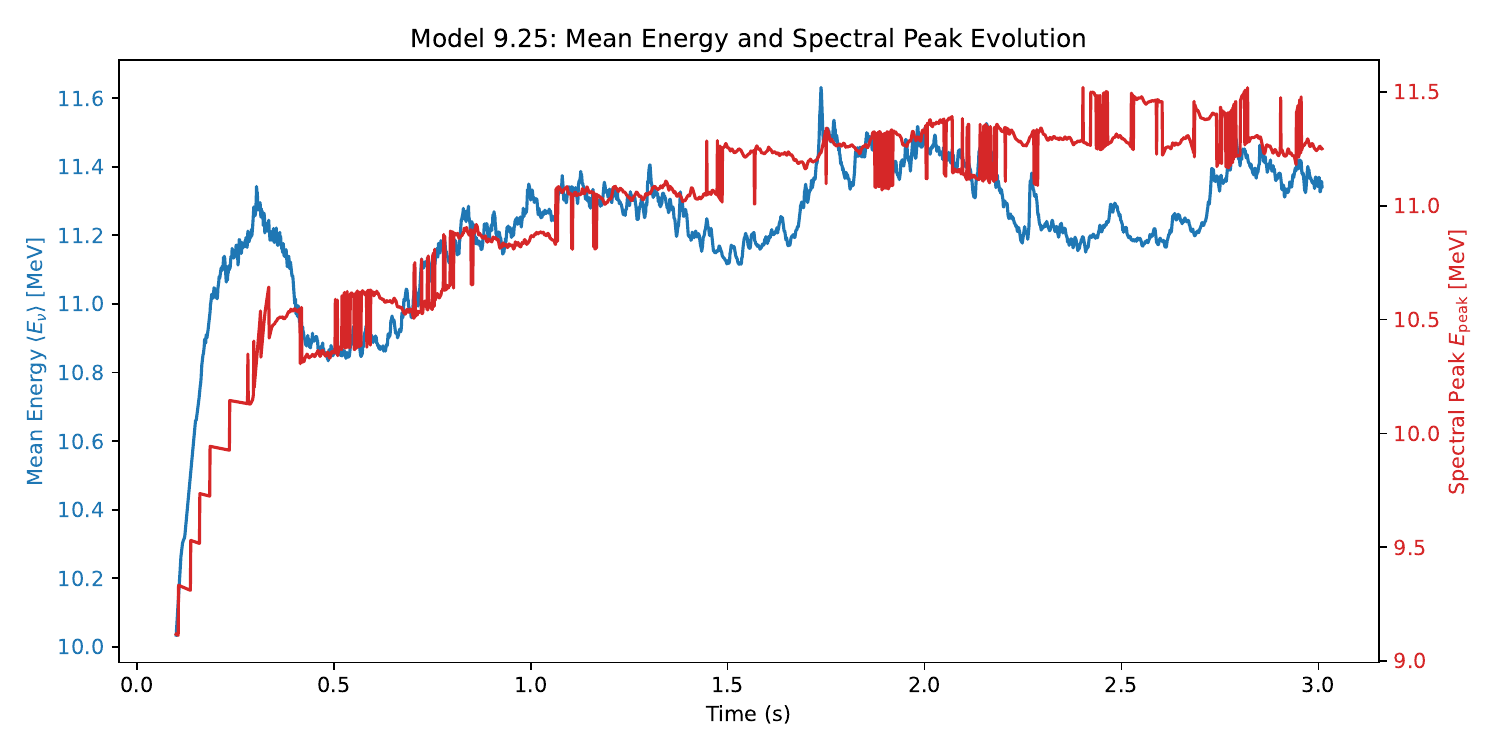}
    \hfill 
    \includegraphics[width=0.48\textwidth]{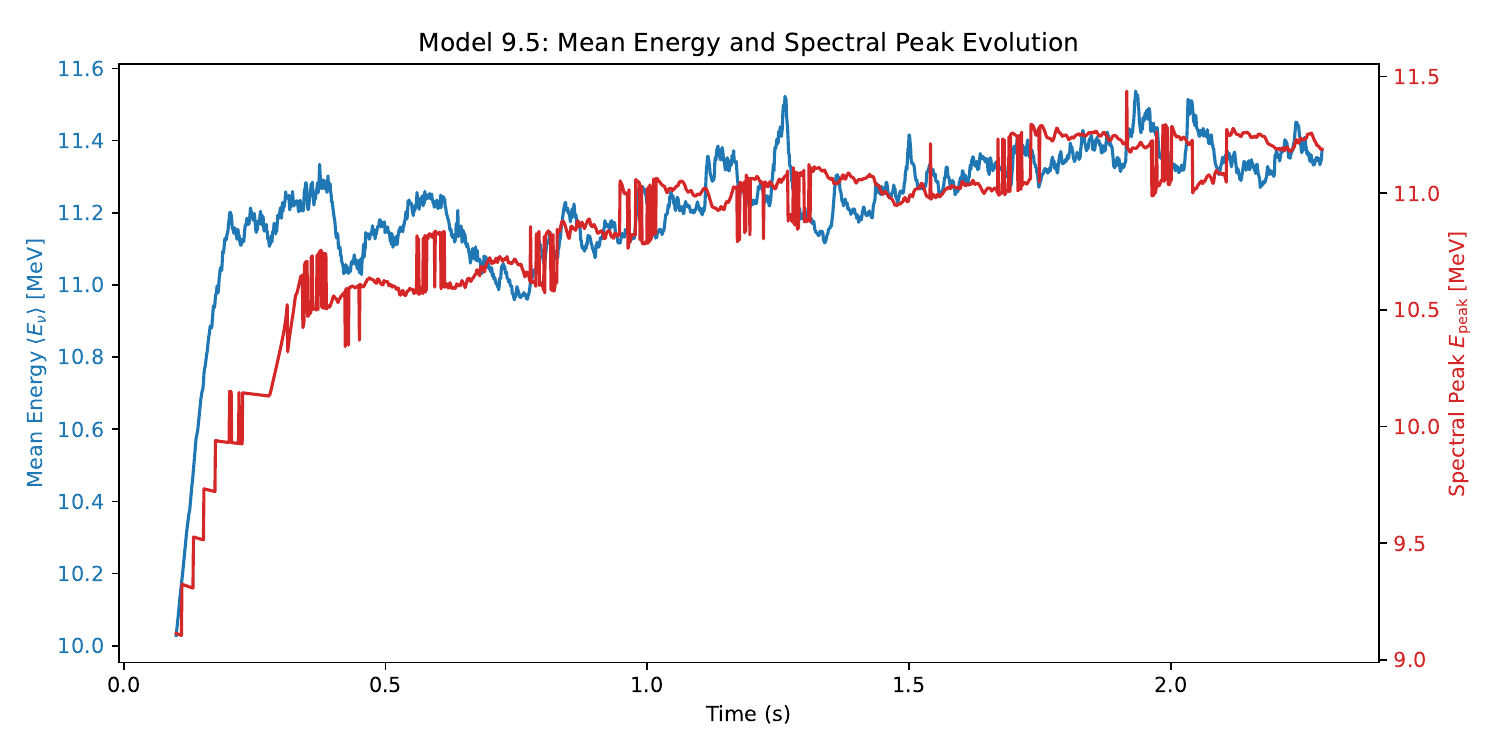}

    \vspace{5mm} 

    \includegraphics[width=0.48\textwidth]{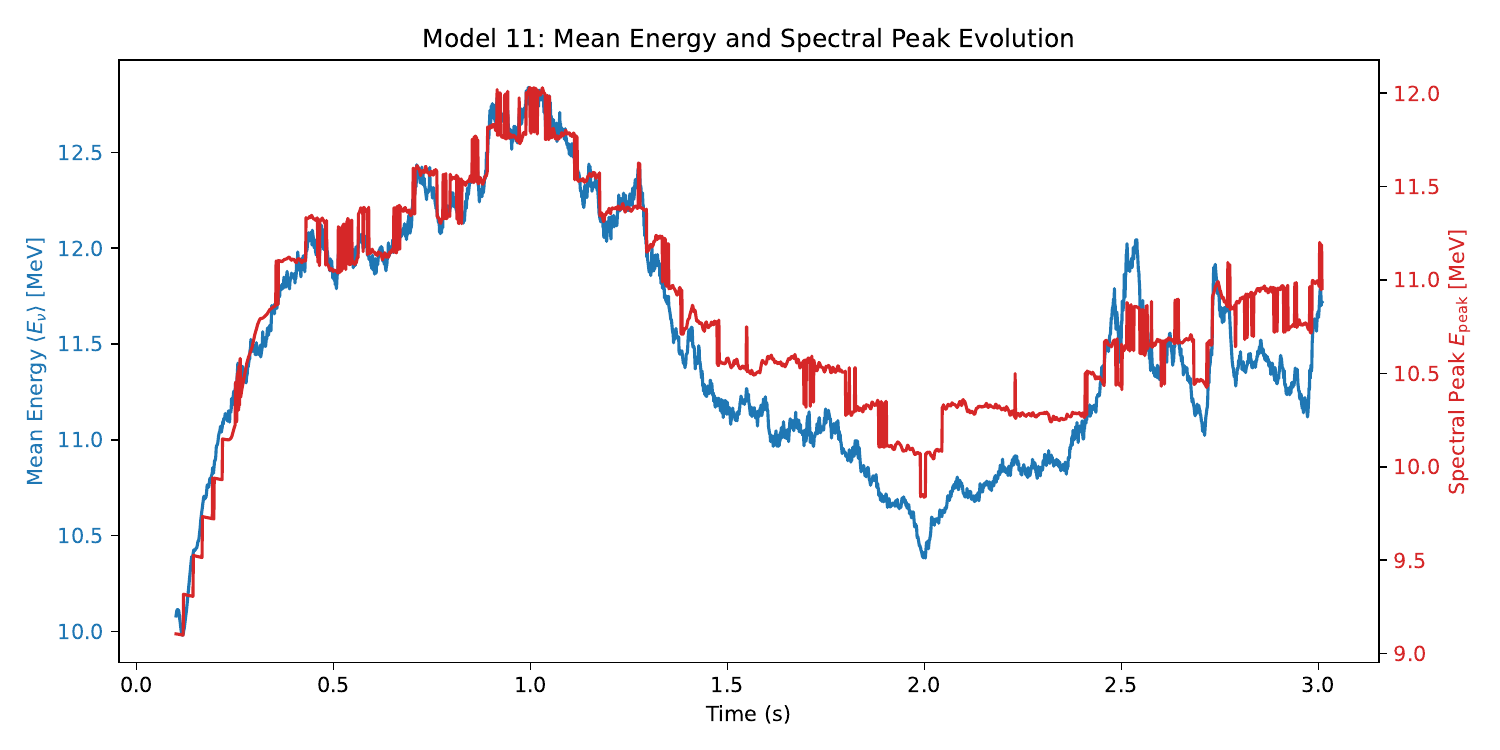}
    \hfill 
\includegraphics[width=0.48\textwidth]{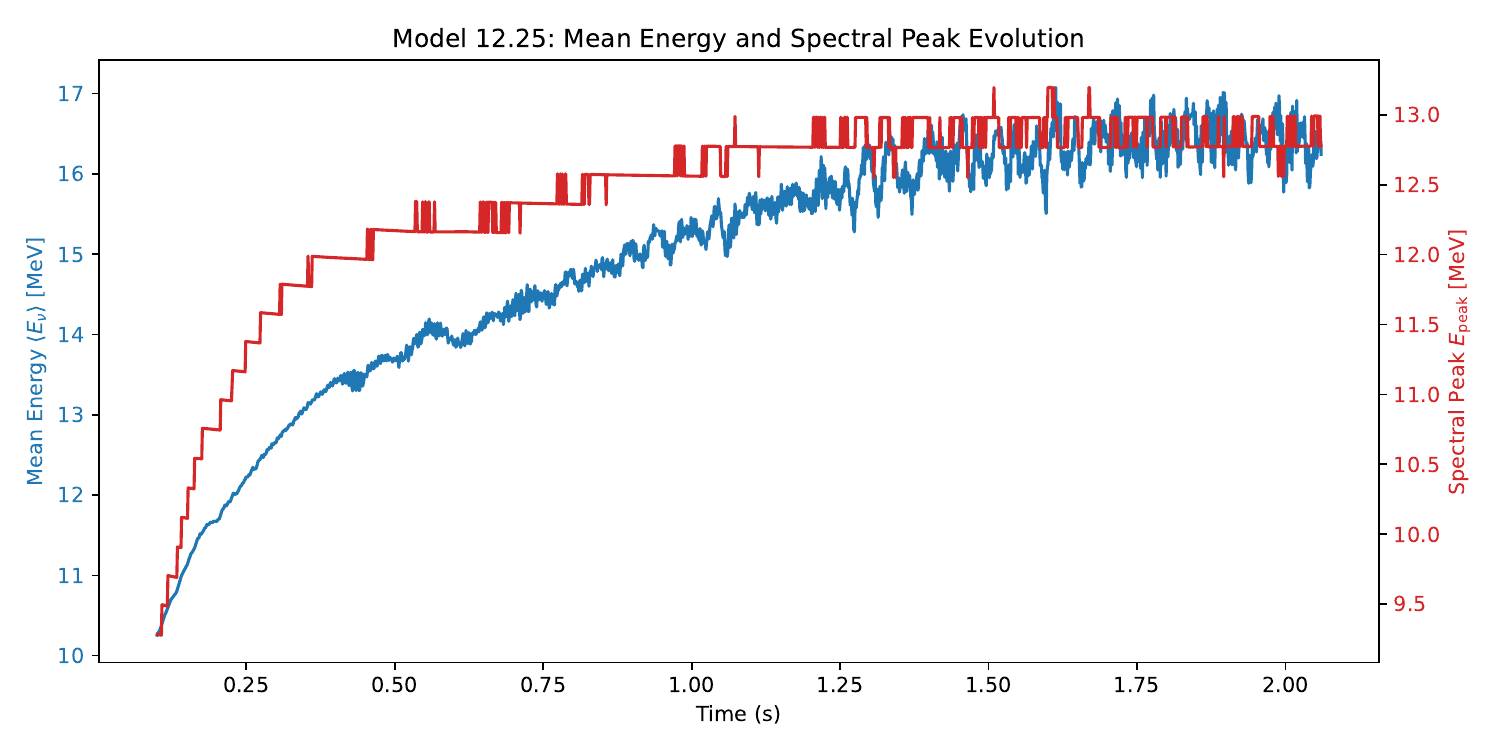}

    \vspace{5mm} 

    \includegraphics[width=0.48\textwidth]{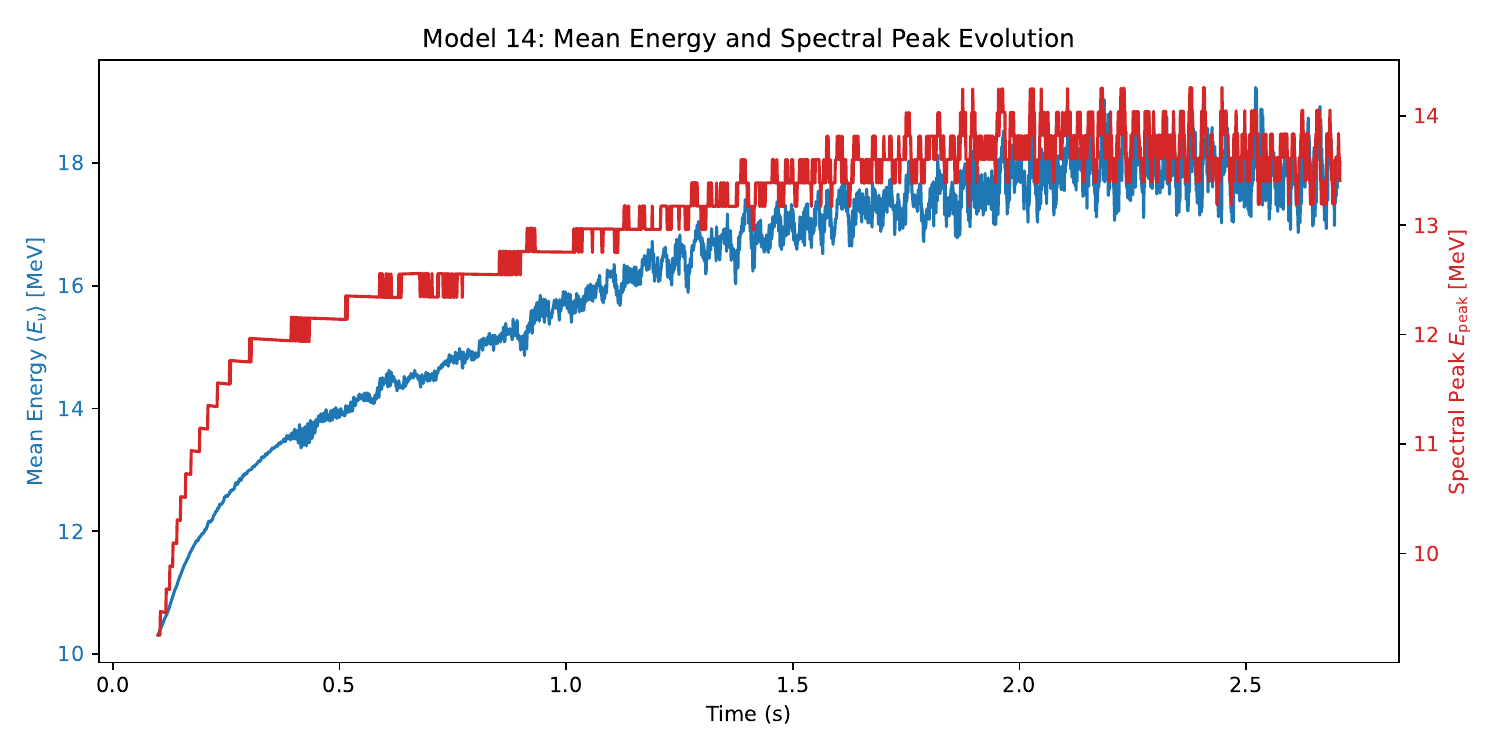}
    \hfill 
\includegraphics[width=0.48\textwidth]{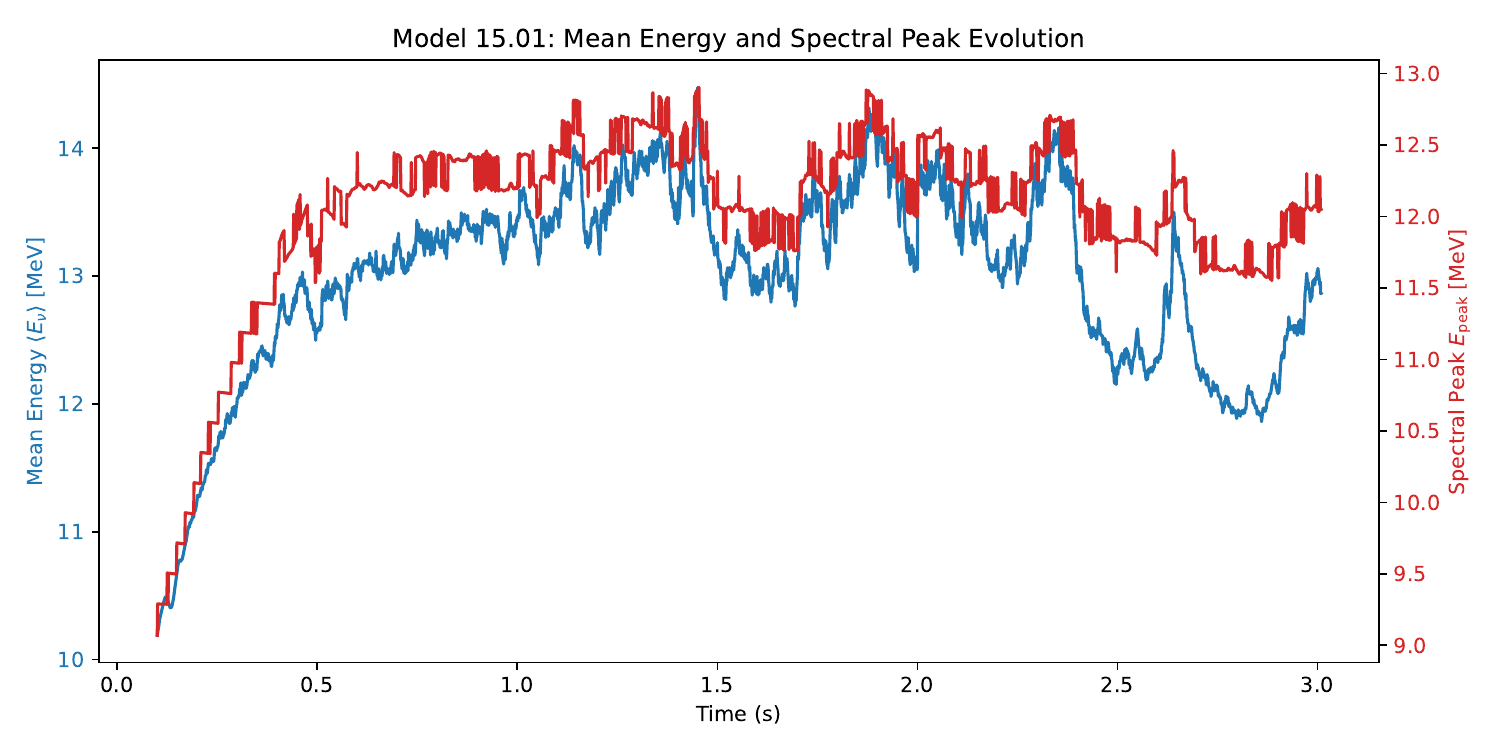}
    \hfill 

    \caption{The evolution of the average neutrino energy and peak energy of the spectrum in different progenitor star mass models.}
    \label{fig:all}
\end{figure*}

\begin{figure*}[h]
    \centering 

    \includegraphics[width=0.48\textwidth]{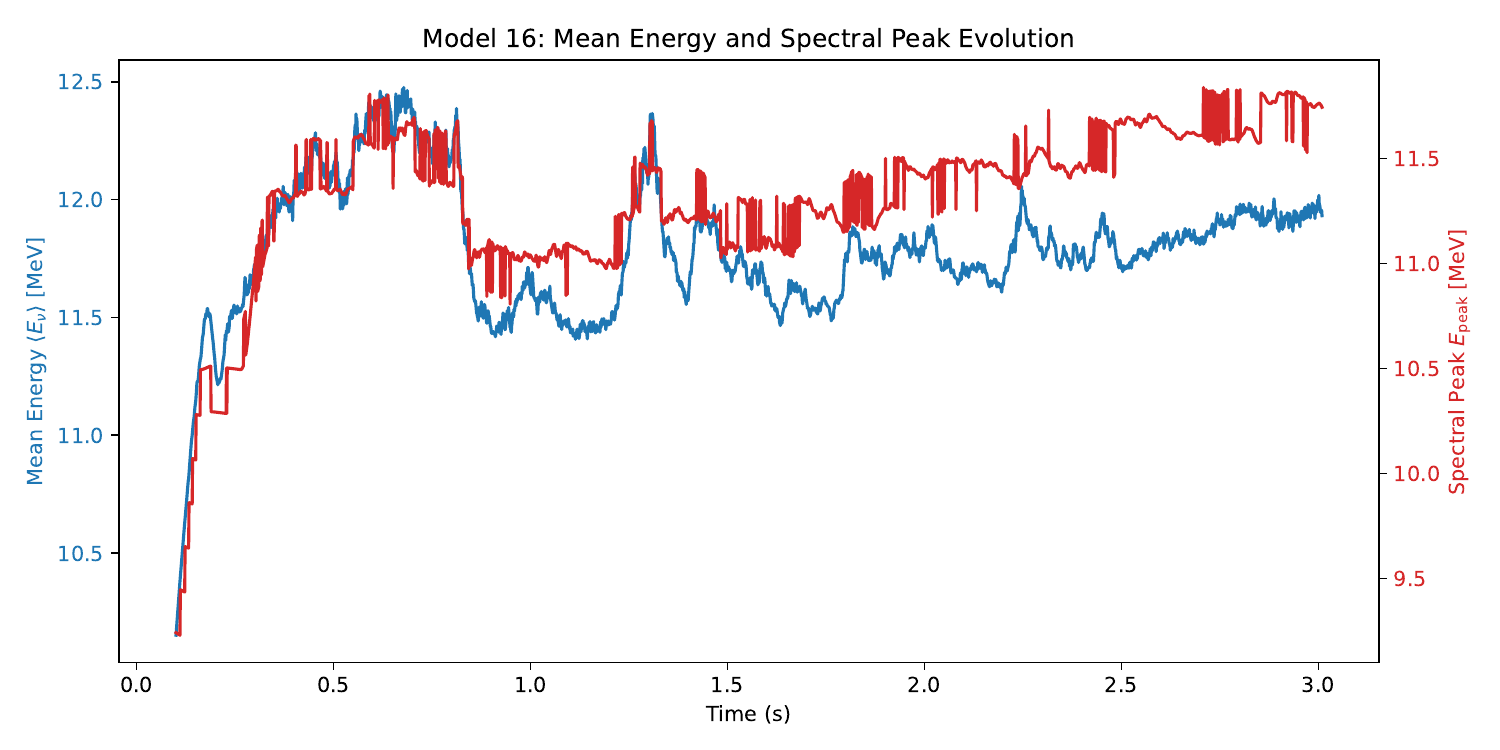}
    \hfill 
    \includegraphics[width=0.48\textwidth]{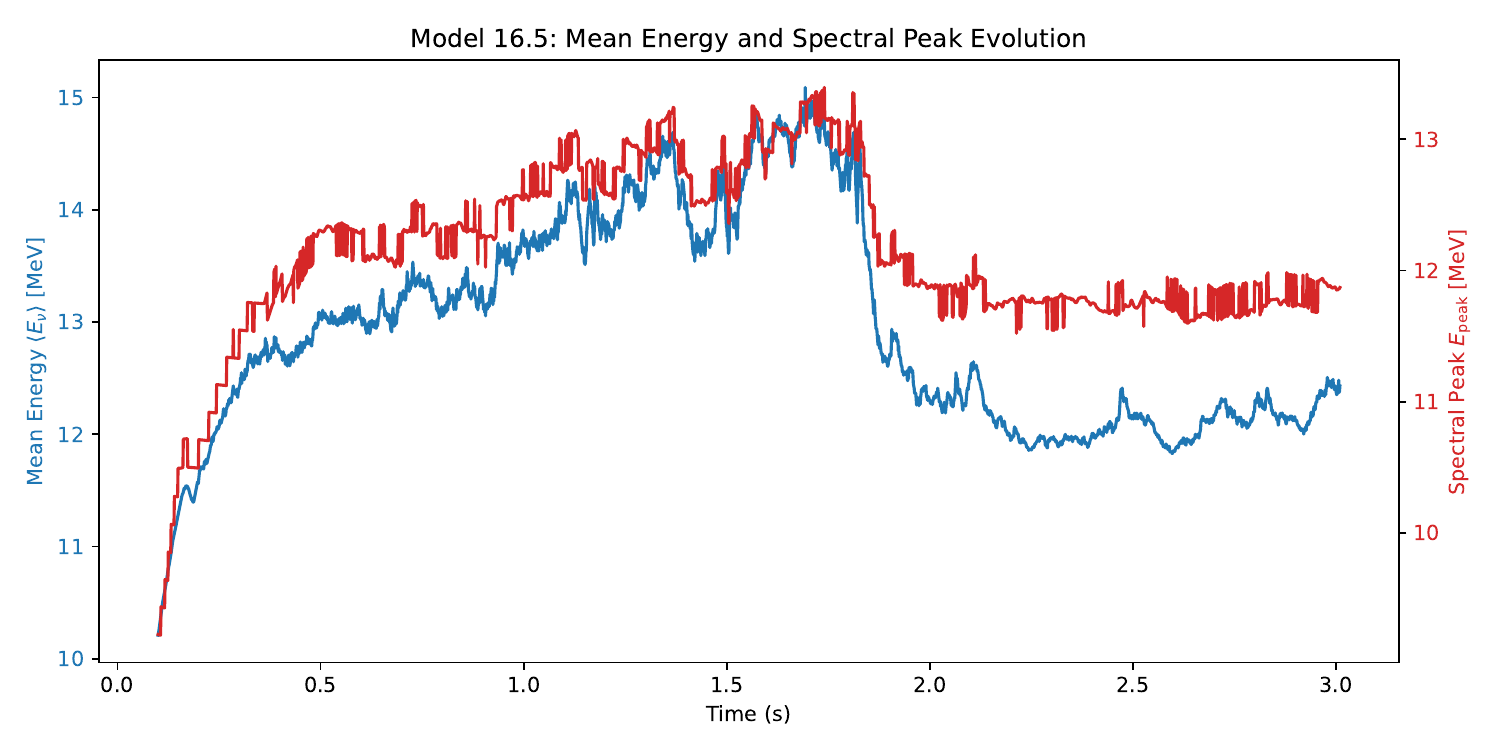}

    \vspace{5mm} 

    \includegraphics[width=0.48\textwidth]{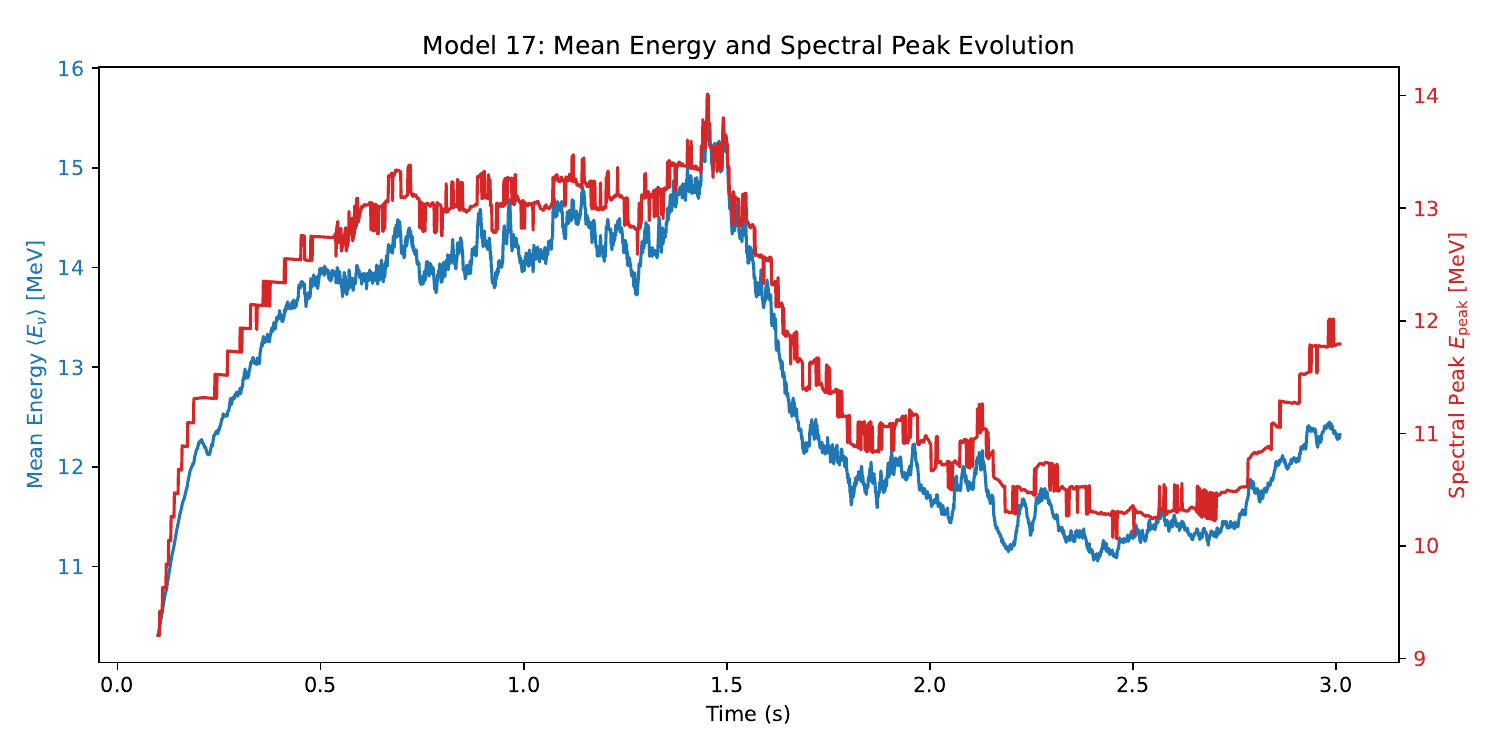}
    \hfill 
\includegraphics[width=0.48\textwidth]{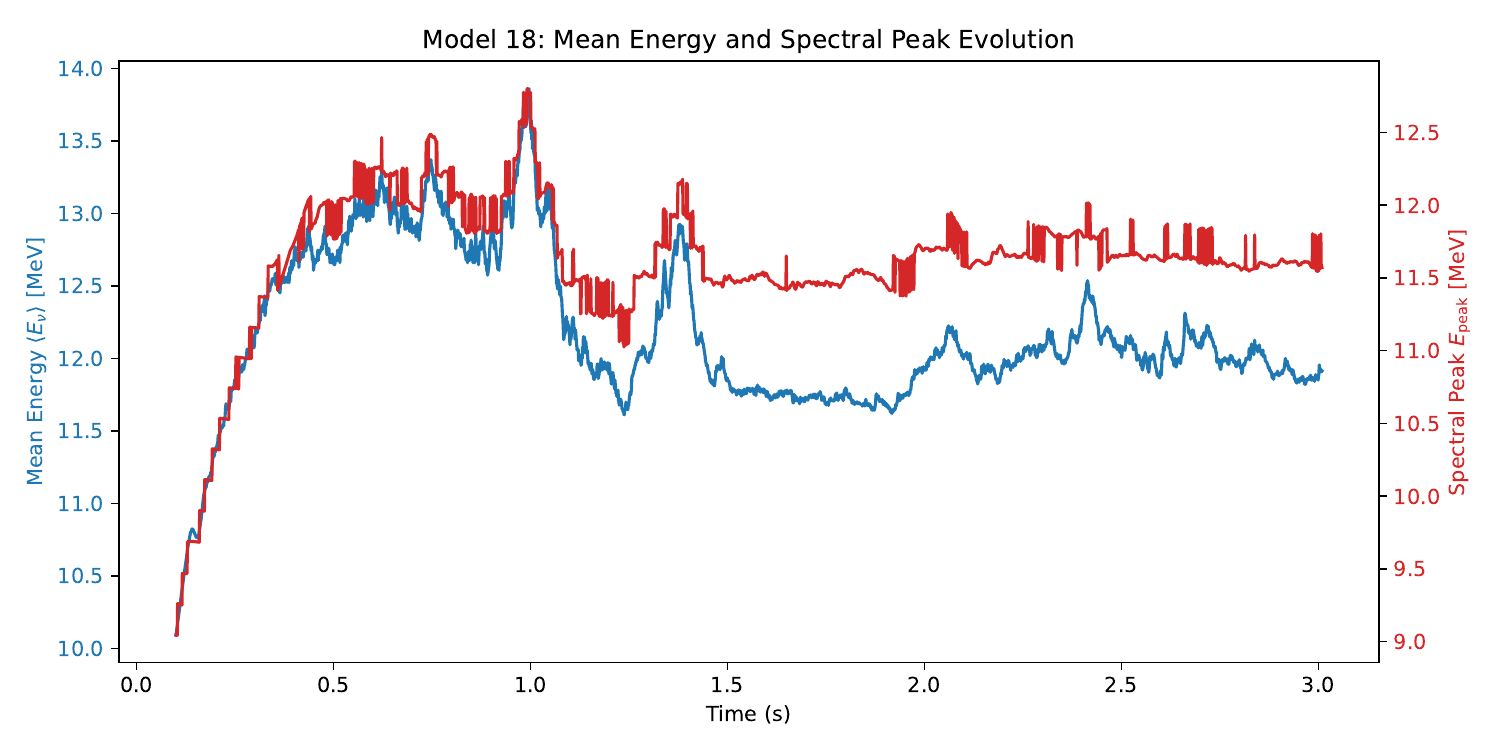}

    \vspace{5mm} 

    \includegraphics[width=0.48\textwidth]{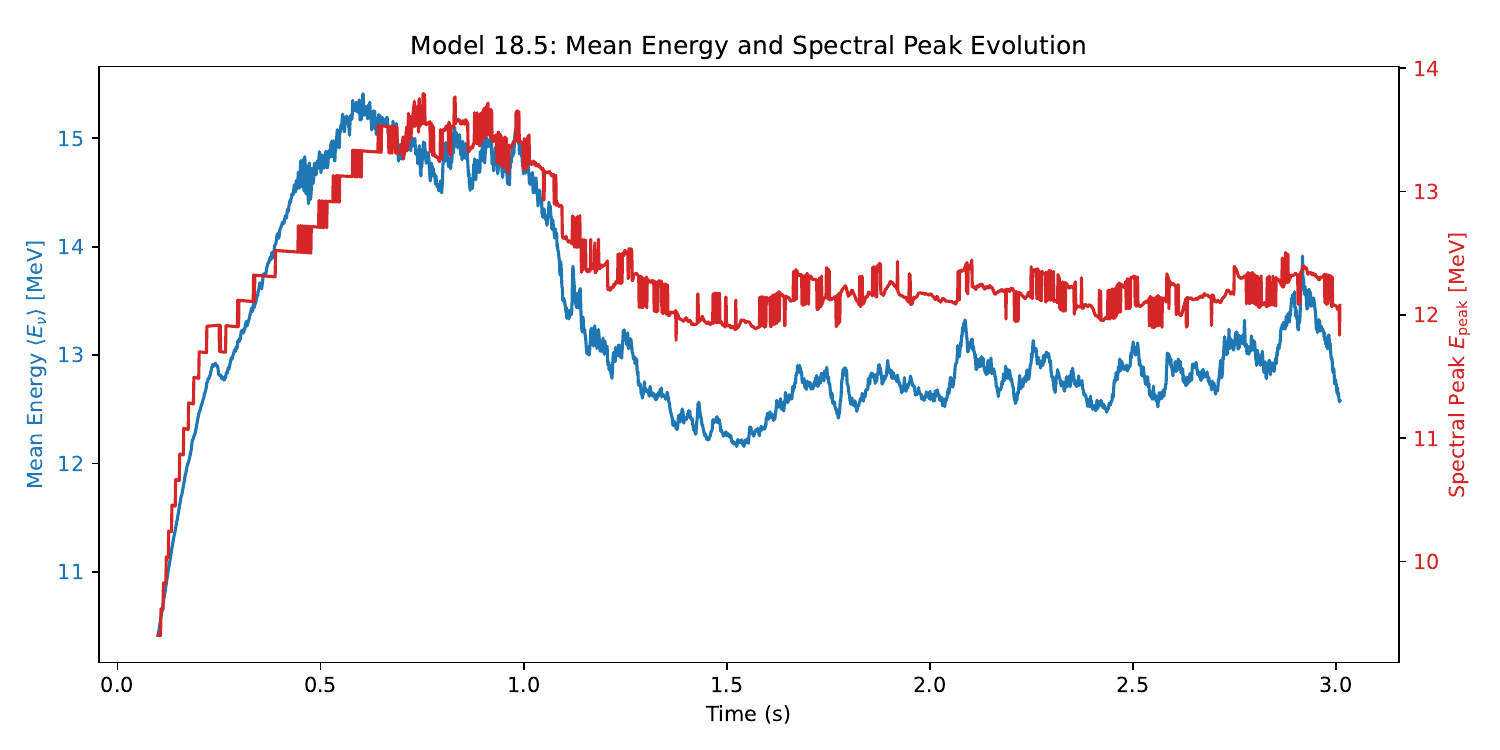}
    \hfill 
\includegraphics[width=0.48\textwidth]{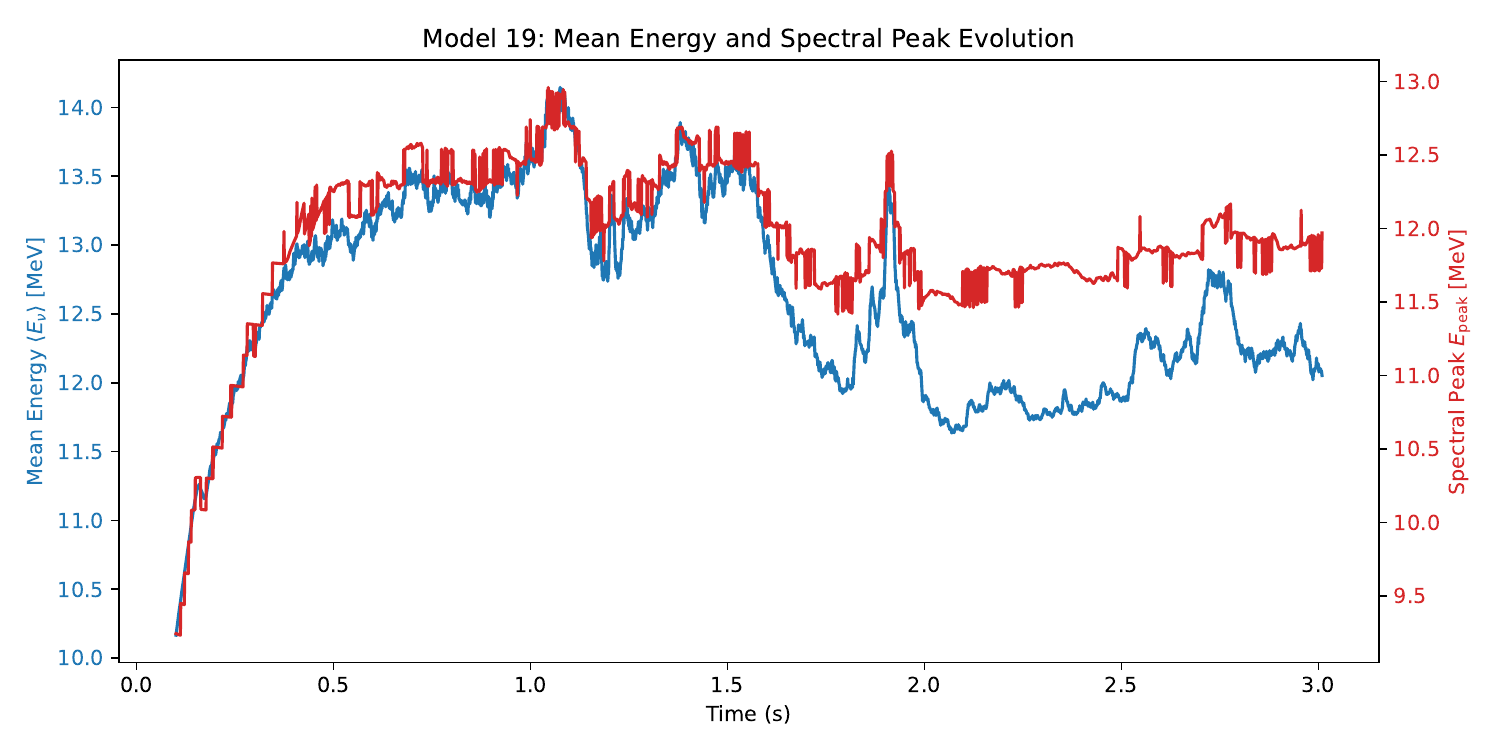}
    \hfill 

    \caption{The evolution of the average neutrino energy and peak energy of the spectrum in different progenitor star mass models.}
    \label{fig:all}
\end{figure*}

\begin{figure*}[h]
    \centering 

    \includegraphics[width=0.48\textwidth]{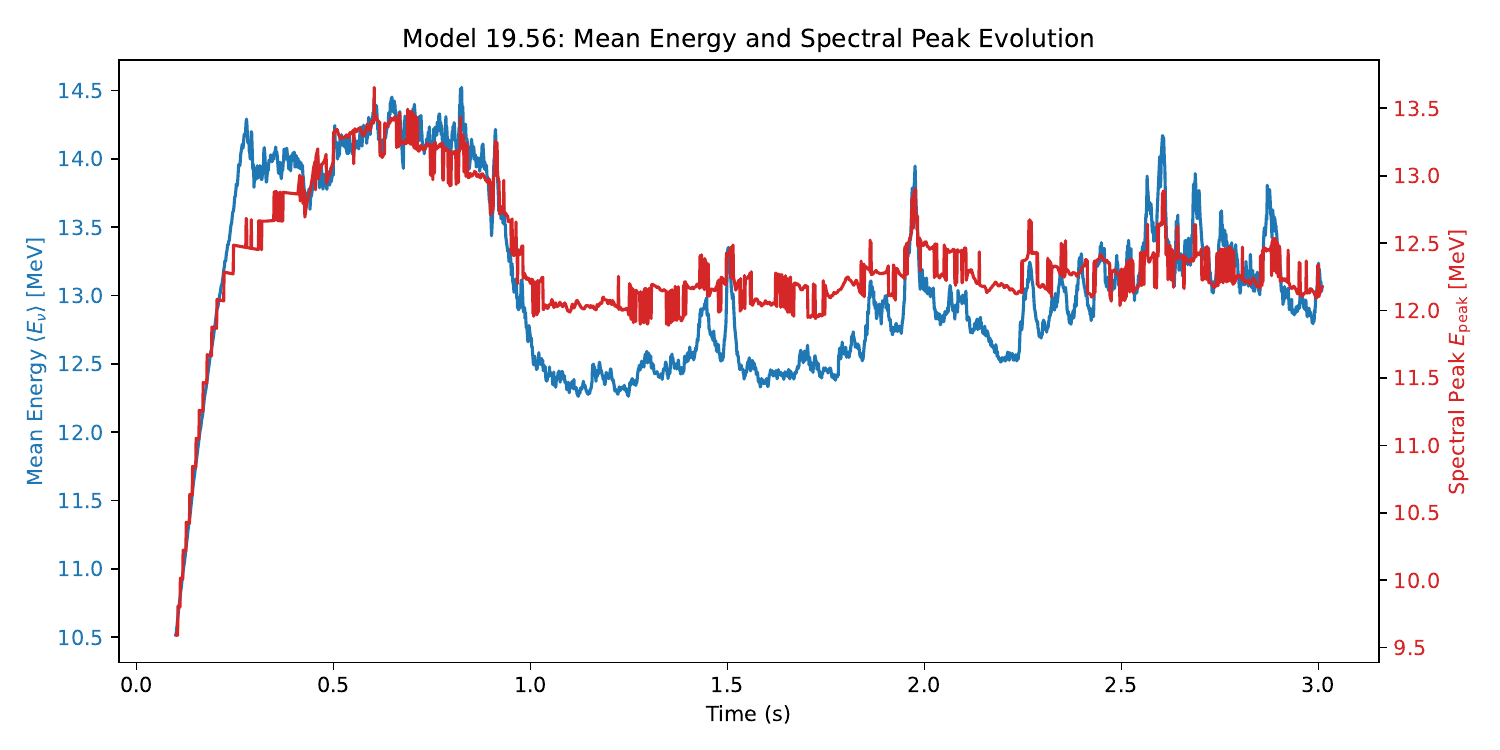}
    \hfill 
    \includegraphics[width=0.48\textwidth]{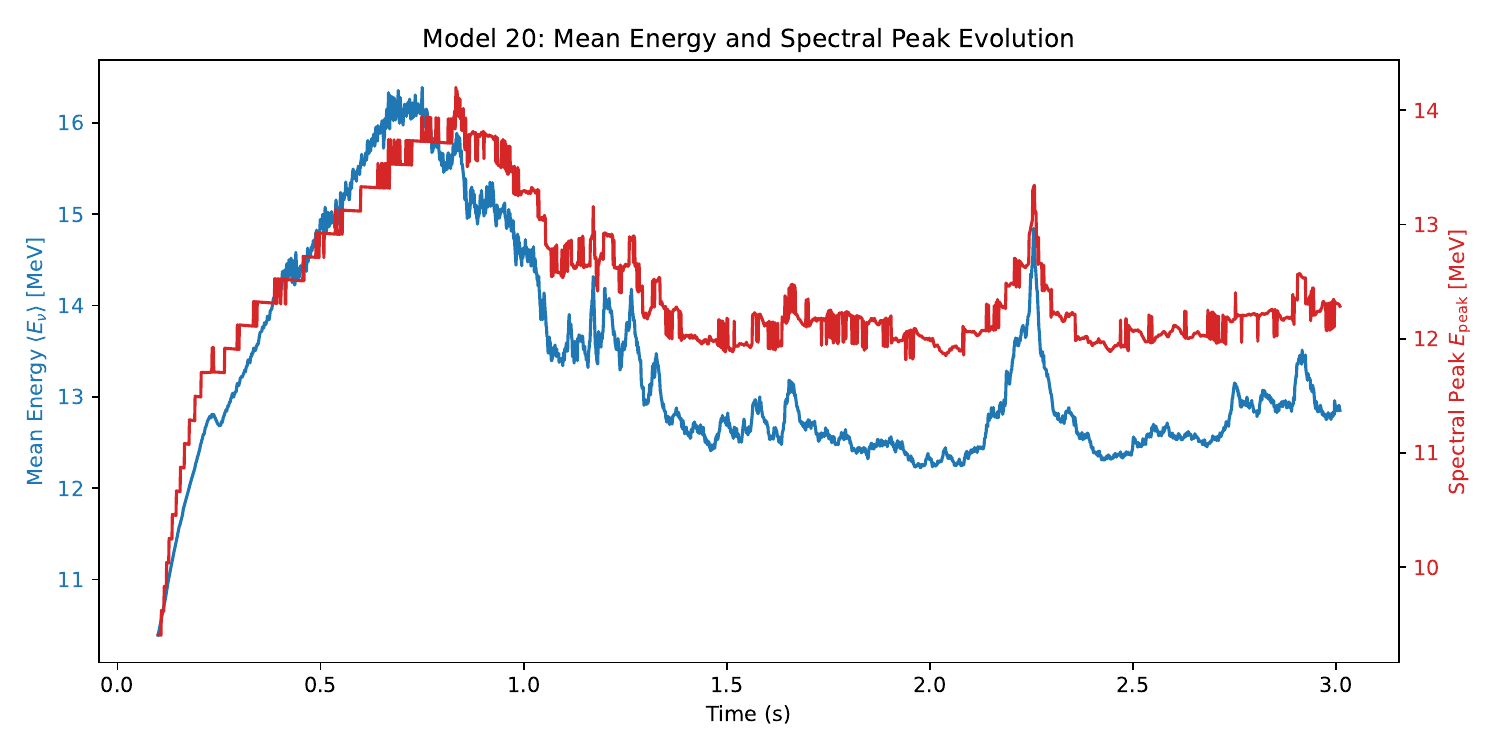}

    \vspace{5mm} 

    \includegraphics[width=0.48\textwidth]{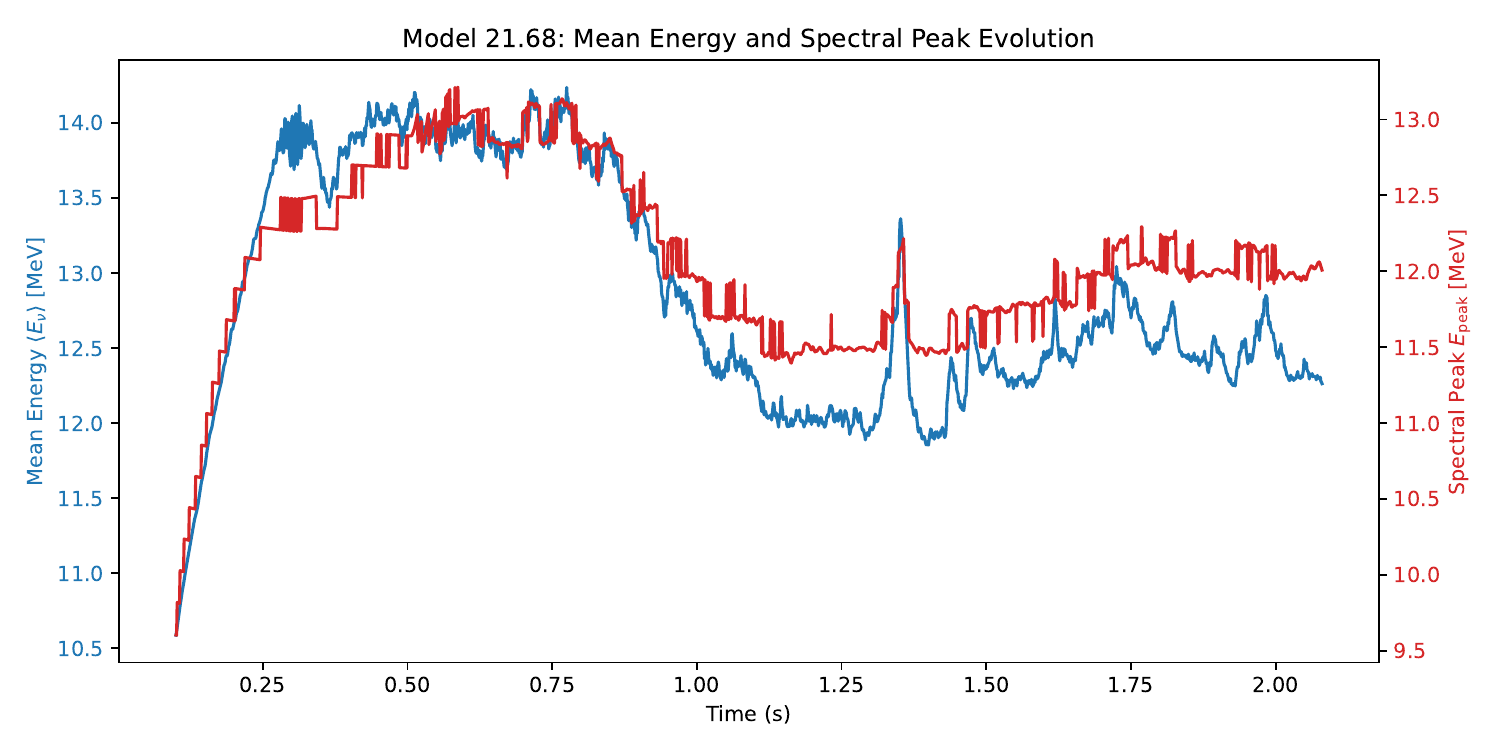}
    \hfill 
\includegraphics[width=0.48\textwidth]{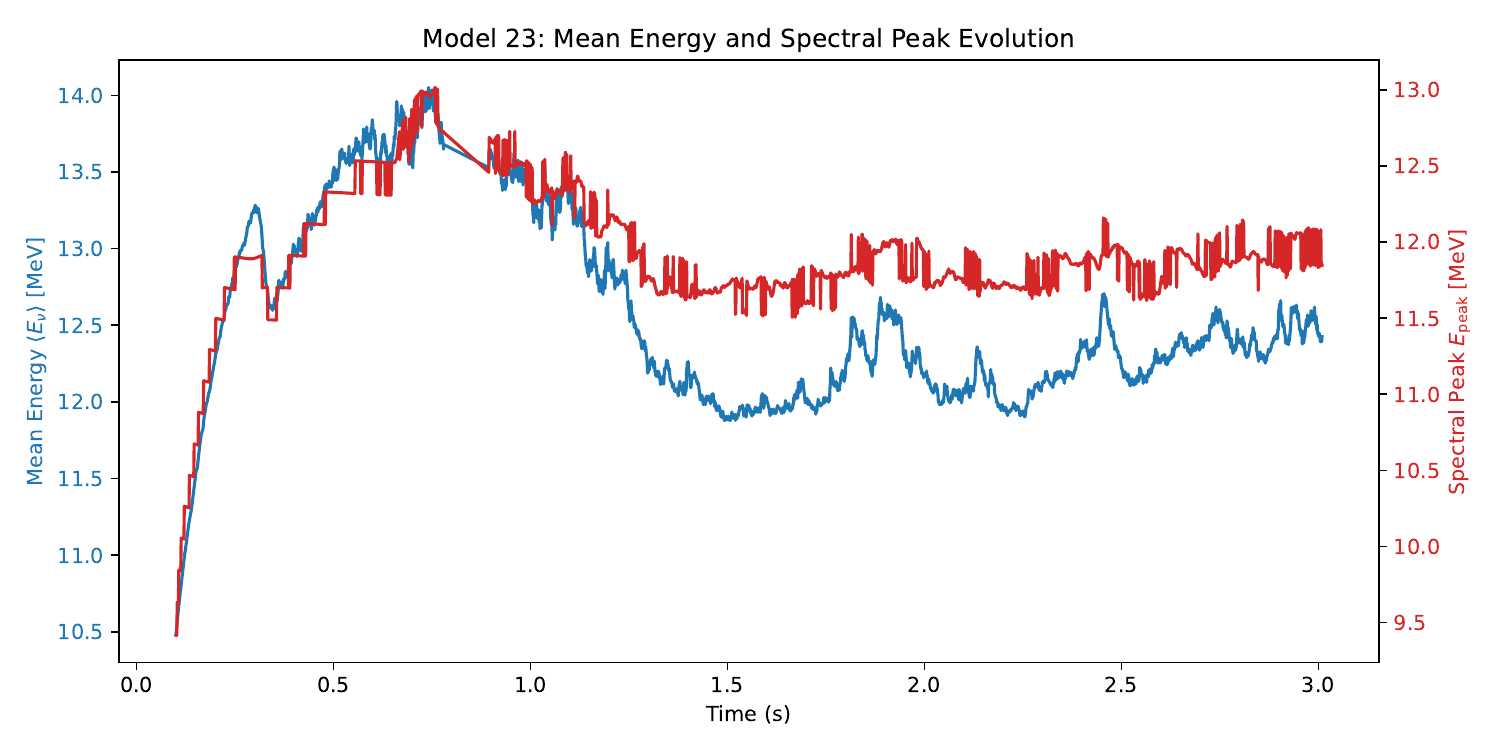}

    \vspace{5mm} 

    \includegraphics[width=0.48\textwidth]{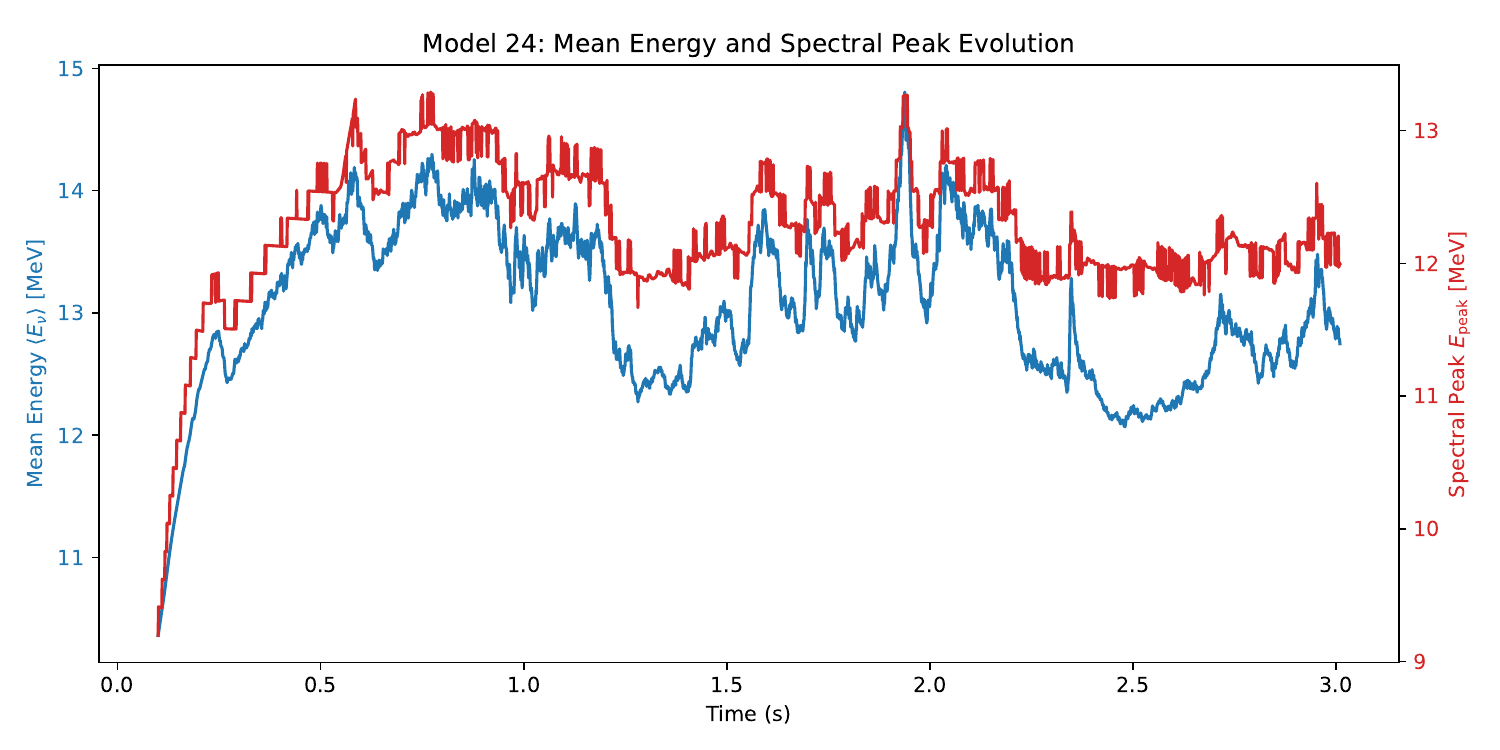}
    \hfill 
\includegraphics[width=0.48\textwidth]{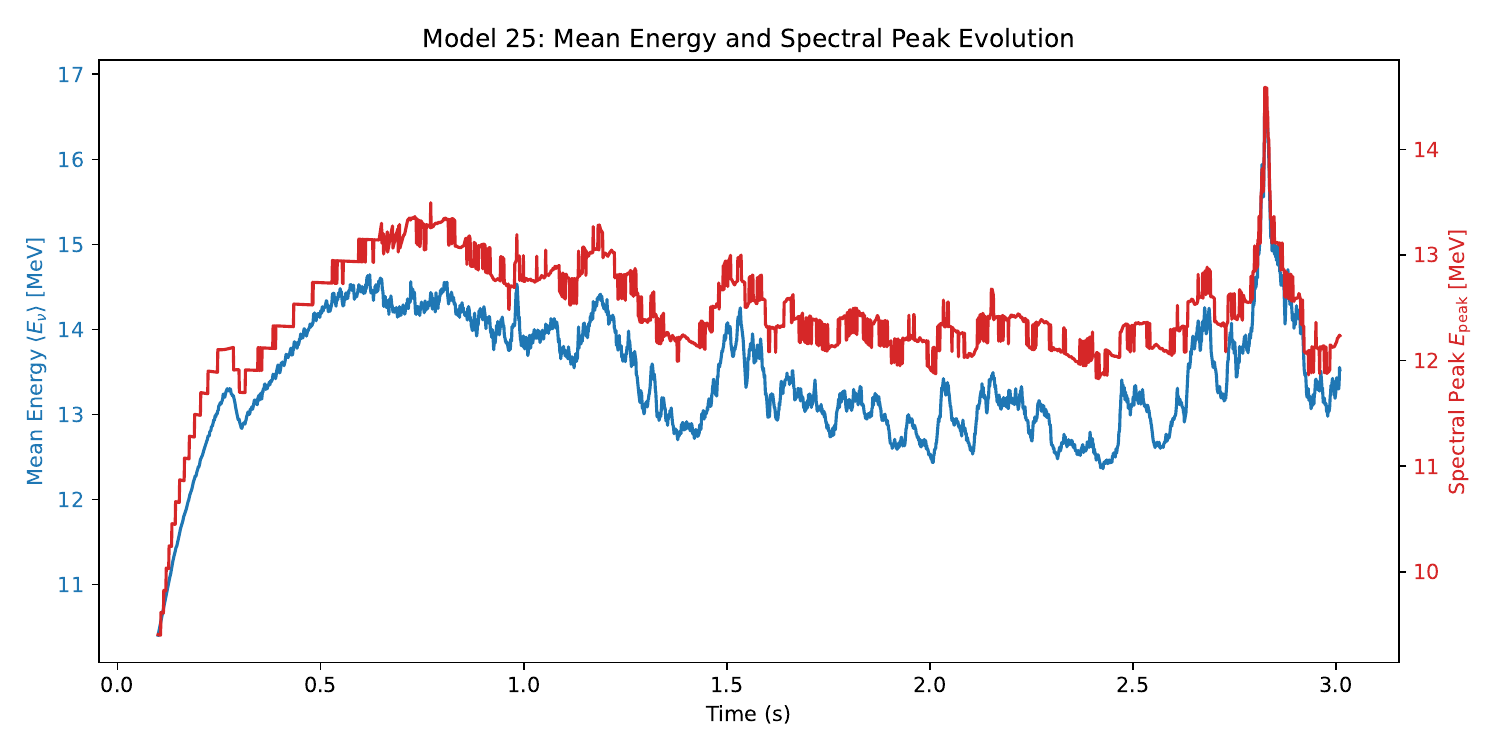}
    \hfill 

    \caption{The evolution of the average neutrino energy and peak energy of the spectrum in different progenitor star mass models.}
    \label{fig:all}
\end{figure*}

\begin{figure*}[h]
    \centering 

    \includegraphics[width=0.48\textwidth]{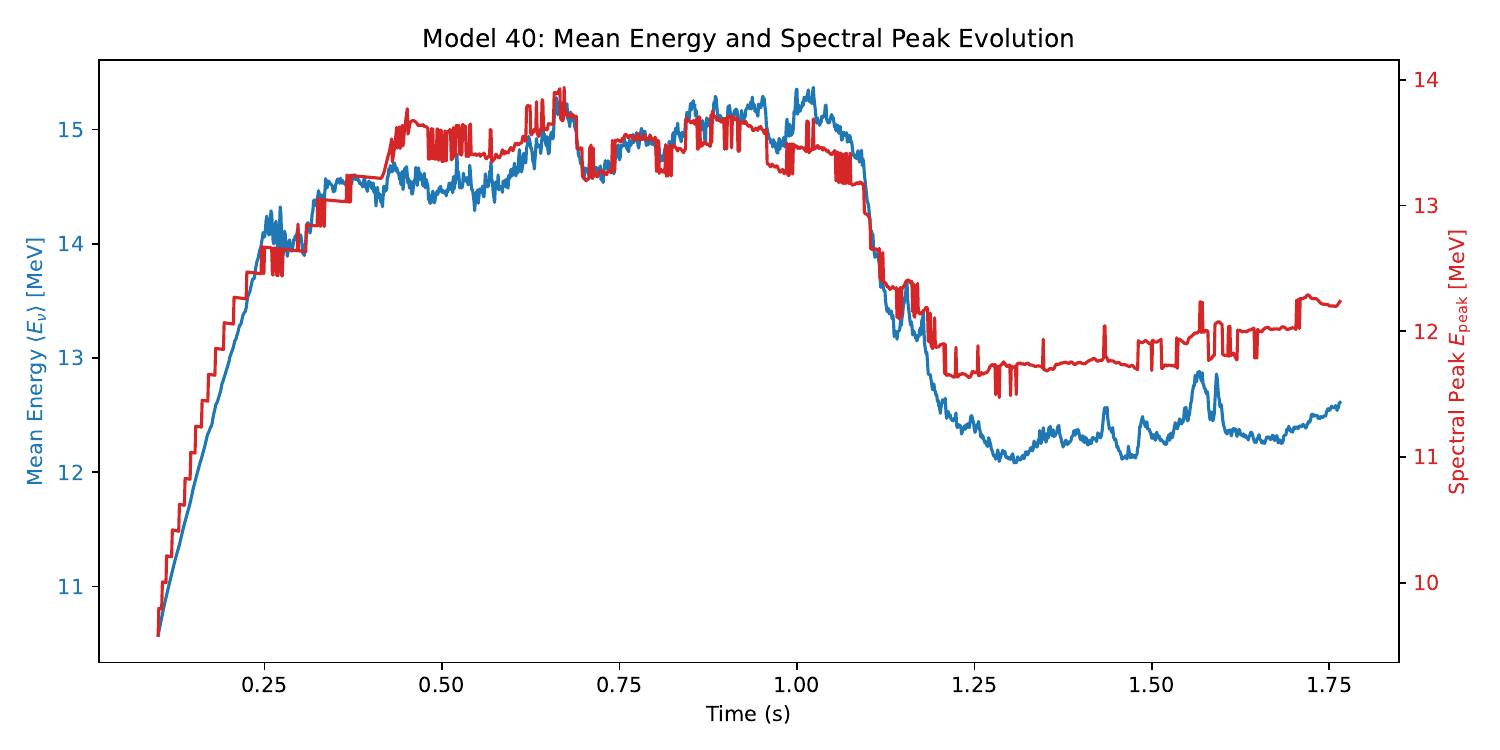}
    \hfill 
    \includegraphics[width=0.48\textwidth]{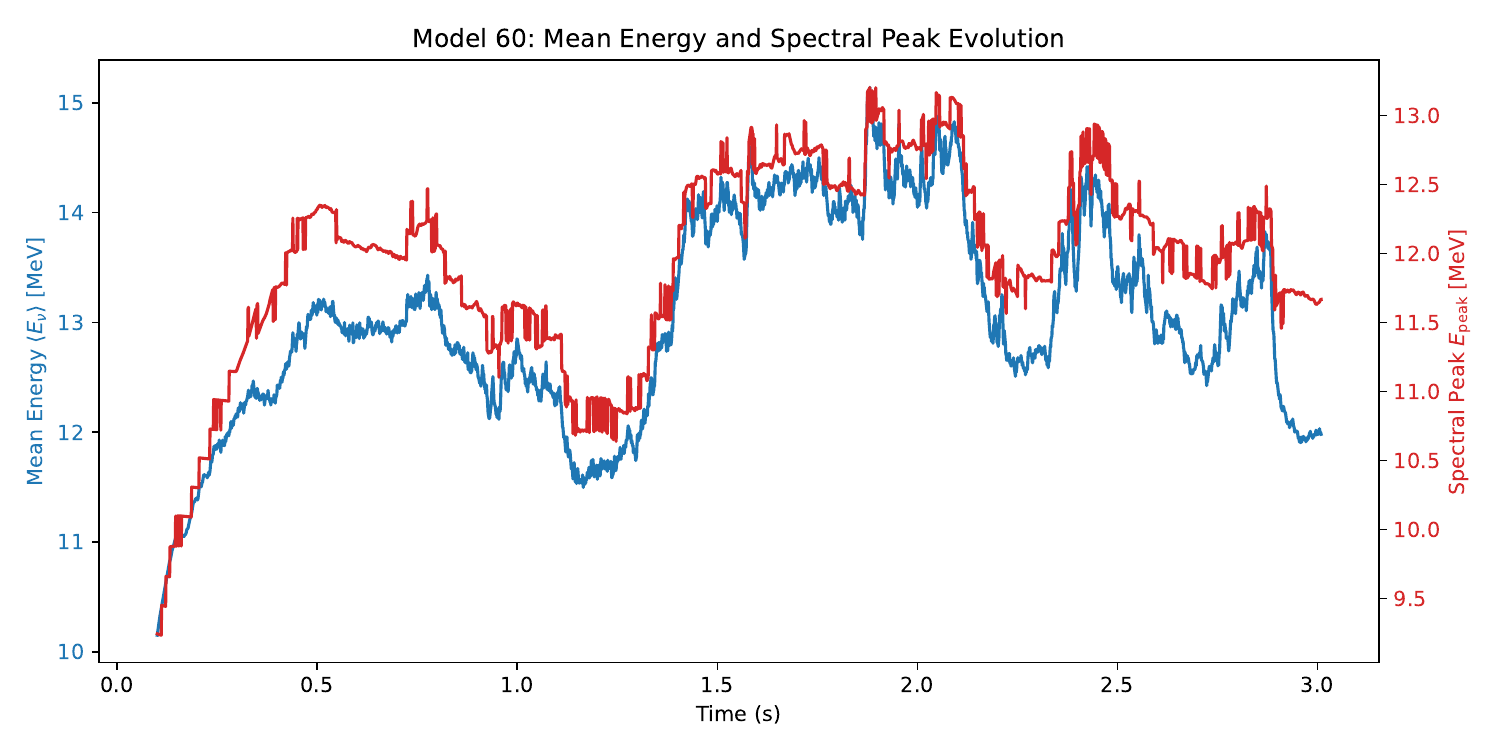}

    \vspace{5mm} 

    \includegraphics[width=0.48\textwidth]{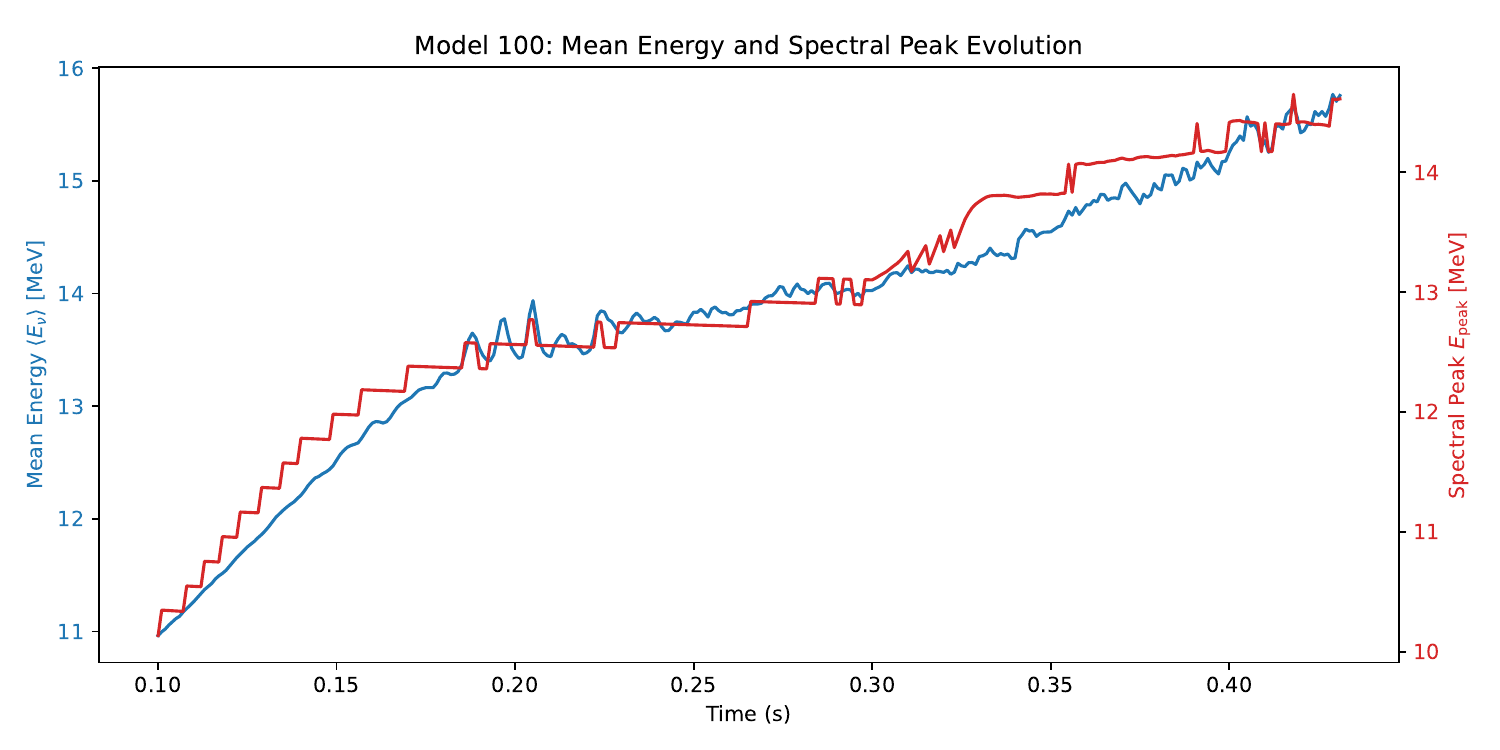}
    \hfill 

    \caption{The evolution of the average neutrino energy and peak energy of the spectrum in different progenitor star mass models.}
    \label{fig:all}
\end{figure*}



\clearpage
\newpage




\bibliography{sample701}{}
\bibliographystyle{aasjournalv7}
\end{document}